\documentclass[a4paper,11pt]{article}
\pdfoutput=1
\usepackage{jheppub}
\usepackage[usenames,dvipsnames]{xcolor}
\usepackage{slashed}
\usepackage[T1]{fontenc}
\usepackage[utf8]{inputenc}
\usepackage{amsmath}
\usepackage{amssymb}
\usepackage{graphicx}
\usepackage{diagbox}
\usepackage{color}
\usepackage{booktabs}
\usepackage{bm}
\usepackage{bbm}
\definecolor{darkblue}{cmyk}{0.9,0.9,0,0}
\definecolor{darkgreen}{rgb}{0,0.55,0}
\usepackage{minitoc}
\usepackage[normalem]{ulem} 
\usepackage{enumitem}
\usepackage{tikz-cd}
\usepackage{float}

\newcolumntype{C}[1]{>{\centering\arraybackslash}m{#1}}
\makeatletter

\newcommand{\comAI}[1]{(*{\textbf{\textcolor{orange}{#1}}}*)}

\newcommand{\comment}[1]{}

\newcommand{\beq}{\begin{equation}}
\newcommand{\eeq}{\end{equation}}
\newcommand{\beqq}{\begin{equation*}}
\newcommand{\eeqq}{\end{equation*}}
\newcommand\beqa{\begin{eqnarray}}
\newcommand\eeqa{\end{eqnarray}}
\newcommand\beqaa{\begin{eqnarray*}}
	\newcommand\eeqaa{\end{eqnarray*}}
\newcommand\bea{\begin{array}}
	\newcommand\eea{\end{array}}

\def\XXint#1#2#3{{\setbox0=\hbox{$#1{#2#3}{\int}$ }
		\vcenter{\hbox{$#2#3$ }}\kern-.5\wd0}}

\def\XXint#1#2#3{{\setbox0=\hbox{$#1{#2#3}{\int}$}
		\vcenter{\hbox{$#2#3$}}\kern-.5\wd0}}

\newcommand{\nn}{\nonumber}

\newcommand{\neqa}{\nonumber\end{eqnarray}}
\newcommand{\la}[1]{\label{#1}}

\newcommand{\eq}[1]{(\ref{#1})}

\def\tr{{\rm tr~}}

\newcommand{\hs}{\frac{\sqrt{3}}{2}}
\renewcommand{\d}{\partial}

\newcommand{\<}{{\langle}}
\renewcommand{\>}{{\rangle}}

\newcommand{\cA}{{\cal A}}

\newcommand{\cC}{{\cal C}}
\newcommand{\cD}{{\cal D}}

\newcommand{\re}{\relax{\rm I\kern-.18em R}}

\renewcommand{\sp}{p\hspace{-.40em}/}

\def\su2{{SU(2)}}

\def\eps{{\epsilon}}

\def\[{\left[}
\def\]{\right]}

\def\s{\sigma}

\def\b{\Bethe}

\def\({\left(}
\def\){\right)}
\def\[{\left[}
\def\]{\right]}

\def\<{\langle}
\def\>{\rangle}

\def\cO{{\cal O}}

\def\cC{{\cal C}}
\def\cW{{\cal W}}
\def\cP{{\cal P}}

\def\s*{\ *_{\!\!\!\!\!\!\!\!\!\,_{\,_\text{\scriptsize{sym}}}}}
\def\hs*{\ \hat{*}_{\!\!\!\!\!\!\!\!\!\,_{\,_\text{\scriptsize{sym}}}}}
\def\d{\partial}

\def\i2{\frac{i}{2}}

\def\bQ{{\bf Q}}
\def\bP{{\bf P}}

\def\bq{\mathbbm{q}}

\def\cJ{{\cal J}}
\def\spi{\relax{\rm \pi\kern-0.5em /}}
\def\sA{\relax{\rm A\kern-0.5em /}}
\def\sp{\relax{\rm p\kern-0.5em /}}
\def\sd{\relax{\rm \d\kern-0.5em /}}
\def\sk{\relax{\rm k\kern-0.5em /}}
\def\sn{\relax{\rm n\kern-0.5em /}}
\def\sl{\relax{\rm l\kern-0.5em /}}
\def\sP{\relax{\rm P\kern-0.7em /}}
\def\sBethe{\relax{\rm \Bethe\kern-0.5em /}}
\def\cN{{\cal N}}

\def\One{\mathbb{I}}

\def\be#1\ee{\begin{equation}\begin{aligned}
#1
\end{aligned}
\end{equation}}

\newcommand{\ii}{\mathrm{i}}
\newcommand{\dd}{\mathrm{d}}

\numberwithin{equation}{section}
\usepackage{varioref}
\usepackage{makeidx}
\usepackage{cleveref}
\makeindex

\title{Quark Anti-Quark Fusion and Walking RG Flows}
\author[a]{Filipp Chernikov}
\author[a,b]{Nikolay Gromov}
\author[c]{Amit Sever}%
\affiliation[a]{%
Mathematics Department, King's College London,
The Strand, London WC2R 2LS, UK
}%
\affiliation[b]{St.Petersburg INP, Gatchina, 188 300, St.Petersburg,
Russia}
\affiliation[c]{
School of Physics and Astronomy, Tel Aviv University, Ramat Aviv 69978, Israel}

\emailAdd{filipp.chernikov@kcl.ac.uk}
\emailAdd{nikolay.gromov@kcl.ac.uk}
\emailAdd{amit.sever@tauex.tau.ac.il}
\abstract{
We study the fusion of two conjugate conformal line defects on the sphere. At small separation, their spectrum is governed by a universal Fusion Master Equation. Below a critical coupling, the fused defect has two conformal fixed points; at criticality, they collide and move into the complex plane, producing walking RG behaviour. Although individual energy levels then drift with the UV scale and are scheme dependent, the $SL(2,\mathbb{R})$ Casimir continues to commute with the Hamiltonian below that scale. This organises the spectrum into conformal families and fixes a universal, scheme-independent density of states. We derive this structure in the planar ladder model and obtain an exact finite-coupling description of conjugate $1/2$-BPS Wilson-line fusion in planar ${\cal N}=4$ SYM using the Quantum Spectral Curve. We test our results against perturbation theory and semiclassical string theory.
}

\begin{document} 
\maketitle
\newpage

\section{Introduction and Summary of Results}

Conformal line defects are renormalisation-group fixed points associated with point-like impurities. A prominent example is the external quark in ${\cal N}=4$ SYM theory, described by the $1/2$-BPS Wilson line. Two such defects can be expanded, or ``fused'', into a sum of conformal defects. In this paper, we solve the problem of quark--antiquark fusion in the ’t Hooft large-$N$ limit.

Our central result is that above a critical coupling, whose existence was first suggested in \cite{Klebanov:2006jj}, the fused defect ceases to flow to a conformal fixed point. Instead, the system undergoes a \emph{walking} RG flow. We show that this walking regime nevertheless exhibits a universal spectral structure governed by an emergent $SL(2,\mathbb{R})$ Casimir equation and a Fusion Master Equation, which we derive and solve exactly at all values of the coupling. We begin by drawing an analogy with local operators.

By now, local operators in conformal field theories (CFTs) are well understood.
Their correlators are strongly constrained by the operator product expansion (OPE), which expresses the product of two nearby operators as a sum over local operators
\beq
\cO_1(x)\cO_2(0)=\bigoplus_k c_{12k}(x)\cO_k(0)\,.
\eeq
It can then be used recursively to reduce higher-point correlation functions to lower-point ones.

The operator content of a CFT also includes extended, or defect, operators. Among the simplest examples are line defects, which will be the focus of this paper. Despite their importance, our understanding of such operators remains comparatively limited; see \cite{Andrei:2018die} for a review. Much like quantum field theories themselves, line defects undergo defect renormalisation-group (DRG) flows, whose fixed points define the conformal defects introduced above. These play a role for defect theories analogous to that of CFTs in ordinary quantum field theory. In particular, conformal line defects preserve an $SL(2,\mathbb{R})\times SO(d-1)$ subgroup of the full conformal symmetry.

As for local operators, the product of two conformal line defects can be expanded in terms of conformal line defects. This expansion is known as \emph{fusion}; see, for example, \cite{Bachas:2007td,Bachas:2013nra,Konechny:2015jna,Soderberg:2021kne,
SoderbergRousu:2023zyj,Kravchuk:2024qoh,Diatlyk:2024zkk,Cuomo:2024psk,Diatlyk:2024qpr}. In contrast to the case of local operators, conformal symmetry imposes weaker constraints, leading to a considerably richer structure.

A first complication is that there is no unique way to bring two line defects together. Ideally, one would like to preserve as much symmetry as possible in the process. One natural setup consists of two infinite parallel lines separated by a transverse distance $r$. This configuration preserves translations along the lines -- an $SL(2,\mathbb{R})$ generator with a single fixed point -- as well as an $SO(d-2)$ rotational symmetry. In the limit $r\to0$, the pair of conformal lines can be replaced by a sum over conformal defects,
\beq\la{flatfusion}
\lim_{r\to0}\cD_1(r)\cD_2(0)=\bigoplus_n\left({\cal D}_n(0)\,  e^{-\int d\tau{C_{12n}\over r}}+\text{corrections}\right)\,.
\eeq
Here, $\cD_k(r)$ denotes a straight conformal line defect extending along the Euclidean time direction $\tau$ and localised at transverse distance $r$ from the origin. The coefficients $C_{12n}$ are the fusion Casimirs. They are scheme-independent and admit a natural interpretation as the energy cost of replacing the two defects on the left-hand side with a single defect on the right-hand side. Moreover, when the two defects are conjugate to one another, $2=\bar 1$, one can show that the lowest Casimir appearing in (\ref{flatfusion}) satisfies $C_{1\bar 1 k}\le 0$, \cite{Kravchuk:2024qoh,Cuomo:2024psk,Diatlyk:2024qpr}.
In other words, a line defect and its anti-defect in their ground state can only attract.

\begin{figure}[t]
\centering
\includegraphics[width =14cm]{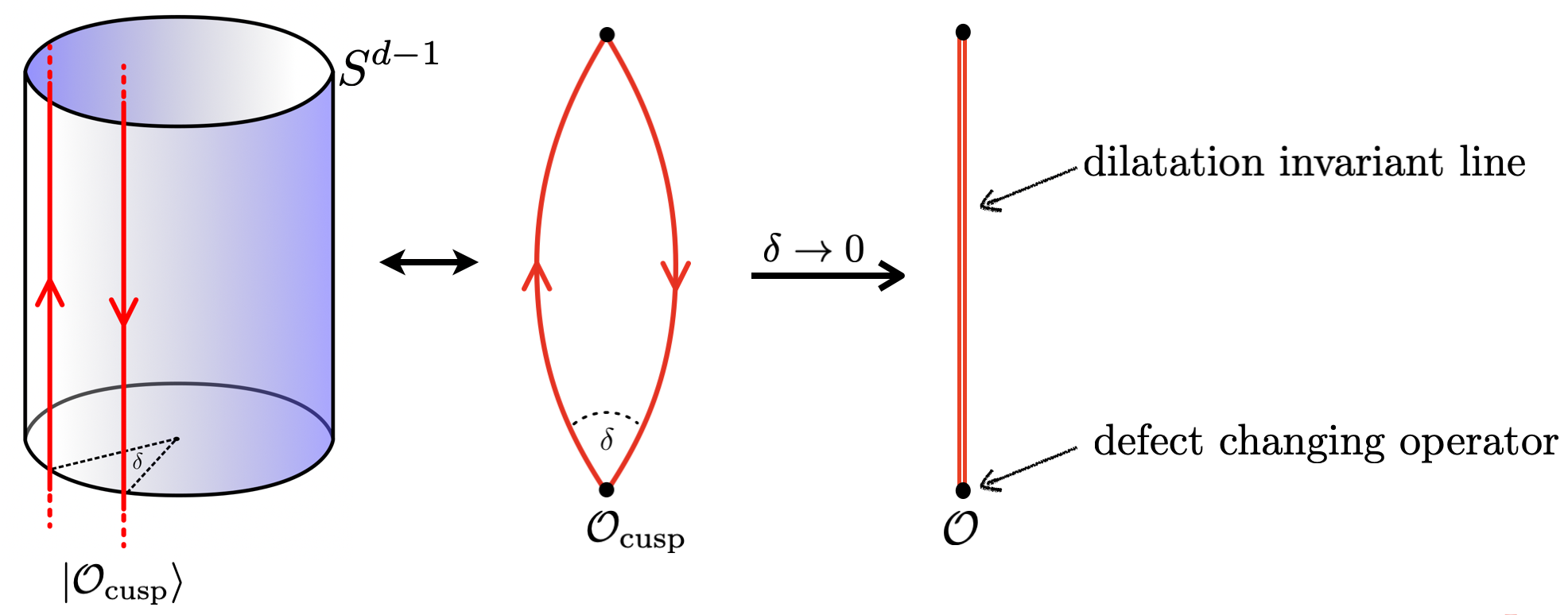}
\caption{Fusion of parallel conformal line and anti-line defects on $S^{d-1}\times\mathbb{R}$. Under the conformal map from the cylinder to the plane, this configuration is equivalent to two rays meeting at a pair of cusps in $\mathbb{R}^d$. For any cusp angle, the configuration preserves the dilatation symmetry centred between the two cusps, together with an $SO(d-2)$ rotational symmetry. In the limit where the cusp angle is taken to zero, $\delta\to0$, the configuration reduces to a single fused line stretched between two defect-changing operators, $\mathcal{O}$.}\label{cylindermap}
\end{figure}
Another way to fuse two conjugate conformal lines is to consider the same limit as in (\ref{flatfusion}), but on the sphere. More precisely, we place the CFT on $S^{d-1}\times\mathbb{R}_\tau$, let the lines extend along the time direction, and separate them by an angle $\delta$ on the sphere. For small separations, this angle can be identified with the transverse distance between the lines in units of the sphere radius, $\delta\simeq{r\over R}\ll1$. 
This configuration preserves time translations -- an $SL(2,\mathbb{R})$ generator with two fixed points -- as well as an $SO(d-2)$ rotational symmetry. Under the conformal map from the cylinder to the plane, it is mapped to a closed contour with two cusps, as shown in Figure \ref{cylindermap}. 
In the fusion limit $\delta\to0$, the configuration reduces to a conformal line defect, analogously to (\ref{flatfusion}). The resulting line now begins and ends on defect-changing operators. Equivalently, the cusp operator $\cO_{\rm cusp}$ becomes a defect-changing operator $\cO$ in the fusion limit. This operator acts on the direct sum of the conformal line defect and the trivial defect, interpolating between the two. Such operators are known as defect-creation and defect-annihilation operators.

Cusp operators have been studied extensively; see, for example, \cite{Polyakov:1980ca,Drukker:1999zq,Makeenko:2006ds,Drukker:2007dw,Drukker:2011za,Correa:2012nk,Correa:2012hh,Drukker:2012de,Gromov:2015dfa,Gromov:2016rrp,Cavaglia:2018lxi,Grozin:2015kna}. In this paper, we exploit this understanding to study the fusion of line defects. A cusp operator is characterised by its cusp anomalous dimension, or equivalently by the energy of the corresponding state on the sphere. This energy is extracted from the expectation value of the cusped configuration, normalised by the expectation value of the same configuration without a cusp, corresponding to $\delta=\pi$,
\beq\la{cusp2pf}
{\langle\overline{\cal O}_\text{cusp}\big[{\cal D}_q(\delta){\cal D}_{\bar q}(0)\big] {\cal O}_\text{cusp}\rangle\over\langle{\cal D}_q(\pi){\cal D}_{\bar q}(0)\rangle}\simeq e^{-\Gamma_\text{cusp}(\delta,\cO)T}\,.
\eeq
Here, $T\gg1$ is a large cutoff in Euclidean time $\tau$, measured in units of the sphere radius. Under the conformal map from the cylinder to the plane, it is related to the UV and IR energy cutoffs through $T=\log{\Lambda_{UV}\over\Lambda_{IR}}$. At small $\delta$, we can identify the angular separation with the defect RG scale in units of the sphere radius. In this limit, the cusp dimension takes the form \cite{Cuomo:2024psk,Kravchuk:2024qoh,Diatlyk:2024zkk}
\beq\la{gammacusp}
\Gamma_\text{cusp}(\delta,{\cal O})={C_{\cal O}\over\delta}+\Delta_{\cal O}+\sum_{m=1}\alpha_m\delta^m{\cal R}^m+\alpha\, \delta^{\Delta_\delta-1}+\dots\,.
\eeq
Here, the leading term in (\ref{gammacusp}) is the corresponding Casimir energy, $C_\cO=C_{1\bar 1 k}$. The conformal defect $\cD_k$ appearing in the fusion expansion (\ref{flatfusion}) is determined by the choice of cusp operator. The first subleading term, $\Delta_\cO$, is the scaling dimension of the defect-changing operator inserted at the endpoints of the conformal line.

The remaining terms, involving positive powers of $\delta$, can be understood by starting from the fused conformal line and perturbing in $\delta$ using conformal perturbation theory. Integer powers of $\delta$ arise from curvature corrections associated with the sphere geometry, whereas fractional powers are generated by irrelevant primary operators living on the line defect. We denote the dimension of the leading such operator -- the irrelevant primary with smallest dimension -- by $\Delta_\delta>1$. In the conformal frame where the line extends from the origin to infinity, the corresponding deformation of the conformal defect action takes the form
\beq
\delta S_\text{defect}^\text{flat}=\delta^{\Delta_\delta-1}\int\dd x\,x^{\Delta_\delta-1}{\mathbb O}_\delta(x)+\ldots\,,
\eeq
where $x$ is the flat-space coordinate along the fused line. It is related to the cylinder time coordinate by $x = R\,e^\tau$. The coefficient $\alpha$ appearing in (\ref{gammacusp}) is determined by the corresponding three-point function
\beq\la{correctionintro}
{\<\bar\cO(\infty){\mathbb O}_\delta(y)\cO(0)\>\over\<\bar\cO(\infty)\cO(0)\>}={\alpha/2\over y^{\Delta_\delta}}\,.
\eeq
Here, $\cO$ is the defect-creation operator for the fused line defect, while $\bar\cO$ is its conjugate defect-annihilation operator.

The $\delta\to0$ limit describes an RG flow on the line in a particular scheme, where $R\delta\simeq r$ plays the role of the RG scale. We will distinguish between two types of such flows. 

The first occurs when $\cD_n$ is a nontrivial conformal line defect. Such a flow terminates at a zero of the defect beta functions. The defect-changing operators obtained from the folded cusps organise into highest-weight representations of $SL(2,\mathbb{R})$.\footnote{This follows from the fact that these are defect operators acting on the direct sum of the fused line and the trivial defect, while a direct sum of conformal defects is itself conformal.} 
At finite $\delta$, however, the $SL(2,\mathbb{R})$ symmetry is broken, and the spectrum of cusp operators is generically no longer evenly spaced by integers. We will refer to the set of cusp operators that share the same $SL(2,\mathbb{R})$ Casimir in the $\delta\to0$ limit as a \emph{conformal family}. At the fixed point, such a family consists of a primary operator together with its descendants, but the notion of a conformal family can be continuously extended to finite $\delta$.

In this paper, we study the fusion of conformal line defects in two examples. We begin with the planar ladder theory, which provides a simple, solvable setting that illustrates the general structure. Our main example is the fusion of two $1/2$-BPS Wilson lines in planar ${\cal N}=4$ SYM theory, which we solve exactly using integrability techniques.

To summarise our results, let us first introduce some notation and clarify our objectives. Let ${\mathbb O}_\Delta$ be a defect-creation operator of dimension $\Delta$ belonging to a given conformal family, though not necessarily a conformal primary in the $\delta\to0$ limit. One way to characterise such an operator is through its three-point function with two defect primary operators of dimensions $\Delta_1$ and $\Delta_2$,
\beq\la{3pf}
G(x_0;x_1,x_2)\equiv\<{\mathbb O}_\Delta(x_0)\cO_{\Delta_1}(x_1)\cO_{\Delta_2}(x_2)\>\,.
\eeq
\begin{figure}[t]
\centering
\includegraphics[width =11cm]{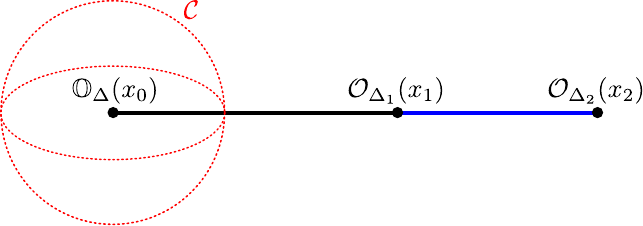}
\caption{An example of the defect three-point function (\ref{3pf}). At $x_0$ sits a defect-creation operator of dimension $\Delta$, for a nontrivial conformal defect (shown in black). This operator may be either a primary or a descendant. The operators inserted at $x_1$ and $x_2$ are primary defect-changing operators of dimensions $\Delta_1$ and $\Delta_2$, respectively. The operator at $x_1$ interpolates between the black and blue conformal defects, while the operator at $x_2$ interpolates between the blue defect and the trivial defect. We act on the operator at $x_0$ -- or equivalently on the pair of operators at $x_1$ and $x_2$ -- with the quadratic Casimir operator $\cC$ of $SL(2,\mathbb{R})$ (shown in red). This leads to the differential equation (\ref{casimireq}).}\label{3pffig}
\end{figure}
We choose these two primary defect operators so that, for $x>x_{1,2}>x_0$, the defect is again trivial. For instance, the operators inserted at $x_1$ and $x_2$ may themselves be defect-changing operators; see figure \ref{3pffig}. We then define the corresponding ``wavefunction'' $\varphi(\sigma)$ associated with the operator ${\mathbb O}_\Delta(0)$ by
\beq\la{scalefactor}
G(0,e^s,e^t)=e^{(\Delta-\Delta_1-\Delta_2){s+t\over2}}{e^{(\Delta_2-\Delta_1){s-t\over2}}\over\(e^{s-t\over2}-e^{t-s\over2}\)^{\Delta_1+\Delta_2}}\,\varphi(s-t)\,.
\eeq
Here, the prefactor $e^{(\Delta-\Delta_1-\Delta_2){s+t\over2}}$ accounts for the fact that the operator of dimension $\Delta$ inserted at the origin is probed by operators of dimensions $\Delta_1$ and $\Delta_2$. The additional dependence on $(s-t)$ in the prefactor is included for later convenience. In terms of the wavefunction $\varphi(\sigma)$, the quadratic Casimir equation of $SL(2,\mathbb{R})$ takes the form of the following second-order differential equation,
\beq\la{casimireq}
\[\d_\sigma^2-\frac{\mathbf C/2}{\cosh\sigma-1}-\frac{\Delta^2+(\Delta_1-\Delta_2)^2}{4}-{\Delta(\Delta_1-\Delta_2)\over2}\coth{\sigma\over2}
\] \varphi(\sigma)=0\,,
\eeq
where ${\bf C}$ is the quadratic Casimir labelling the conformal family. As we will show, this equation, together with the appropriate boundary condition at the UV RG scale, implies the standard relation between the Casimir and the spectrum of the corresponding conformal family, $\Delta_n=n+\frac{1}{2}(1\pm\sqrt{1-4{\bf C}})$ with $n=0,1,2,\ldots$.

The second type of flow we study exhibits walking behaviour between a pair of complex-conjugate zeros of the beta function. Such ``walking'' or ``quasi-conformal'' behaviour, in which the RG trajectory passes near complex-conjugate fixed points, has been studied in the context of bulk quantum field theories; see \cite{Gorbenko:2018ncu,Gorbenko:2018dtm}. In these flows, the finite-energy spectrum of defect operators drifts with the RG scale towards smaller values and becomes scheme-dependent.

\paragraph{Summary of Results.}

\begin{figure}[t]
\centering
\includegraphics[width = 0.73\textwidth]{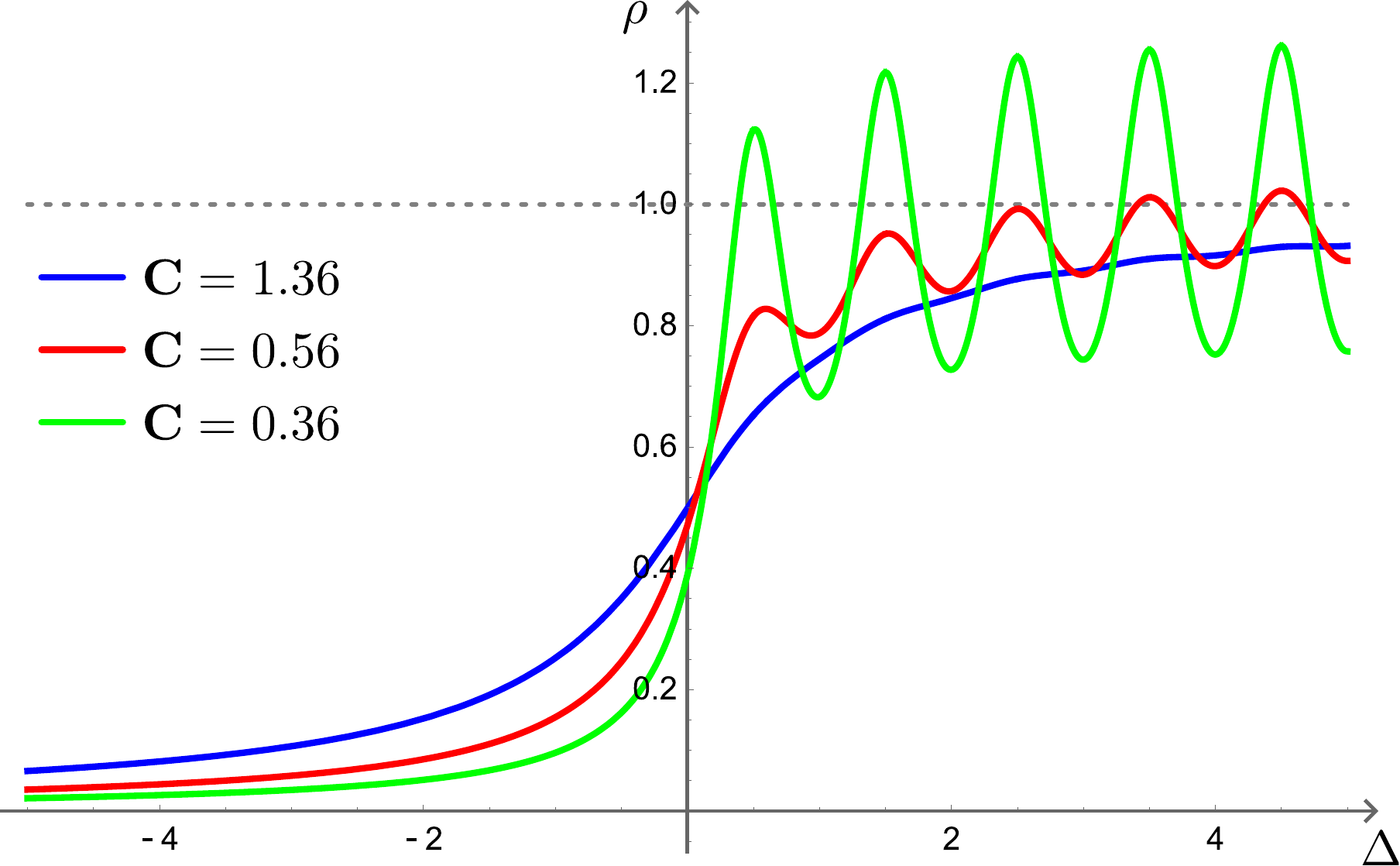}
\caption{The density $\rho(\Delta)$ in (\ref{density}) for several values of ${\bf C}$. As ${\bf C}$ approaches $1/4$ from above, the density develops oscillations, eventually approaching a sum of equally spaced delta functions.}\label{densityfig}
\end{figure}

In all cases exhibiting walking behaviour on the line, we find that the Casimir equation (\ref{casimireq}) and its associated boundary condition continue to hold. In other words, even though the line defect is no longer conformal and its spectrum drifts with the RG scale without remaining evenly spaced, the Casimir operator nevertheless commutes with the Hamiltonian on the sphere at distances much larger than the UV RG scale. This property fixes the density of states within each conformal family as the UV RG scale is removed, which in our scheme corresponds to taking $\delta\to0$. Remarkably, this density takes a universal, scheme-independent form
\beq\la{density}
\rho(\Delta)=\frac{i}{2\pi} \[\psi ^{(0)}\Big(\frac{1}{2}-\Delta-{i\over2}\sqrt{4{\bf C}-1}\Big)-\psi ^{(0)}\Big(\frac{1}{2}-\Delta +{i\over2}\sqrt{4{\bf C}-1}\Big)\]\,,
\eeq
where $\psi^{(0)}(x)={\Gamma'(x)\over\Gamma(x)}$ is Euler's psi function. In figure \ref{densityfig}, we plot this density for several values of the Casimir. At large negative energies, it decays exponentially, while at large positive energies, it approaches the constant $1$. As ${\bf C}$ approaches $1/4$ from above, the density develops oscillations and, at ${\bf C}=1/4$, approaches a sum of equally spaced delta functions.

In the planar ladder model, the quadratic Casimir is ${\bf C}=4\hat g^2$, where $\hat g$ is the ladders coupling. The model has two conformal fixed points for $\hat g<1/4$, which collide at $\hat g=1/4$ and give rise to walking behaviour above this value.

More generally, we find that the complete spectrum of each conformal family at small $\delta$ is determined by a universal Fusion Master Equation, derived in section~\ref{fusionlimit}; see equation~(\ref{gluing}). All model-dependent information entering this equation is encoded in two quantities intrinsic to the fused defect: the Casimir parameter
$\vartheta=\frac12\sqrt{1-4{\bf C}}$
and a connection coefficient ${\mathbb S}$. Both are independent of $\delta$ and of the excitation number. Once $\vartheta$ and ${\mathbb S}$ are known, the Fusion Master Equation reconstructs the full tower of cusp dimensions, including their behaviour at small but finite $\delta$.

In planar ${\cal N}=4$ SYM, there are infinitely many such conformal families. Each family has its own critical value of the 't~Hooft coupling, above which $\vartheta$ becomes imaginary, and the corresponding spectrum enters the walking regime.

To solve the fusion problem in ${\cal N}=4$ SYM, we show that the cusp Quantum Spectral Curve (QSC) of \cite{Gromov:2015dfa} separates into two regimes in the limit $\delta\to0$; see section~\ref{sec: QSC}. In the far regime, the QSC reduces to the Casimir equation (\ref{casimireq}) with $\Delta_1=\Delta_2=0$, and the Baxter functions are identified with Mellin transforms of the three-point function (\ref{3pf}). The near regime is independent of both $\delta$ and $\Gamma_{\rm cusp}$ and determines directly the two quantities
$\vartheta$ and ${\mathbb S}$
entering the Fusion Master Equation.

\begin{figure}[ht]
    \centering
    \includegraphics[width=0.8\linewidth]{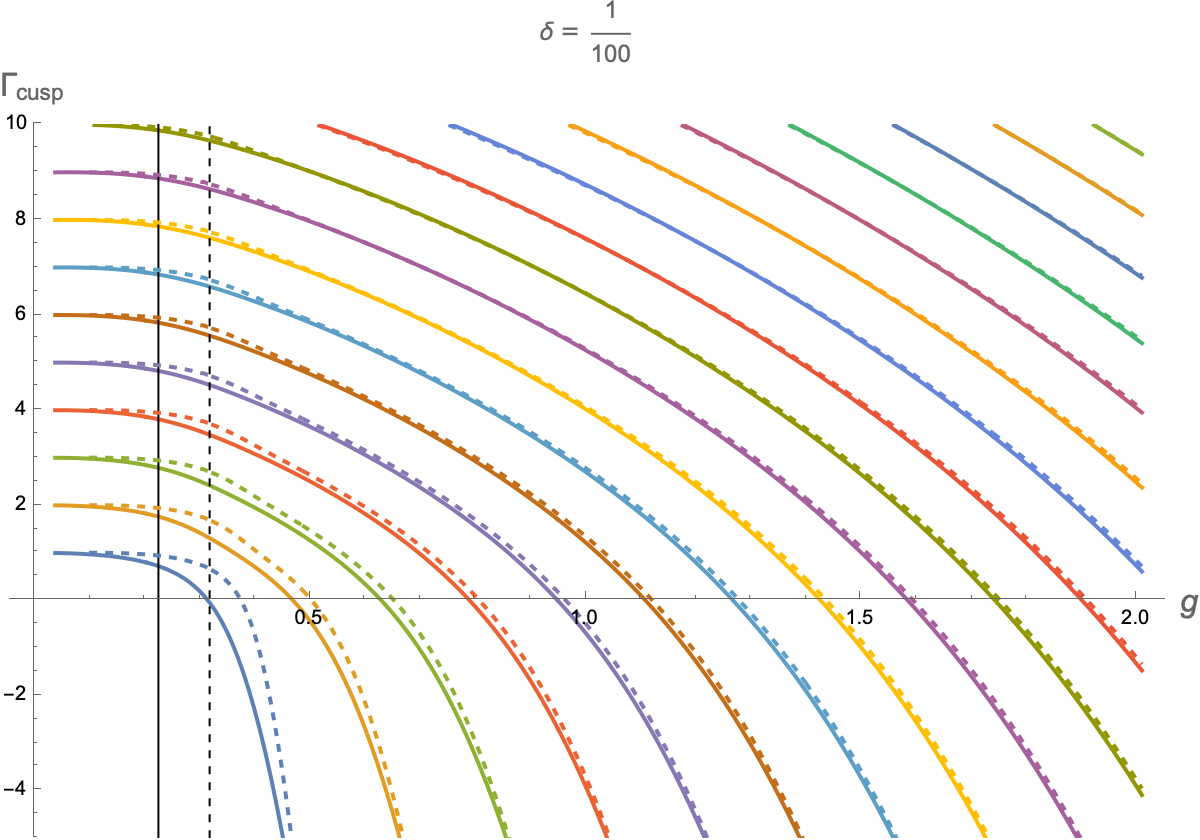}
    \caption{The full spectrum of $\Gamma_{\rm cusp}$ for two conformal families: even (solid) and odd (dashed) as defined in the main text, generated from our numerical values of $\vartheta(g)$ and ${\mathbb S}(g)$ at $\delta=1/100$ for $\theta=3\pi/10$, where $\theta$ is an internal cusp angle. The seed data generated by the near-region QSC controls the spectrum of infinitely many states at small $\delta$. The vertical black lines show the critical values of the coupling for the two families. At large $g$, these two families become almost indistinguishable, as we find that both are described by the same classical solution.}
    \label{fig:families}
\end{figure}
Thus, rather than solving separately for every excited cusp state, the fused QSC determines two functions from which the complete small-$\delta$ spectrum of a conformal family can be reconstructed (see figure~\ref{fig:families}). We solve this system numerically at finite coupling for two types of families that we will refer to as {\it even} and {\it odd}, determine their critical couplings, and verify directly that the Fusion Master Equation reproduces the finite-$\delta$ spectrum. We also obtain weak-coupling expansions through six loops for the even family and five loops for the odd family; see section~\ref{sec: QSCWeak}.

At strong coupling the system lies in the walking regime. The scaling dimensions and the Casimir behave as
$\Delta^2,\,{\bf C}\propto g^2=g_{\rm YM}^2N/(4\pi)^2$,
and the density (\ref{density}) reduces to
\beq\la{dsdensity}
\rho(\Delta)={1\over2}-{\ii\over2\pi}\log{\sqrt{\bf C}+\ii\Delta\over\sqrt{\bf C}-\ii\Delta}+O(1/g)\,.
\eeq
In sections~\ref{foldedstringsec} and~\ref{AdSsec}, we reproduce this result from a semiclassical time-dependent folded string in global $AdS_5$, stretched through a finite angle on $S^5$. The semiclassical quantisation reproduces not only the density of states, but the full classical limit of the Fusion Master Equation, including the leading strong-coupling behaviour of both $\vartheta$ and ${\mathbb S}$.

Finally, let us comment on the stability of conformal defects and on $1/N$ corrections away from the planar limit. For a conformal fixed point with ${\bf C}<1/4$ to be stable, all defect operators--including defect-changing operators--must be irrelevant. However, a primary defect-creation operator with ${\bf C}>0$ has dimension $\Delta_0=\frac{1}{2}(1\pm\sqrt{1-4{\bf C}})$ and is therefore relevant. Deforming the defect action by such an operator generates a sequence of defect RG flows that can be solved explicitly in the planar limit \cite{Nagar:2024mjz}. The endpoint of these flows can be understood as effectively removing the trivial defect sector from the Hilbert space, together with the associated defect-changing operators, leaving behind the same conformal fixed point with which we started. In this sense, the stable conformal defect may be viewed as being ``full of holes'' already at planar order. This picture suggests that $1/N$ corrections should remain small below the critical coupling. We leave a detailed analysis of these corrections for future work. 

It is important to place the present work in the context of earlier results. In \cite{Klebanov:2006jj}, the flat-space spectrum of quark-antiquark states was studied, and it was argued that above a critical 't Hooft coupling, the quark-antiquark potential supports infinitely many bound states. Here, we determine this critical coupling to an arbitrary numerical precision and show that it corresponds to the point ${\bf C}=1/4$, where two defect fixed points collide and move into the complex plane. The flat-space spectrum studied in \cite{Klebanov:2006jj} is related to the cusp dimensions considered here through the limit in which $E=\Delta/r$ is held fixed as $\delta\simeq r/R\to0$. Since the spacing between positive-energy states is set by the inverse sphere radius, the spectrum becomes continuous in this limit, and much of the structure that we emphasise in this paper is washed out. More recently, \cite{Alday:2025pmg} exploited this flat-space structure to bootstrap planar scattering amplitudes on the Coulomb branch of ${\cal N}=4$ SYM theory, and a qualitative estimate for the critical coupling was first made. Finally, \cite{Aharony:2022ntz,Aharony:2023amq} studied a closely related example in which two defect fixed points merge and move into the complex plane. The primary focus there was the formation of a screening cloud around the defect. Investigating the corresponding condensate far from the fused defect lies beyond the scope of the present paper.

\paragraph{Organisation of the paper.}
Section~\ref{sec:ladders} solves the fusion problem in the planar ladder model: we derive the Casimir equation, the Fusion Master Equation, and the density of states, and analyse both the conformal and walking regimes, including the strong-coupling limit via a semiclassical folded string in $AdS_2$. Section~\ref{DCORGsec} discusses the chain of defect RG flows triggered by defect-changing operators and explains how these flows remove the trivial defect sector, leaving the stable conformal fixed point. Section~\ref{sec: QSC} contains the main integrability analysis: we take the fusion limit of the cusp QSC, derive the near- and far-regime equations that determine $\vartheta(g,\theta)$ and ${\mathbb S}(g,\theta)$, and present numerical results for the even and odd conformal families over a range of couplings and internal angles. Section~\ref{pertsec} gives a direct perturbative definition of the fused line defect and compares Feynman-diagram and QSC results at weak coupling through two-loop order. Section~\ref{AdSsec} reproduces the strong-coupling predictions semiclassically from a time-dependent folded string in global $AdS_3$ with a finite internal angle $\theta$ on $S^5$, recovering both the density of states and the full classical limit of the Fusion Master Equation. Open problems are collected in Section~\ref{sec:future}. Technical details of the QSC framework and higher-order perturbative expansions are given in the appendices.

\section{Planar Ladder}\label{sec:ladders}

The simplest model of two conjugate defects meeting at a cusp is the so-called ladder limit, introduced in \cite{Erickson:2000af} via a double-scaling limit of the cusped Wilson line in ${\cal N}=4$ SYM. Despite its simplicity, this model captures much of the essential quark--antiquark physics using only a single dynamical degree of freedom; see \cite{Erickson:2000af,Pineda:2007kz,Klebanov:2006jj,Correa:2012nk,Gromov:2016rrp}. In this limit, only a restricted class of planar diagrams survives, namely ladder diagrams built from scalar propagators stretched between the two rays forming the cusp; see Figure~\ref{specfig}.a\,.

Concretely, let $\vec x_1(s)$ and $\vec x_2(t)$ denote the two semi-infinite rays meeting at the origin with geometric cusp angle $\delta$. The effective interaction between the rays is mediated by a scalar propagator connecting a point on one ray to a point on the other. In Euclidean signature, this propagator takes the form
\begin{equation}\la{Pst}
P(s,t)=(4\pi\hat g)^2|\dot{\vec x}_1(s)|\,|\dot{\vec x}_2(t)|\,\<\Phi(x_1(s))\Phi^\dagger(x_2(t))\>=4\hat g^2\frac{|\dot{\vec x}_1(s)|\,|\dot{\vec x}_2(t)|}{|\vec x_1(s)-\vec x_2(t)|^2}\, ,
\end{equation}
where $8\pi\hat g$ denotes the effective coupling of the free scalar to the line, which we take to be positive without loss of generality. A convenient parametrisation of the rays is $\vec x_i(s)=\hat n_i\,e^s$, where $\hat n_1\cdot\hat n_2=\cos\delta$. In this parametrisation,
\beq\la{Pst2}
P(s,t)={2\hat g^2\over\cosh(s-t)-\cos\delta}\,.
\eeq

At $L$ loops, one sums all diagrams containing $L$ such propagators, nested between the two rays in every possible planar configuration. To resum these contributions to all orders, one considers the sum of all ladder diagrams whose outermost scalar insertions lie between the cusp and two fixed endpoints, $x_1(s)$ and $x_2(t)$. The resulting object, $G(s,t)$, can be interpreted as a three-point function analogous to (\ref{3pf}), where defect-annihilation operators are inserted at $x_1(s)$ on one ray and at $x_2(t)$ on the other. Differentiating with respect to the positions of these endpoints along the rays leads to the Dyson equation
\begin{equation}\la{BSeq}
\partial_s \partial_t G(s,t) \;=\; P(s,t)\,G(s,t)\, .
\end{equation}
The configuration with two rays is mapped to itself under the dilatation $s,t\to s+\alpha,t+\alpha$. This is reflected in the fact that the equation \eq{BSeq} is invariant under the dilatation, which allows us to split the variable into the part dependent on $s+t$ and the part dependent on $s-t$. If, furthermore, we assume that the cusp has a definite dimension $\Gamma_{\text cusp}$, we get (similarly to (\ref{scalefactor}))
\beq\la{Gst}
G(s,t)=e^{-{s+t\over2}\Gamma_\text{cusp}}\,\psi(s-t)\,.
\eeq
Substituting this ansatz into (\ref{BSeq}) yields the one-dimensional Bethe--Salpeter equation
\beq\la{BSPlad}
\(-\d_\sigma^2-{2\hat g^2\over\cosh\sigma-\cos\delta}+{\Gamma_\text{cusp}^2\over4}\)\psi_\Delta(\sigma)=0\,.
\eeq
This equation is a Schr\"{o}dinger problem with an effective negative energy $E=-\Gamma_\text{cusp}^2/4 \leq 0$, as follows directly from \eqref{BSPlad}. It is important to emphasise, however, that $\psi$ is not a genuine quantum-mechanical wavefunction, and therefore $|\psi|^2$ does not admit a probabilistic interpretation. Consequently, normalisability is not the appropriate quantisation condition. Instead, the spectrum is fixed by the boundary conditions. 

Let us now explain that, for states with negative $\Gamma_{\rm cusp}$, normalisability is equivalent to those boundary conditions, whereas the wavefunctions corresponding to positive $\Gamma_{\rm cusp}$ are in general non-normalisable.

There are two types of physical solutions. The first consists of ``bound states,'' namely states with $\Gamma_\text{cusp}<0$ that are localised near the cusp. Their wavefunctions decay exponentially at large separations $|s-t|$ as
\beq\la{bsqcondition}
\lim_{|\sigma|\to\infty}\psi_\text{bound}(\sigma)=e^{{|\sigma|\over2}\Gamma_\text{cusp}}\,,\qquad\Gamma_\text{cusp}<0\,, 
\eeq
and are therefore also normalisable  in the Schr\"{o}dinger sense. For large $|\Gamma_\text{cusp}|$, such states are localised near $s=t$. 
The second class consists of ``excited'' states, with $\Gamma_\text{cusp}>0$. Their quantisation condition at large $|\sigma|$ is such that only the exponentially growing solution is allowed,
\beq\la{qcondition}
\lim_{|\sigma|\to\infty}\psi_\text{excited}(\sigma)=e^{{|\sigma|\over2}\Gamma_\text{cusp}}\(1+O(e^{-n|\sigma|})\)\,,\qquad\Gamma_\text{cusp}>0\,.
\eeq
The key here is that the corrections go in integer powers $e^{-|\sigma|}$, so that for some sufficiently large $n$, we have
$\Gamma_\text{cusp}-2n<-\Gamma_\text{cusp}$ and thus we can define the pure-exponentially growing asymptotic unambiguously.\footnote{In contrast to, say, the square well potential, where corrections are power-like, and there is no practical way to define a unambiguous purely exponentially growing solution.}
As in the case of the normalisable solution, requiring purely exponentially growing asymptotics at both infinities (\ref{qcondition}) defines a discrete spectrum. These solutions are delocalised in $s-t$. One can check (see e.g.\cite{Cavaglia:2018lxi}) that all states with ${\Gamma}_{\rm cusp}>0$ become negative at sufficiently large $\hat g^2$.
In this sense, the quantisation condition \eq{qcondition} is an analytic continuation of the bound state energies to the second sheet of the spectral plane (with the continuum cut at $E\ge 0$) where they become resonances when we decrease $\hat g^2$ \cite{Cavaglia:2018lxi}.

\subsection{The Fusion Limit}\la{fusionlimit}

\begin{figure}[t]
\centering
\includegraphics[width =\textwidth]{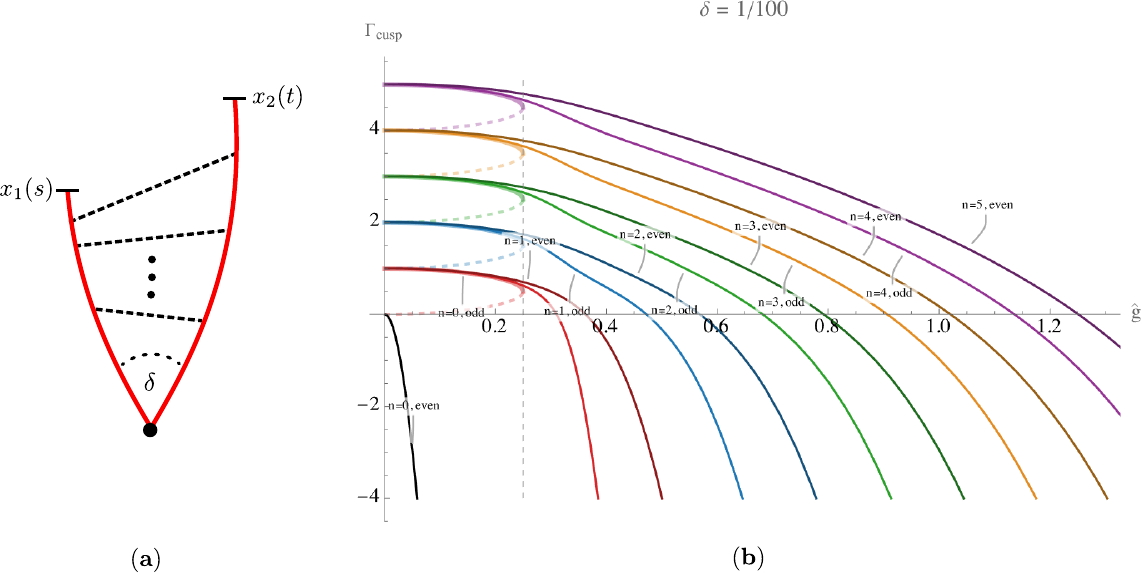}
\caption{${\bf a})$ Ladder diagrams contributing to the expectation value of a line defect with a cusp of opening angle $\delta$. We sum over all planar scalar propagators stretched between the two rays, with endpoints ranging up to $x(s)$ on one ray and up to $x(t)$ on the other. ${\bf b})$ The cusp spectrum as a function of $\hat g$ at $\delta=10^{-2}$. The gap between the excited states remains of order one, whereas the gap between the bound states grows parametrically, scaling as $1/\delta$. Above a critical value of $\hat g$, the splitting between even and odd excited states also becomes of order one. The black lines are the asymptotic $\delta\to0$ limit of the spectrum. Below, we will find that it is given by $\Delta_m=\Delta_0+m$, with $\Delta_0={1\over2}(1+\sqrt{1-16\hat g^2})$, and is double degenerate.}\label{specfig}
\end{figure}
Here, we are interested in the fusion limit $\delta\simeq{L\over R}\to0$. In this limit, the energies of the bound states scale as $\Delta\sim-1/\delta$, while the energies of the excited states remain finite at sufficiently small $\hat g$; see figure \ref{specfig}. The bound states correspond to configurations with nonzero Casimir energy between the quark and antiquark. For $\hat g<1/4$, there is a single such bound state, whereas for $\hat g>1/4$, there are infinitely many \cite{Klebanov:2006jj}. All of these bound states carry distinct Casimir energies and dress the trivial defect.
For general $\delta$, the system (\ref{BSeq})--(\ref{qcondition}) does not admit an analytic solution. Here, however, we obtain an exact solution for the finite-energy states. For $\hat g<1/4$, these states organise into the conformal family of a new conformal defect, while for $\hat g>1/4$ they exhibit walking behaviour. To demonstrate this, we analyse the problem (\ref{BSPlad}), (\ref{qcondition}) for finite $\Gamma_\text{cusp}>0$ in the fusion limit $\delta\to0$.
As we explain below, one should consider two different regions of $\sigma$: Far region - where we send $\delta\to 0$, while keeping $\sigma$ fixed, and Near region - where $\sigma/\delta$ is fixed in the limit.

\subsubsection{Far Region}

At $|\sigma|\gg\delta$ we can simply set $\delta=0$ in (\ref{BSPlad}) and the problem reduces to
\beq\la{largex}
\(\d_\sigma^2+{2\hat g^2\over\cosh\sigma-1}-{\Gamma_\text{cusp}^2\over4}\)\psi_\text{far}(\sigma)=0\,.
\eeq
This is the regime in which the two lines effectively appear fused together, and information on $\delta$ is naively gone. 
Intriguingly, we can recognise in (\ref{largex}) the quadratic Casimir equation (\ref{casimireq}), by identifying  $\Delta=\Gamma_\text{cusp}$, $\Delta_1=\Delta_2$, where the role of the quadratic Casimir eigenvalue is played by the coupling
\beq\la{Cdef}
{\bf C}=4\hat g^2\,.
\eeq
Note that the potential diverges at the origin as $1/\sigma^2$, resulting in an essential singularity for the solution.
For positive $\sigma$, the exact solution satisfying the condition (\ref{qcondition}) can be found analytically
\beq\la{far}
\psi_\text{far}(\sigma)=e^{\frac{\sigma}{2}\Gamma_\text{cusp}} \left(1-e^{-\sigma}\right)^{{1\over2}+\vartheta}\,_2F_1\Big({1\over2}+\vartheta,{1\over2}+\vartheta-\Gamma_\text{cusp};1-\Gamma_\text{cusp};e^{-\sigma}\Big)\,,
\eeq
where we defined
\beq\la{varthetadef}
\vartheta(\hat g)\equiv{1\over2}\sqrt{1-16 \hat{g}^2}={1\over2}\sqrt{1-4{\bf C}}\,.
\eeq
The second solution has an exponent with an opposite sign and can be obtained by using the $\Gamma_\text{cusp}\to -\Gamma_\text{cusp}$ symmetry of  \eq{largex}, but as it does not satisfy the correct quantisation condition \eq{qcondition} we will not need it here.

So far, the solution exists for any $\Gamma_{\rm cusp}$. In order to deduce the quantisation condition, we have to resolve the singularity at $\sigma=0$ by matching with the regime $\sigma\sim\delta$. When $\sigma\ll1$ (but still $\sigma\gg \delta$), the potential in (\ref{largex}) simplifies to
\beq\la{singpotential}
V(\sigma)=-{2g^2\over\cosh\sigma-1}\simeq-{4\hat g^2\over\sigma^2}\,,\qquad \sigma\ll1\,.
\eeq
Correspondingly, the solution (\ref{far}) should then reduce to
\beq\la{psifarsmallx}
\psi_\text{far}(\sigma)\simeq c_+^\text{far}\,\sigma^{\Delta_+}+c_-^\text{far}\,\sigma^{\Delta_-}\,,\qquad\Delta_\pm={1\over2}\pm\vartheta\,,
\eeq
with
\beq\la{cpocm}
{c_+^\text{far}\over c_-^\text{far}}=-{\Gamma\big(\Delta_+\big)\Gamma\big(2\Delta_-\big)\over\Gamma\big(\Delta_-\big)\Gamma\big(2\Delta_+\big)}\times{\Gamma\big(\Delta_+ -\Gamma_\text{cusp}\big)\over\Gamma\big(\Delta_--\Gamma_\text{cusp}\big)}\,.
\eeq

If, instead of the asymptotic equation (\ref{largex}), we considered the Casimir equation (\ref{casimireq}) with $\Delta_1\neq\Delta_2$ and $\Delta=\Gamma_\text{cusp}$, then this ratio would be modified to
\beq
{c_+^\text{far}\over c_-^\text{far}}\ \to\ -\frac{\Gamma (2\Delta_-)\Gamma \left(\Delta_++\Delta _1-\Delta _2\right)}{\Gamma(2\Delta_+)\Gamma\left(\Delta_-+\Delta _1-\Delta _2\right)}\times{\Gamma\left(\Delta_+ -\Gamma_\text{cusp}\right)\over\Gamma\left(\Delta_--\Gamma_\text{cusp}\right)}\,.
\eeq
We see that the dependence on $\Gamma_\text{cusp}$ enters only through the final ratio of gamma functions, which remains unchanged.

As we explained above, to determine the allowed values of $\Gamma_\text{cusp}$, we must match this solution, which is valid for $\sigma\gg\delta$, to the solution of (\ref{BSPlad}) in the region $\sigma\lesssim\delta$. Before carrying out this matching procedure, let us show that
we can associate $\delta$ with the UV RG scale
and derive the corresponding beta function, but in a different, slightly more transparent regularisation scheme.

\subsubsection{The Defect Beta Function}
   
In this section, we change the logic a bit. Instead of studying the ``microscopic'' definition of the conformal defect,  given by the limit $\delta\to 0$ in \eq{BSPlad}, we can follow the logic of the renormalisation group and regularise the $1/\sigma^2$ singularity of the potential (\ref{singpotential}) with mixed Neumann--Dirichlet boundary conditions at a new UV scale $\rho$,\footnote{There is a vast literature on the $1/\sigma^2$ potential. For a textbook treatment, see \cite{Landau:1991wop}; for early references, see \cite{Case:1950an,Camblong:2000in}; and for relatively recent RG treatments, see \cite{Beane:2000wh,Barford:2002je,Bawin:2003dm,Barford:2004fz,Hammer:2008ra} and the references therein. There is, however, one important difference between those works and ours concerning the definition of the positive-energy problem. Previous works considered the standard Schr\"{o}dinger problem, whereas in our case the positive-energy states satisfy a different quantisation condition; see (\ref{qcondition}).}
\beq\la{bc}
\left.\[\sigma\d_\sigma-\eta(\rho)\]\psi_\text{far}(\sigma)\right|_{\sigma=\rho}=0\,,
\eeq 
which is ignorant of the actual details of the original definition at $\sigma\sim\delta$.
Here $\eta(\rho)$ plays the role of the coupling associated with a counterterm localised at $\sigma=\rho$.

We then tune $\eta=\eta(\rho)$ so that $\psi$ becomes independent of the arbitrary UV scale $\rho$. This is achieved by requiring that
\beq
\eta(\rho)={c_+\Delta_+\rho^{\Delta_+}+c_-\Delta_-\rho^{\Delta_-}\over c_+\rho^{\Delta_+}+c_-\rho^{\Delta_-}}\,.
\eeq
The corresponding $\beta$-function is given by
\begin{figure}[t]
\centering
\includegraphics[width = 0.45\textwidth]{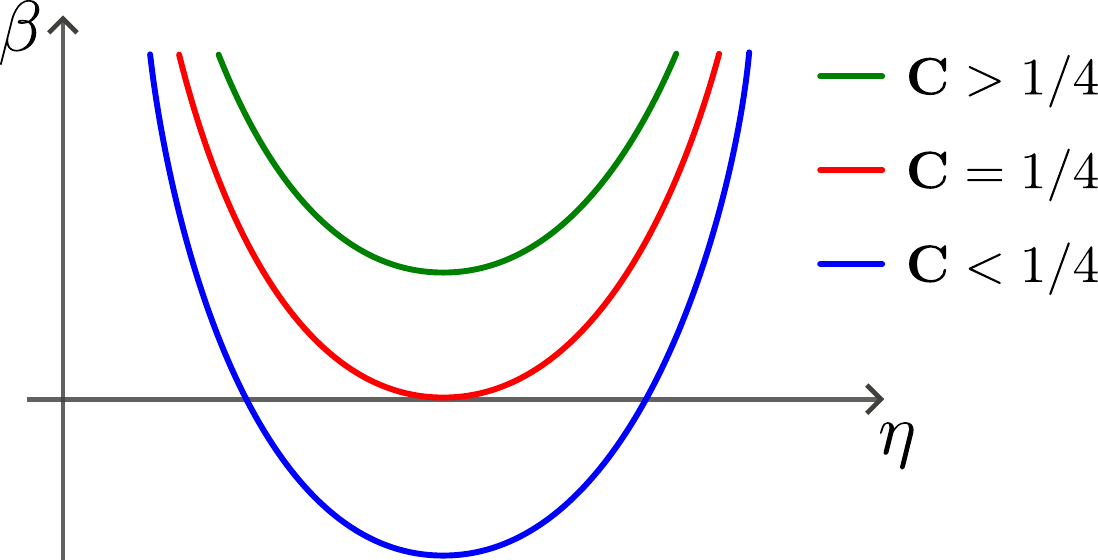}
\caption{The defect beta function $\beta_\eta$ in (\ref{betaeta}). For ${\bf C}<1/4$, there are two fixed points, one stable and one unstable. They collide at the critical value ${\bf C}=1/4$. For ${\bf C}>1/4$, the fixed points move into the complex $\eta$-plane, leaving no real fixed points.}\label{beta}
\end{figure}
\beq\la{betaeta}
\beta_\eta=-\rho\d_\rho\eta=(\eta-\Delta_+)(\eta-\Delta_-)\,.
\eeq
We see that for $\hat g<1/4$, corresponding to real $\vartheta$, there are two fixed points; see figure \ref{beta}. The stable fixed point lies at $\eta=\Delta_+$, where $c_-/c_+=0$, and the spectrum is given by $\Gamma_\text{cusp}=\Delta_++n$ with $n=0,1,2,\ldots$. Similarly, the unstable fixed point lies at $\eta=\Delta_-$, where $c_+/c_-=0$, and the spectrum becomes $\Gamma_\text{cusp}=\Delta_-+n$. At the critical value $\hat g=1/4$, corresponding to $\vartheta=0$, the two fixed points collide. For $g>1/8$, where $\vartheta$ becomes imaginary, they move into the complex plane.

By imposing the boundary condition (\ref{bc}) on the solution (\ref{psifarsmallx}), together with (\ref{cpocm}), we can extract the spectrum of defect-changing operators. We now return to the cusp regularisation, in which the parameter $\delta$ provides the physical UV boundary condition, and carry out the explicit matching between the near and far regions to derive the quantisation condition for $\Gamma_\text{cusp}$. We will see that for $\hat g<1/4$, the resulting RG flow always terminates at the stable fixed point and thus $\Gamma_{\rm}=\Delta_++n,\;n=0,1,\ldots$. This is to be expected, since for any $\eta>\Delta_-$ the flow terminates at the stable fixed point, whereas reaching the unstable fixed point requires fine-tuning.

\subsubsection{Near Region and Gluing Condition}

Here we analyse the Bethe--Salpeter equation (\ref{BSPlad}) in the regime of small separations, $\sigma\sim\delta$, and we introduce $y=\sigma/\delta$. Fixing $y$ and expanding at small $\delta$ we get
\beq
\(\d_y^2+{4\hat g^2\over y^2+1}-{1\over4}(\delta\,\Gamma_\text{cusp})^2\)\psi_\text{near}(y)=0\,.
\eeq
This is precisely the Bethe--Salpeter equation for parallel lines on the plane, as in (\ref{flatfusion}). It has previously been used to determine the Casimir energies, equivalently, the quark--antiquark potential, in the regime where $\Gamma_\text{cusp}\sim1/\delta$; see \cite{Erickson:2000af,Pineda:2007kz,Klebanov:2006jj,Correa:2012nk,Gromov:2016rrp}. For finite $\Gamma_\text{cusp}\sim1$, however, the last term becomes negligible, and the equation reduces to 
\beq\la{smallx}
\(\d_y^2+{4\hat g^2\over y^2+1}\)\psi_\text{near}(y)=0\,.
\eeq
This region encodes the boundary conditions at small $\sigma$ and is independent of $\Gamma_\text{cusp}$. The two linearly independent solutions of (\ref{smallx}) can be obtained analytically and are given by
\beq
\psi_\text{near}^\text{even}(y)=\ \, _2F_1\Big(-{\Delta_+\over2},-{\Delta_-\over2};\frac{1}{2};-y^2\Big)\,,\quad
\psi_\text{near}^\text{odd}(y)=y \, _2F_1\Big({\Delta_-\over2},{\Delta_+\over2};\frac{3}{2};-y^2\Big)\,,
\eeq
where we have separated them into even and odd solutions, since at finite $\delta$ they have different cusp anomalous dimensions.

In order to glue with small $\sigma$ in the far regime we have to take $y\gg1$, where the potential reduces to $-4\hat g^2/y^2$, and the solutions thus match the form (\ref{psifarsmallx}) e.g.
\beq
\psi^{\rm even/odd}(y) = c_+^{\rm even/odd}
(\delta\; y)^{\Delta_+}+c_-^{\rm even/odd}
(\delta\; y)^{\Delta_-}
\eeq
with
\beq\la{cpmeven}
{c_+^\text{even/odd}\over c_-^\text{even/odd}}=\delta^{-2\vartheta}{\Gamma\left(\mp{\Delta_\pm\over2}\right)\Gamma\left(1\mp{\Delta_\pm\over2}\right)\Gamma(\Delta_+-{1\over2})\over
\Gamma\left(\mp{\Delta_\mp\over2}\right)\Gamma\left(1\mp{\Delta_\mp\over2}\right)\Gamma(\Delta_--{1\over2})}\,,
\eeq
with the upper sign for even and the lower for odd.
Note that so far, the factor of $\delta$ is a matter of conventions, as the equation itself does not depend on $\delta$ explicitly. However, as at an intermediate scale $\sigma\sim\sqrt{\delta}$, both regimes overlap and thus the asymptotics (\ref{bsqcondition}) and (\ref{qcondition}), lead to a non-trivial quantisation condition
\beq
{c_+^\text{far}\over c_-^\text{far}}={c_+^\text{even/odd}\over c_-^\text{even/odd}}\,.
\eeq
Substituting the expressions (\ref{cpmeven}), (\ref{psifarsmallx}) into these conditions, we obtain one of the key equations of this paper, to which we refer as the Fusion master equation (FME)
\beq\la{gluing}
\boxed{\log\frac{\Gamma\left(\Delta_+-\Gamma_\text{cusp}\right)}{\Gamma\left(\Delta_--\Gamma_\text{cusp}\right)}+2\vartheta\log{\delta\over8}+2\pi \ii n+\log {\cal S}=O(\delta)}\,,
\eeq
or in product form
\beq\la{gluingprod}
\boxed{
(\delta/8)^{2\vartheta}
\frac{\Gamma\left(\Delta_+-\Gamma_\text{cusp}\right)}
{\Gamma\left(\Delta_--\Gamma_\text{cusp}\right)} {\cal S}=1+O(\delta^{2\vartheta+1})}\,.
\eeq
In \eq{gluing}, the integer $n$ labels different branches of the multi-valued logarithm in \eqref{gluing}, corresponding to distinct radial excitations of the cusp operator. The factor ${\cal S}$ is independent of $\Gamma_\text{cusp}$ and encodes the details of the short-distance (near-region) dynamics.
In the ladder case for even and odd states, we have
\beq \la{SpmSh}
{\cal S}_\pm(\vartheta)=\frac{\Gamma\left(\frac{3}{4}\pm\frac{\vartheta}{2}\right)\Gamma \left(\frac{1}{4}\mp\frac{\vartheta}{2}\right)\Gamma\left(+\vartheta-\frac{1}{2}\right)\Gamma(-\vartheta)^2}{\Gamma\left(\frac{3}{4}\mp\frac{\vartheta}{2}\right)\Gamma \left(\frac{1}{4}\pm\frac{\vartheta}{2}\right)\Gamma\left(-\vartheta-\frac{1}{2}\right)\Gamma(+\vartheta)^2}\,,
\eeq
where the upper (lower) sign corresponds to the even (odd) states, and we used $\Delta_\pm={1\over2}\pm\vartheta$. Let us give their weak coupling expansion explicitly, as it will be useful for comparison with the more general ${\cal N}=4$ case
\beqa\la{Sweak}
{\cal S}_+&=&\frac{2}{\pi  \hat g^2}-\frac{16 (-1+\gamma +4 \log2)}{\pi }\\
\nn &-&\frac{8 \hat  g^2 \left(-84+72
   \gamma -24 \gamma ^2+\pi ^2-384 \log ^2 2+288 \log2-192 \gamma  \log2\right)}{3
   \pi }+O\left(\hat g^4\right)\,,\\
{\cal S}_-&=&8 \pi  \hat g^2-64 \hat g^4 (\pi  (-2+\gamma +4 \log2))\\
\nn &+&\frac{32}{3} \pi  \hat g^6 \left(192-120
   \gamma +24 \gamma ^2+\pi ^2+384 \log ^2 2-480 \log2+192 \gamma  \log 2\right)+O\left(\hat g^8\right)\,.
\eeqa
We also clearly see the key differences between the two states: for the even state, there is a divergence at small $\hat g$ in ${\cal S}_+$. We will see below that this is responsible for the non-commutativity of the $\delta\to0$ and $\hat g\to 0$ limits for this state. 

Let us briefly analyse the result \eq{gluing}. 
First, we see that there are two crucially different cases $\hat g<1/4$ and $\hat g>1/4$.
For $\hat g<1/4$, we have that $\vartheta$ is real and positive. This means that the term depending on delta tends to $-\infty$, which then forces $\Gamma_{\rm cusp}\to\Delta_++n,\;n=0,1,\dots$ in order for the pole of the $\Gamma$ function to compensate. At the same time, for $\hat g>1/4$, $\vartheta$ is purely imaginary, and thus we can always find sufficiently large $n$ to compensate for this divergence. However, in this case, there is no way to get rid of the dependence of \comAI{on} delta completely, as there will always be a finite part which depends on the precise value of $\delta$.

Below we study these two cases in more detail, but we can already conclude that $\hat g<1/4$ and $\Gamma_\text{cusp}>0$ indicate a discrete tower of defect operators with dimensions $\Gamma_\text{cusp}=\Delta_++n$, with $n=0,1,2,\ldots$ (twice degenerate as in the limit where even and odd states give the same result). For $\hat g>1/4$, the spectrum instead exhibits walking behaviour.

\subsection{Below Critical Coupling}
In this section, we consider the case $\hat g<1/4$. In this regime, $\vartheta={1\over2}\sqrt{1-16\hat g^2}$ is real. The quadratic Casimir related to $\hat g$ via \eq{Cdef} satisfies ${\bf C}<1/4$, as expected for a unitary highest-weight representation of $SL(2,\mathbb{R})$. In the fusion limit $\delta\to0$, the term $\vartheta\log\delta$ in the FME~(\ref{gluing}) tends to $-\infty$.  Below, we explore the consequences of this.

\subsubsection{Positive $\Gamma_\text{cusp}$}
For $\Gamma_\text{cusp}>0$, the divergent term can only be balanced by approaching one of the poles of $\Gamma(\Delta_+-\Gamma_\text{cusp})$. Expanding (\ref{gluing}) around this limit, we obtain
\beq\la{tower}
\Gamma_\text{cusp}={1\over2}+\vartheta+m-\frac{\Gamma (m+2\vartheta+1)}{m!\Gamma (-2\vartheta) \Gamma (2\vartheta+1)}{\cal S}_\pm(\vartheta)\({\delta\over8}\)^{2\vartheta}+O(\delta^{4\vartheta},\delta^{2\vartheta+1})\,,
\eeq
where $m=0,1,2,\ldots$.
Comparing this expression with (\ref{gammacusp}), we conclude that the fusion limit flows to a non-trivial conformal defect with a conformal family of defect operators of dimensions $\Delta_m=\Delta_++m$. 
The corresponding primary operator has dimension $\Delta_0$, and its wavefunction is given by (one can simply use \eq{far} for this value of $\Gamma_{\rm cusp}=\Delta_+$)
\beq\la{psievenodd}
\psi_0^\text{even}(\sigma)=\(\sinh{|\sigma|\over2}\)^{\Delta_+}\,,\qquad\psi_0^\text{odd}(\sigma)={\rm sign}(\sigma)\(\sinh{|\sigma|\over2}\)^{\Delta_+}\,.
\eeq
Once combined with the prefactor in (\ref{Gst}), it reproduces the three-point function of primary operators on the line with conformal dimensions $\{\Delta_0,0,0\}$. At $\hat g=0$, the corresponding defect-changing operator reduces to the adjoint scalar of the free theory. The remaining operators in the family are its conformal descendants, obtained by acting with $m$ longitudinal derivatives on the primary. 

Next, comparing (\ref{tower}) with \eq{correctionintro}, namely 
the first correction in $\delta$, we further find that it should correspond to an irrelevant operator and also read off the structure constant
\beq\la{alphapm}
\Delta_\text{irr}=2\Delta_+\,,\qquad\alpha_\pm=-\frac{\Gamma (m+2\vartheta+1)}{m!\Gamma (-2\vartheta) \Gamma (2\vartheta+1)}{\cal S}(\vartheta)\,.
\eeq
Here, the $m$-dependent factor, $\frac{\Gamma (m+2\vartheta+1)}{m!\Gamma (2\vartheta+1)}$, precisely reproduces the change in the ratio (\ref{correctionintro}) between the primary operator $\cO_0$ and its $m$th descendant.

This irrelevant operator is expected \cite{Nagar:2024mjz} to factorise into a product of two conjugate defect-changing operators, the defect-creation operator $\cO_0$ and its conjugate defect-annihilation operator $\cO_0^\dagger$
\beq\la{Odelta}
{\mathbb O}_\delta\propto\cO_0^\dagger\times\cO_0\,.
\eeq
This expectation follows from the fact that, in the planar limit in which we work, the dimension of such a composite operator is simply the sum of the dimensions of its constituent operators, in agreement with $\Delta_\text{irr}=2\Delta_+$. Furthermore, recall that ${\mathbb O}_\delta$ is the least irrelevant operator near the stable fixed point, with $\delta$ playing the role of the RG scale. The flow from the unstable to the stable defect is therefore expected to terminate in a neighbourhood of the stable fixed point controlled by this operator. Such flows were analysed in \cite{Nagar:2024mjz}, providing further support for the identification (\ref{Odelta}).

\subsubsection{The orphan state $\Gamma_\text{cusp}<0$}

\begin{figure}
    \centering
    \includegraphics[width=\linewidth]{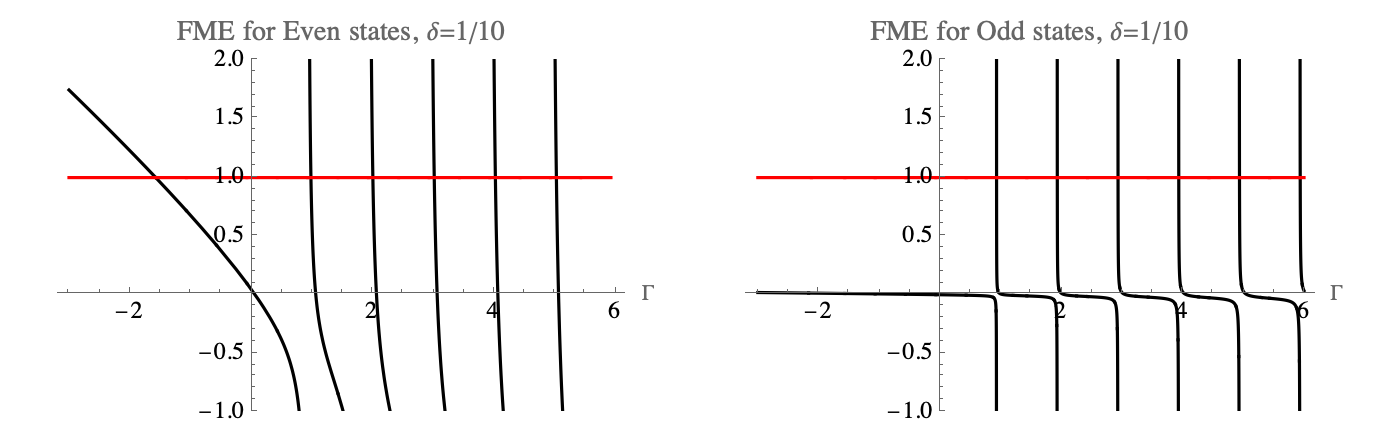}
    \caption{The exponential form of the FME, (\ref{gluingprod}) for the even and odd sectors at $\hat g=1/8$ and $\delta=1/10$. The intersections of the black curve (left-hand side of \eq{gluingprod}) with the red curve (right-hand side of \eq{gluingprod}) determine the physical values of $\Gamma_\text{cusp}$. We see that, whereas the even states are located near positive integers, the odd sector contains an additional ``orphan'' state, which we argue corresponds to the quark--antiquark potential (equivalently, to the bare cusp without insertions).}\label{fig:FMEevenodd}
\end{figure}

\begin{figure}
    \centering
    \includegraphics[width=0.5\linewidth]{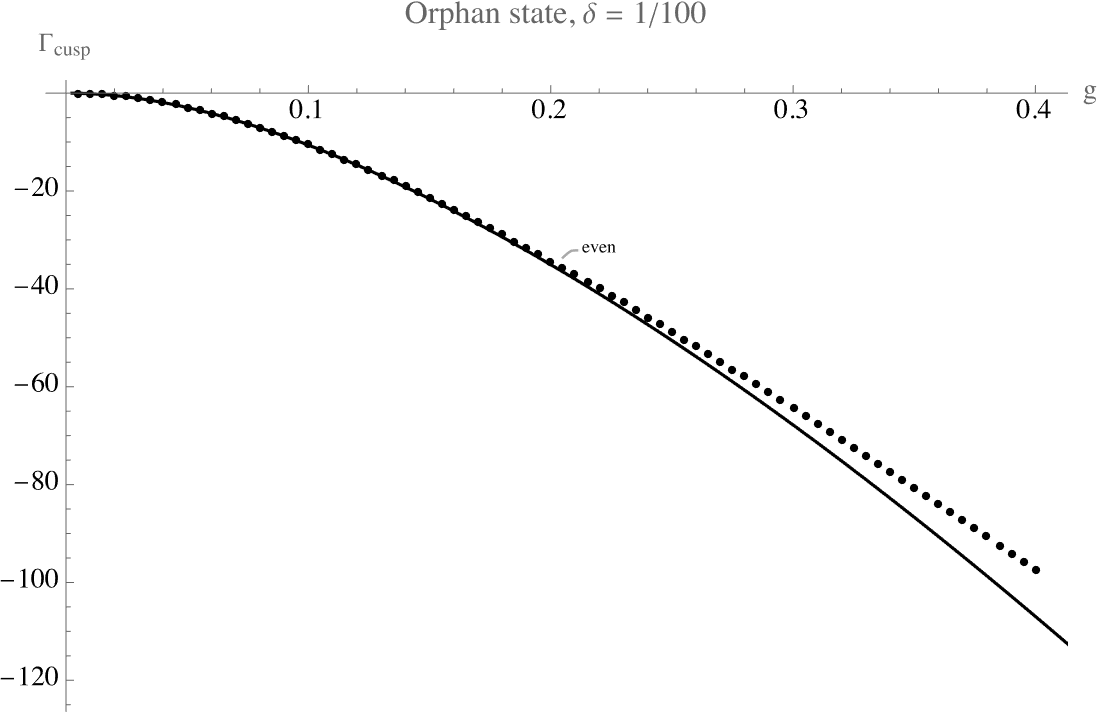}
    \caption{For the ground-state the energy scales as $\Gamma_{\rm cusp}\sim 1/\delta$, meaning that in principle FME should not apply as we assumed $\Gamma_{\rm cusp}\sim 1$ in the derivation. Nevertheless it still manages to capture it quite well, with error increasing with larger $g$. Solid line is solution of FME and dots correspond to the numerically exact solution at finite $\delta=1/100$.}
    \label{fig:FMEorphan}
\end{figure}

By looking at the plot of the exponential form of the FME equation, depicted in Figure~\ref{fig:FMEevenodd}, one can see that for the even case there is an additional solution which has $\Gamma_{\rm cusp}<0$\footnote{In fact, odd states also have an orphan solution but the key difference is that it diverges to infinity at weak coupling. For $g=1/8$, such a solution would be of order $-3000$, which is well beyond the applicability range of FME.}.
Note that there are no poles in  $\Gamma(\Delta_+-\Gamma_\text{cusp})$ to compensate for the $\log\delta$ term in FME \eq{gluing} at negative values. The only remaining way to balance the divergent factor proportional to $\vartheta\log\delta$ is therefore to take $\Gamma_\text{cusp}$ to be large and negative, in which case
\beq\la{GoGlimit}
\lim_{\Gamma_\text{cusp}\to-\infty}\log\frac{\Gamma\left(\Delta_+-\Gamma_\text{cusp}\right)}{\Gamma\left(\Delta_--\Gamma_\text{cusp}\right)}=2\vartheta\log(-\Gamma_\text{cusp})\[1+O(1/\Gamma_\text{cusp}^2)\]\,.
\eeq
Hence, the only solution to (\ref{gluing}) with $\Gamma_\text{cusp}<0$ arises in the regime where $\Gamma_\text{cusp}$ becomes parametrically large and negative
\beq
\Gamma_\text{cusp}=-{a(\hat g)\over\delta}\[1+O(\delta^2)\]\,,\qquad n=0\,.
\eeq
That is, we have a state with non-zero Casimir energy and zero mode number. That isolated state can only dress the trivial line.

Recall, however, that (\ref{gluing}) is not valid in the regime of parametrically large $|\Gamma_\text{cusp}|$. Consequently, it cannot be trusted for determining the coefficient $a(\vartheta)$, or even for establishing whether such a state exists. In fact, the exact finite-coupling solution of the cusp problem shows that a state with nonzero Casimir energy exists only in the even sector of wavefunctions; see figure \ref{specfig}.

The coefficient $a(\vartheta)$ determines the prefactor of the quark--antiquark potential, which has been studied extensively in the literature \cite{Pineda:2007kz,Correa:2012nk}, culminating in a full non-perturbative solution and analytically computed $7$ loops at weak coupling in  \cite{Gromov:2016rrp}. 

Let us denote by $\tilde a(\vartheta)$ the coefficients obtained by formally extrapolating (\ref{gluing}) beyond its regime of validity, that is, by naively continuing to use the equation at large negative $\Gamma_\text{cusp}$. In this way, we find that
\beq\la{apm}
\tilde a(\hat g)=8\[{\cal S}_+(\hat g)\]^{-{1\over2\vartheta}}\,.
\eeq

To compare $\tilde a(\vartheta)$ with the known perturbative expansion of $a(\vartheta)$, we expand both quantities as
\beq
a(\vartheta)=c_0+\hat\lambda c_1+\hat\lambda^2 c_2+\hat\lambda^3 c_3+\dots\,,\qquad \tilde a_+(\vartheta)=\tilde c_0+\hat\lambda\tilde  c_1+\hat\lambda^2\tilde c_2+\hat\lambda^3\tilde c_3+\dots\,,
\eeq
where $\hat\lambda\equiv(4\pi)^2\hat g^2$. The coefficients $c_i$ have been computed to high loop order in \cite{Pineda:2007kz,Correa:2012nk,Gromov:2016rrp}, where it was found that
\beqa\la{ladderseven}
c_0\!&=&\!0\,,\qquad c_1={1\over4\pi}\,,\qquad
c_2={1\over8\pi^3}\[\log{e^{\gamma_E}\hat\lambda\over2\pi}-1\]\,,\\
c_3\!&=&\!{1\over32\pi^5}\[\log{e^{\gamma_E}\hat\lambda\over2\pi}\(\log{e^{\gamma_E}\hat\lambda\over2\pi}+1\)-{7\over2}-{\pi^2\over12}\]\,,\nn\\
c_4\!&=&\!{1\over192\pi^7}\Bigg[\log^2{e^{\gamma_E}\hat\lambda\over2\pi}\(\log{e^{\gamma_E}\hat\lambda\over2\pi}+6\)-{1\over4}\(18+5\pi^2\)\log{e^{\gamma_E}\hat\lambda\over2\pi}+{13\over2}\zeta_3+{19\over12}\pi^2-15\Bigg]\,.\nn
\eeqa
Comparing this expression with the perturbative expansion of (\ref{apm}), we find that
\beq
\tilde c_0=c_0\,,\quad\tilde c_1=c_1\,,\quad\tilde c_2=c_2\,,\quad\tilde c_3=c_3+{1\over2^8\pi^3}\,,\quad\tilde c_4=c_4+{1\over92^9\pi^5}\Big(33\log{e^{\gamma_E}\hat\lambda\over2\pi}-35\Big)\,.
\eeq
We see that (\ref{apm}) captures most of the structure correctly. In particular, it reproduces the leading and subleading logarithms exactly. 

At this point, one can ask why this works at all, as we explicitly neglected the term $\delta \Gamma$ in the derivation of FME. The reason is that $c_0=0$, implying that, in fact, the correction to FME at weak coupling would be regular and of order $\hat g^2$. That is why it first only affects regular terms and higher orders in $\hat g^2$. At the same time, if we naively repeated the same procedure for the odd case, by replacing ${\cal S}_+$ with ${\cal S}_-$ in \eq{apm}, we find $\tilde a_-(\hat g)\simeq \frac{1}{g^2}+{ O}(g^0)$, which would immediately invalidate FME even at weak coupling. Even though FME fails at order $g^6$, it still seems to be useful to organise the perturbation theory by adding higher order terms in $g^2$ to FME.

\begin{figure}
    \centering
    \includegraphics[width=0.7\linewidth]{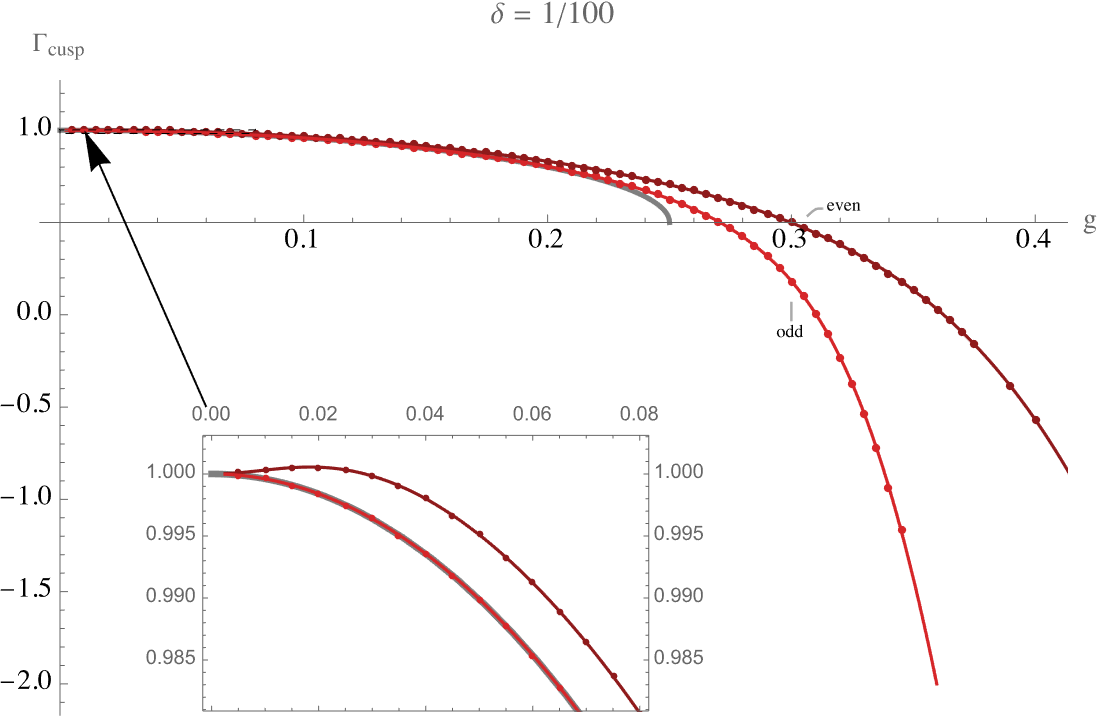}
    \caption{The high precision numerical data (dots) for two states with bare dimension $1$. The leading order spectrum is given by $\Gamma_{\rm cusp}=1/2+\vartheta$, and even for $\delta=1/100$ the difference is visible in the figure. Using the FME gives a much better approximation (solid lines) for all coupling regimes, even after the critical point. One of the reasons for good agreement is that FME captures all $1/\log\delta$ corrections at finite coupling and resums all divergences in $\delta$ in perturbation theory for the even state.}
    \label{fig:FMEperfection}
\end{figure}

\subsubsection{Non-commutativity of the Order of Limits}\la{noncommutativity}
As is common in theoretical physics, one frequently exchanges the order in which different limits are taken. It is therefore instructive to examine cases in which these limits fail to commute. In this subsection, we discuss the non-commutativity of the $\delta\to0$ and $\hat g\to0$ limits. For simplicity, we restrict attention to the two states with $\Gamma_\text{cusp}=1$ at $\hat g=0$. Their dimensions are known for arbitrary $\delta$ to two-loop order in the weak-coupling expansion; see, e.g., \cite{Cavaglia:2018lxi}. In the conventions of that reference, the even and odd dimensions read
\begin{align}
\Delta_{{\rm even}}&=1+4\,\hat g^2-16\,\hat g^4\left((\pi -\delta )\,\cot\left(\delta/2\right)+1\right)+{O}(\hat g^6)
\,,\\
\Delta_{{\rm odd}}&=1-4\,\hat g^2+16\,\hat g^4\left((\pi -\delta )\,\tan\left(\delta/2\right)-1\right)+{O}(\hat g^6)\nn\,.
\end{align}

We already see that the one-loop term of the even state $\Delta_{\rm even}$ has the opposite sign to the prediction of \eq{tower}, which gives $\tfrac12+\vartheta = 1-4\hat g^2-16\hat g^4+\cdots$ for both states at leading order in $\delta$. Thus, the one-loop result already signals the non-commutativity of the two limits. The issue becomes more pronounced at two loops, where the coefficient of $\hat g^4$ in $\Delta_{\rm even }$ diverges in the limit $\delta\to0$.\footnote{In section \ref{pertsec} we will interpret $\delta$ as an RG scale and cancel such terms with counter terms. This procedure will result in the correct perturbative expansion of $\Delta_{1,+}$.} To make this explicit, let us expand first in $\hat g$ and only then in $\delta$, retaining the first two nontrivial terms in the small-$\delta$ expansion
\begin{align}\label{deltaplus1}
\Delta_{\rm even}=&\,1 + 4\hat{g}^{2}
+\hat g^4\,\Big[-\frac{32\,\pi }{\delta }+16+O(\delta )\Big]+O(\hat g^5)
+ \ldots\,,\\
\Delta_{\rm odd}=&\,1-4\,\hat g^2+\hat g^4\,\big[-16+8\,\pi \,\delta +O(\delta ^2)\big]+O(\hat g^5)
+ \ldots\,.\nn
\end{align}
We see that $\Delta_{\rm odd}$ is regular in the limit $\delta\to0$ and smoothly approaches the result obtained when the fusion limit is taken first.

Let us now demonstrate that the fusion master equation \eq{gluing} not only reproduces the linear-in-$\delta$ correction to $\Delta_{\rm odd}$, but also correctly captures the non-commutativity of the $\delta\to0$ and $\hat g\to0$ limits in the case of $\Delta_{\rm even}$. To this end, we substitute the definition of $\vartheta(\hat g)$ from \eq{varthetadef}, together with a regular perturbative ansatz for the even state,
\beq
\Gamma_\text{cusp}=1+\gamma_1\hat g^2+\gamma_2\hat g^4+\ldots\,,
\eeq
into the master equation \eq{gluing}, and expanding at small $\hat g$ with $\delta$ held fixed, the leading order yields the condition
\beq
-\frac{\left(\gamma^{\rm even}_1-4\right)\,\delta }{4\,\pi \left(\gamma^{\rm even}_1+4\right)\,\hat g^2}=O(\hat g^0)\,.
\eeq
This can be satisfied only if $\gamma^{\rm even}_1=4$, which cancels the $1/\hat g^2$ divergence. This correctly reproduces the one-loop result and, in particular, the sign flip relative to the case where $\delta$ is taken to zero first. Expanding next to order $\hat g^0$ with $\gamma^{\rm even}_1=4$, we find
\beq
-\frac{\left(\gamma^{\rm even}_2-16\right)\,\delta }{32\,\pi } = 1+O(\hat g^2)\,,
\eeq
that is, $\gamma^{\rm even}_2=16-\tfrac{32\pi}{\delta}$, again in agreement with \eq{deltaplus1}. The same procedure also produces a prediction at the next perturbative order, namely
\beq
\gamma^{\rm even}_3=\frac{128\,\pi ^2}{\delta ^2}-\frac{128\,\pi\left(2\log(2\delta)-1\right)}{\delta }+O(\delta^0)+\ldots\,,
\eeq
which would be very interesting to compare against a direct perturbative computation. Similarly, for the odd state we find
\begin{align}
\gamma_1^{\rm odd}=&-4\,,\\
\gamma_2^{\rm odd}=&-16+8\pi\delta+\ldots\,,\nn\\
\gamma_3^{\rm odd}=&-128-32\,\pi\,\delta\left(2\log(2\delta)-3\right)+\ldots\,,\nn
\end{align}
again in agreement with the two-loop expression \eq{deltaplus1}. We conclude that the FME correctly resums the leading infrared divergences of perturbation theory and smoothly interpolates between the regimes in which the $\delta\to0$ and $\hat g\to0$ limits are taken in different orders.

\subsection{Above Criticality: The Density of States}

Consider now the case $\hat g>1/4$. In this regime $\vartheta$ becomes purely imaginary, so we write ${\rm Im}\,\vartheta = \tilde\vartheta\equiv{1\over2}\sqrt{16\hat g^2-1}>0$. The Casimir ${\bf C}=4\hat g^2>1/4$ no longer corresponds to a unitary highest-weight representation of $SL(2,\mathbb{R})$. The fusion master equation (\ref{gluing}) then takes the form
\beq\la{gluing2}
\boxed{\log\frac{\Gamma\left(1/2+\ii\tilde\vartheta-\Gamma_\text{cusp}\right)}{\Gamma\left(1/2-\ii\tilde\vartheta-\Gamma_\text{cusp}\right)}+2\ii\tilde\vartheta\log{\delta\over8}+2\pi\ii n+\log {\cal S}_\pm(\ii\tilde\vartheta)=O(\delta)}\,.
\eeq
For finite real $\Gamma_\text{cusp}$, one can no longer approach a pole of the gamma function, because of the imaginary shift by $\ii\tilde\vartheta$ in its argument. Consequently, the only way to balance the divergent phase is to take the mode number large,
$n\sim-{\tilde\vartheta\over\pi}\log\delta\to\infty$.
We see that when we change the regulator $\delta\to\delta'$, the states drift logarithmically, the mode number being shifted by $n\to n+\frac{1}{\pi}\tilde\vartheta\log \frac{\delta'}{\delta}$. Equivalently, as is clear from (\ref{gluing2}), the solutions exhibit a discrete scale symmetry\footnote{{A related phenomenon was discussed in \cite{Kaplan:2009kr,Iqbal:2011aj,Aharony:2023amq}.}}
\beq\la{descretesRG}
\delta\ \rightarrow\ e^{\pi\over\tilde\vartheta}\delta\,,\qquad n\ \rightarrow\ n-1\,.
\eeq

On the other hand, if we follow a fixed state, labelled by a fixed value of $n$, as $\delta\to0$, we find that its dimension gradually drifts towards large negative values of $\Gamma_\text{cusp}$. It is therefore natural to define the speed at which the states drift at a fixed energy scale $\Delta$ 
\beq
{\cal V}(\Delta)\equiv\left.-{\d\Gamma_\text{cusp}\over\d\log\delta}\right|_{\Gamma_{\rm cusp}=\Delta}\,.
\eeq
Due to the discrete scale symmetry (\ref{descretesRG}), that speed is proportional to the inverse density of states
\beq\la{rhodef}
\rho(\Delta) \equiv \left.\frac{1}{\partial_n\Gamma_{\rm cusp}(\tilde\vartheta)}\right|_{\Gamma_{\rm cusp}=\Delta}={\tilde\vartheta\over\pi}{1\over {\cal V}(\Delta)}=
\frac{\psi \left(\mathrm{i}\,\tilde\vartheta-\Delta +\frac{1}{2}\right)-\psi\left(-\mathrm{i}\,\tilde\vartheta-\Delta +\frac{1}{2}\right)}{2\,\pi \,\mathrm{i}}\,.
\eeq
Even though the spacing between states remains of order $1$, this density is defined as the states span the scale $\delta$. 
Note that it is independent of $\delta$, as well as of the details of the near-region dynamics encoded in ${\cal S}_\pm(\ii\tilde\vartheta)$. The total density is obtained by summing over all conformal families. In the present ladder model, there are only two such families, corresponding to the even and odd sectors.

A further observation is that, in this regime above the critical point, all terms in (\ref{gluing2}) are purely imaginary; in particular, ${\cal S}_\pm$ becomes a pure phase. 

Let us now study the asymptotic behaviour of $\rho(\Delta)$ in (\ref{rhodef}) at large $|\Delta|$,
\beq
\rho(\Delta)\simeq
\begin{cases}
\displaystyle
1+\frac{\sinh\!\left(2\pi \tilde\vartheta\right)}
{\cos\!\left(2\pi\Delta\right)+\cosh\!\left(2\pi \tilde\vartheta\right)},
& \Delta\to+\infty, \\[1.2em]
\displaystyle
\frac{\tilde\vartheta}{\pi|\Delta|},
& \Delta\to-\infty .
\end{cases}
\eeq
This means that, except for small oscillations at large positive $\Delta$, there is one state per unit interval. In the opposite limit, the density decays to zero, implying $\Gamma_{{\rm cusp},n}\sim -e^{-{\pi n\over\tilde\vartheta}}$, so that the states become increasingly sparse. Since the lowest state scales as $1/\delta$, we can estimate the total number of negative-energy states to grow as $\frac{\tilde\vartheta}{\pi}\log\frac{1}{\delta}$. 

As $\hat g$ crosses the critical value $1/4$, the spectrum undergoes a sharp transition. Because of the term $\tilde\vartheta\log\delta$ in (\ref{gluing2}), the limits $\delta\to0$ and $\hat g\to1/4$ do not commute.

\subsubsection{The Density and Smeared Observables}

Finally, let us make an interesting interpretation of the FME equation (\ref{gluing2}) as a momentum quantisation condition for a particle of momentum $\tilde\vartheta$ confined to a box of size $\log\frac{8}{\delta}$. The first and last terms can be interpreted as the phase shifts due to scattering at the walls of the box. In this picture, as the size of the box diverges when $\delta\to 0$, the spectrum becomes denser, and instead of tracking individual states, one can introduce their density. We should keep in mind, however, that in our case $\tilde\vartheta$ is a fixed function of $\hat g$; equation (\ref{gluing2}) is instead a quantisation condition on $\Gamma_{\rm cusp}$. We see, nevertheless, that while the spacing between consecutive solutions for $\Gamma_{\rm cusp}$ remains of order $1$ at fixed $\tilde\vartheta$, their locations are very sensitive to small changes in the coupling.

Based on this, one way to construct a scale-independent object is to consider an observable smeared over a small interval in the coupling or, equivalently, in $\tilde\vartheta$. More specifically, consider a sum over energies of the form
\beq
{\cal F}(\tilde\vartheta)=\sum_n F(\Gamma_{{\rm cusp},n}(\tilde\vartheta))
\eeq
and smear it over a range in $\tilde\vartheta$,
\beq
\langle {\cal F} \rangle\equiv\int \sigma(\tilde\vartheta)
{\cal F}(\tilde\vartheta)\,d\tilde\vartheta \,,
\eeq
where $\sigma$ is normalised, $\int \sigma\,d\tilde\vartheta = 1$, and is narrowly peaked around some value of $\tilde\vartheta$ with a width $\frac{1}{-\log\delta}\ll\Delta\tilde\vartheta\ll 1$. Each term in the sum is a rapidly varying function of $\tilde\vartheta$, advancing by one unit, $n\to n+1$, as $\tilde\vartheta\to \tilde\vartheta+\frac{\pi}{\log(\delta/8)}$. In other words,
\beq
\partial_{\tilde\vartheta}\Gamma_{{\rm cusp},n}(\tilde\vartheta) = \frac{\log(\delta/8)}{\pi\rho(\Delta)}+O(1)\,,
\eeq
which allows us to change the integration variable so that, in the limit $\delta\to 0$, we effectively obtain an integral over $n$,
\beq
\langle{\cal F}\rangle = \int F(\Delta)\,\rho(\Delta)\,d\Delta\,.
\eeq

\subsubsection{A Natural Interpretation of the Density}\la{neturalrhosec}

\begin{figure}
    \centering
    \includegraphics[width=\linewidth]{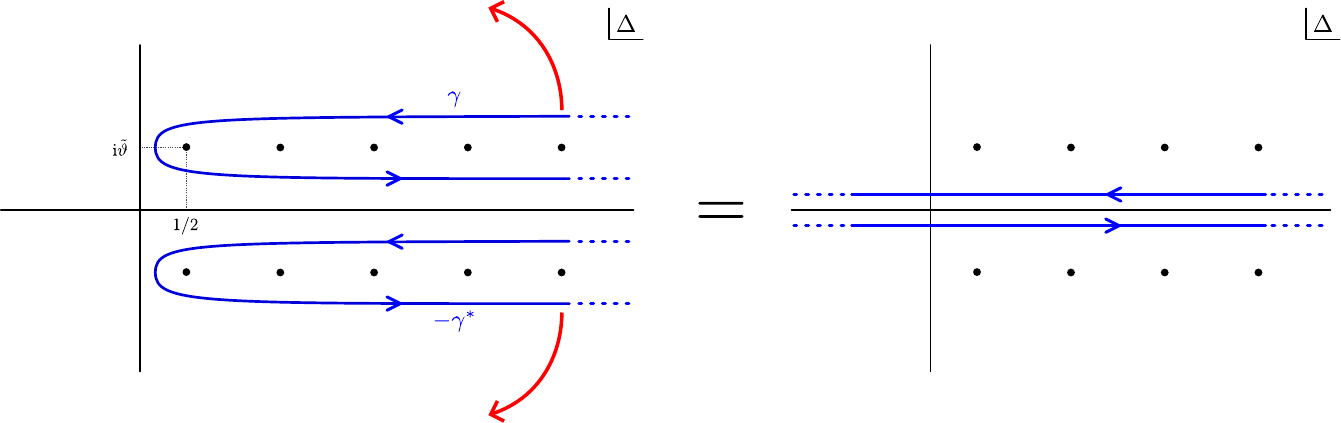}
    \caption{The contours $\gamma$ and $-\gamma^*$ encircle the poles of $\psi^{(0)}(\Delta_+-\Delta)$ and $\psi^{(0)}(\Delta_--\Delta)$, respectively. The upper part of $\gamma$ and the lower part of $-\gamma^*$ are opened through the upper and lower half-planes and extended to the negative real axis. After this contour deformation, the two contours combine into an integral along the whole real axis. Possible singularities of $f(\Delta)$ on the negative real axis are avoided by small indentation contours.}
    \label{psicontour}
\end{figure}

We show below that the density (\ref{rhodef}) emerges naturally when averaging a real observable that only depends on the energy over the two complex-conjugate fixed points.

Consider first the conformal family below the critical coupling, $\hat g<1/4$, where $\Delta_n=\Delta_++n$. Summing an observable $f(\Delta)$ over the states in this family yields
\beq\la{familysum}
F(\Delta_+)\equiv\sum_{n=0}^\infty f(\Delta_++n)\,.
\eeq
We can convert this sum into an integral by exploiting the fact that Euler's $\psi$-function, $\psi^{(0)}(-x)$, possesses simple poles at the non-negative integers, each with unit residue,
\beq\la{familyint}
F(\Delta_+)={1\over2\pi\ii}\subset\hspace{-4.5mm}\int\limits_\gamma\dd\Delta\,\psi^{(0)}(\Delta_+-\Delta)f(\Delta)\,,
\eeq
where the contour $\gamma$ wraps the real axis over the interval $[\Delta_+,+\infty)$ {and we assume that $f$ has no singularities along that interval.}

We now gradually increase $\hat g$. As it crosses the critical value $\hat g_c=1/4$, the poles of the $\psi$-function move into the complex plane, to
\beq
\Delta_n=n+{1\over2}\pm\ii\tilde\vartheta\,,
\qquad
\tilde\vartheta={1\over2}\sqrt{4{\bf C}-1}>0\,,
\eeq
where the sign depends on whether $\hat g$ crosses $1/4$ above or below the square-root branch cut of
$\vartheta={1\over2}\sqrt{1-16\hat g^2}$. The integral representation (\ref{familyint}) then provides an analytic continuation of $F(\Delta_+)$. It can be interpreted as the same observable evaluated at one of the two complex fixed points.

The analytic continuation of (\ref{familyint}) is generally complex, even if the observable obeys the reality condition
$f(\Delta^*)=f(\Delta)^*$. Its real part is
\beq\la{Resum}
{\rm Re}\,F(\widetilde\Delta_+)
=
{1\over4\pi\ii}
\subset\hspace{-4.5mm}\int\limits_\gamma
\dd\Delta\,
\psi^{(0)}(\widetilde\Delta_+-\Delta)f(\Delta)
+c.c.\,,
\eeq
where $\widetilde\Delta_+={1\over2}+\ii\tilde\vartheta$, and $\gamma$ is the analytic continuation of the original contour, deformed so as to encircle the poles of the $\psi$-function. The two terms in (\ref{Resum}) involve the contours $\gamma$ and $-\gamma^*$, the first lying above the real axis and the second below it.

We assume that $f(\Delta)$ is analytic in the regions swept out when $\gamma$ and $-\gamma^*$ are opened through the upper and lower half-planes, except possibly for poles or branch cuts lying on the negative real axis. We furthermore require the contributions from the large arcs at infinity to vanish. Since
$\psi^{(0)}(\Delta)\sim\log\Delta$
at large $|\Delta|$, a simple sufficient condition is
\beq
f(\Delta)=O\left(|\Delta|^{-1-\epsilon}\right),
\qquad
|\Delta|\to\infty\,,
\eeq
uniformly on the upper and lower arcs used in the deformation, for some $\epsilon>0$. 

Under these assumptions, the upper part of $\gamma$ can be opened through the upper half-plane and extended to the negative real axis, while the corresponding part of $-\gamma^*$ is opened through the lower half-plane, as shown in figure~\ref{psicontour}. Together with the parts of the original contours extending towards positive infinity, they combine into an integral along the whole real axis.

If {$f(x+1/2)$} has poles or branch cuts on the negative real axis, the contours must instead be deformed around these singularities. The result then takes the form
\beq\la{rhoaveragesingular}
{\rm Re}\,F(\widetilde\Delta_+)
=
\operatorname{PV}
\int\limits_{-\infty}^{\infty}
\dd\Delta\,\rho(\Delta)f(\Delta)
+
{\cal D}_{-}[f]\,,
\eeq
{where ${\cal D}_{-}[f]$ denotes the contribution arising from the discontinuities of $f$.}

The meaning of the density becomes particularly transparent by using the series representation of the digamma function. One finds
\beq\la{rholorentzian}
\rho(\Delta)
=
{\tilde\vartheta\over\pi}
\sum_{n=0}^{\infty}
{1\over
\left(\Delta-n-\frac12\right)^2+\tilde\vartheta^2}\,.
\eeq
Thus, every pair of complex-conjugate levels
\beq
\Delta_n^\pm
=
n+{1\over2}\pm\ii\tilde\vartheta
\eeq
produces a Lorentzian profile on the real energy axis. In this sense, $\rho(\Delta)$ is the positive real spectral function associated with the two complex-conjugate conformal towers. At the critical point,
\beq
\lim_{\tilde\vartheta\to0^+}\rho(\Delta)
=
\sum_{n=0}^{\infty}
\delta_{\rm D}\left(\Delta-n-\frac12\right)\,,
\eeq
so that (\ref{rhoaveragesingular}) reduces to the original discrete sum (\ref{familysum}).

Finally, the density is related to, but should not be identified with, the spectral measure of the $SL(2,\mathbb R)$ principal series. The pair
\beq
\Delta_\pm={1\over2}\pm\ii\tilde\vartheta
\eeq
forms a shadow pair in the principal series, and the same Gamma-function data appears in both constructions. However, the Plancherel measure is a measure over the principal-series parameter $\tilde\vartheta$, whereas $\rho(\Delta)$ is a density in the real energy $\Delta$ at fixed $\tilde\vartheta$.

\subsection{Strong Coupling and Falling Folded Strings in $AdS$}\la{foldedstringsec}

We now consider the double-scaling limit $\hat g,\Delta\to\infty$, with ${{\cal E}}=\Delta/\hat g$ held fixed. In this regime, the system lies above criticality, and the density (\ref{density}) reduces to the strong-coupling expression (\ref{dsdensity}). Our goal is to compare this density with the corresponding result in ${\cal N}=4$ SYM theory at strong coupling.

Although the ladder model can itself be obtained from ${\cal N}=4$ SYM through a particular double-scaling limit \cite{Correa:2012nk}, we will not take that limit here. Instead, we directly study the fusion of $1/2$-BPS Wilson lines in strongly coupled ${\cal N}=4$ SYM using its dual string-theory description on $AdS_5\times S^5$. This allows us to analyse quark--antiquark fusion semiclassically in terms of strings propagating in $AdS$.

A priori, there is no reason for the strong-coupling limits of the two theories to agree. Recall that the ladder limit is obtained as $g\to0$, while the angle $\theta$ between the scalar directions in R-space is scaled together with the line so that $\hat{g}^2 \equiv g^2\cos\theta$ stays finite. Nevertheless, we will see some structural similarity, and the reason for that will be evident when we consider ${\cal N}=4$ SYM and its relation to the universality of the FME.

However, we have seen that the ladder-theory density (\ref{dsdensity}) follows from the fact that the Casimir operator commutes with the Hamiltonian on the sphere. We will therefore interpret the agreement between the two strong-coupling densities as a first indication that the same commutation property also holds in ${\cal N}=4$ SYM. Later, in section \ref{sec: QSC}, we will establish this property at finite `t Hooft coupling using Quantum Spectral Curve techniques.

In ${\cal N}=4$ SYM theory, we study the fusion of a pair of conjugate $1/2$-BPS Wilson lines. They are defined by \cite{Maldacena:1998im}
\beq\la{MWL}
\cW_q=\cP\,\exp\int_\gamma\(\ii A\cdot\dd x+\Phi_1|\dd x|\)\,,
\eeq
where $\Phi_1$ is one of the six scalar fields of the theory, and $\gamma$ denotes the contour along which the Wilson line is supported. Under the AdS/CFT correspondence, this line operator is dual to a string worldsheet ending on the contour $\gamma$ at the boundary of $AdS_5$ and localised at a point on the $S^5$ factor.

The quark--antiquark fusion problem in flat spacetime is described by a contour $\gamma$ consisting of two antiparallel straight lines on the plane; see (\ref{flatfusion}). The corresponding dual string propagates in Poincar\'e $AdS$ and was studied long ago in \cite{Klebanov:2006jj}. Here, by contrast, we study the fusion problem in the presence of a cusp, for which the contour $\gamma$ consists of two antiparallel straight lines on the sphere. We must therefore revisit the analysis of \cite{Klebanov:2006jj} in global $AdS$.

We parametrise global $AdS_5$ as
\beq\la{ds2}
\dd s^2={R^2\over\sin^2\chi}\(\dd\chi^2-\dd t^2+\cos^2\chi\dd\Omega_3^2\)\,.
\eeq
The fused defect at $\delta=0$ extends along the time direction $t$ at the $AdS$ boundary $\chi=0$. To define such a non-conformal defect, one must introduce an RG scale, for instance, through a nonzero separation between the quark and antiquark lines, $r=R,\delta$. Here, however, we adopt a different regularisation scheme that is technically more convenient. Instead of separating the lines, we introduce a cutoff in the radial direction by restricting to $\chi>\chi_0>0$, and impose identical boundary conditions for the string at $\chi=\chi_0\ll1$.

\begin{figure}[t]
\centering
\includegraphics[width =\textwidth]{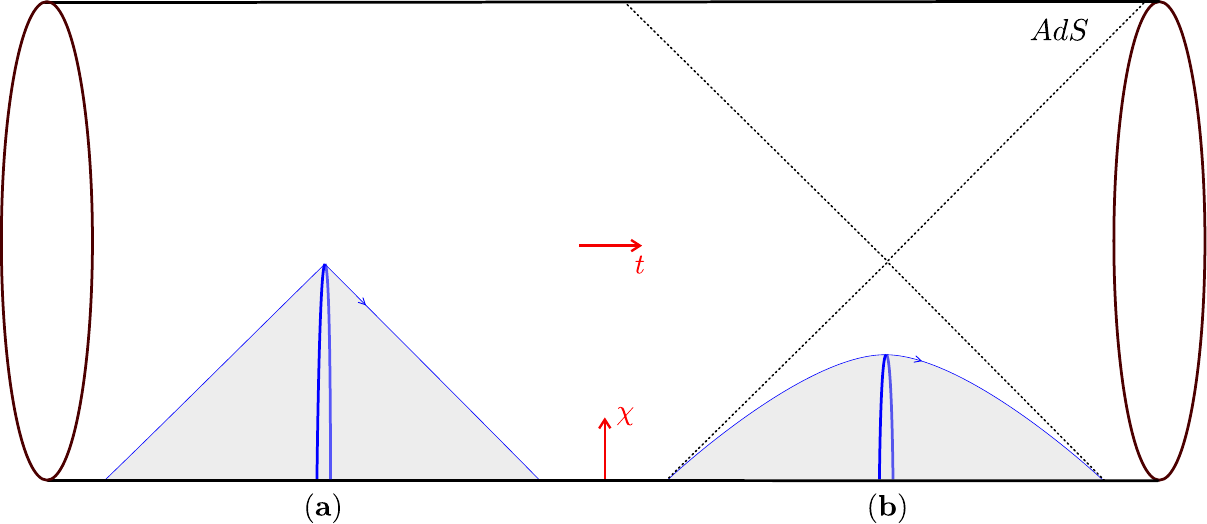}
\caption{${\bf a})$ A folded string in global $AdS$. The tip of the string behaves like a massless particle pulled towards the boundary by the tension of the two folds. ${\bf b})$ The same folded string extended along a finite opening angle in the $S^5$ factor; see section \ref{AdSsec}. Upon reduction on the $S^5$, the string acquires an effective mass density in $AdS$. As a result, the tip of the string, while accelerating towards the boundary, moves at a speed lower than that of light.}\label{fallingstring}
\end{figure}

The time-dependent string solution of interest extends along the time direction $t$ and the radial direction, from the UV cutoff near the boundary at $\chi_0$ to a maximal radial position $\chi(t)$. At that point, the string folds back towards the same boundary line at $\chi_0$ along a second sheet of the worldsheet. The tip of the folded string at $\chi(t)$ can be viewed as a massless particle pulled towards the boundary by the tension of the two string segments; see figure \ref{fallingstring}.a. The solution is entirely contained within an $AdS_2$ slice of $AdS_5$, parametrised by the coordinates $\chi$ and $t$ in (\ref{ds2}).

If we release the fold from rest at some radial position $\chi=\chi_M$, it accelerates towards the boundary. Its energy, evaluated at the moment it is released from rest, is simply twice the length of a string segment extending from the cutoff surface at $\chi_0$ to the turning point at $\chi_M$,
\beq\la{energyfolded}
E=(2g)\[2\int\limits^{\chi_M}_{\chi_0}{\dd\chi\over\sin^2\chi}-\int\limits^{\pi-\chi_0}_{\chi_0}{\dd\chi\over\sin^2\chi}\]=-4g\cot\chi_M\,,
\eeq
where $g={R^2\over4\pi\alpha'}={\sqrt{g_{YM}^2N}\over4\pi}$ is the `t Hooft coupling. This energy is negative because we subtract the area of a reference string stretching all the way from the boundary at $\chi_0$ to the opposite boundary point at $\pi-\chi_0$ on the other side of the $AdS_2$ slice. This subtraction removes the energy of a quark--antiquark pair located at the south and north poles of the sphere, corresponding to the denominator in (\ref{cusp2pf}).

Using energy conservation, the radial momentum of the massless particle at the fold, when it is located at some radial position $\chi<\chi_M$, is given by

\beq\la{pchi}
p_\chi=4g\(\cot\chi-\cot\chi_M\)\,.
\eeq
Using (\ref{pchi}), we can now apply the Born-Sommerfeld quantisation condition to count the number of quantum states that correspond to this classical solution. Following the solution as the fold propagates from the boundary to $\chi_M$ and back, we find
\beq\la{BS}
n={1\over2\pi}\oint p_\chi\dd\chi=%
{4g\over\pi}\[\log{\sin\chi_M\over\sin\chi_0}-(\chi_M-\chi_0)\cot\chi_M\]\,.
\eeq
We see that, as $\chi_0\to0$, the number of states scales as $n\sim{4g\over\pi}|\log\chi_0|$. Comparing this with the ladder-theory result, ${\tilde\vartheta\over\pi}|\log\delta|\to{2\hat g\over\pi}|\log\delta|$, we find agreement provided we identify $\delta\propto\chi_0$ and $\hat g$ with $2g$.

To compute the density of states, we note that
\beq
{\d n\over\d\chi_M}={4g\over\pi}{\chi_M\over\sin^2\chi_M}\,,\qquad {\d E\over\d\chi_M}={4g\over\sin^2\chi_M}\,,
\eeq
and therefore
\beq
\rho(E)={\d n\over\d E}={\chi_M\over\pi}\,.
\eeq 
Provided that $\lim_{E\to-\infty}\rho(E)=0$ and $g>0$, the density can be written as
\beq\la{scdensity}
\rho(E)={\d n\over\d E}={\chi_M\over\pi}=1+{\ii\over2\pi}\log{E+4\ii g\over E-4\ii g}\,.
\eeq    
The global $AdS$ energy should be identified with the dimension of the defect-changing operator, $E=\Delta$. We therefore conclude that (\ref{scdensity}) agrees with the result obtained from the Casimir equation (\ref{dsdensity}), provided we identify the `t Hooft coupling with one half of an effective ladder coupling,
$g=\hat g_\text{eff}/2={\bf C}/8$. In section \ref{sec: QSC}, we will confirm this identification by solving the fusion problem at finite coupling and only then taking the strong-coupling limit; see figure \ref{fig:geffoverg}.

Before concluding this section, let us make one final comment. A closely related string solution is the folded string pulsating about the centre of the global $AdS$. Such pulsating strings in $AdS_2$ were quantised in \cite{Vegh:2024uie} using integrability to resolve quantisation ambiguities. Repeating the same analysis for our solution reproduces the far Schr\"{o}dinger equation (\ref{largex}), or equivalently the Casimir equation (\ref{casimireq}) with $\Delta_1=\Delta_2=0$.

\section{The Defect RG Flow that Removes the Trivial Line%
}\la{DCORGsec}
We have referred to the conformal defect that we found as ``stable''. There are, however, two closely related potential sources of instability. First, for an RG fixed point to be stable, all operators must be irrelevant. In the case of a defect fixed point, this condition also applies to defect-changing operators. The reason is that the direct sum of conformal defects is itself conformal, and defect-changing operators should therefore be viewed as ordinary defect operators living on this enlarged defect theory. For the two fixed points discussed above at $\hat g<1/4$, the lowest-dimension defect-changing operators have dimensions
$\Delta_\pm=(1\pm\sqrt{1-16\hat g^2})/2<1$,
and are therefore relevant.

Another potential source of instability is the following. Since our conformal line arises from the fusion of a line in the fundamental representation with one in the anti-fundamental representation, it decomposes into singlet and adjoint components. At higher orders in the $1/N$ expansion,
loops of adjoint matter generate holes in the conformal flux sourced by the line. One may therefore worry that these holes could eventually screen the defect.

In this section, we show that the net effect of the RG flows triggered by the relevant defect-changing operators is to remove the trivial defect from the line Hilbert space. The endpoint of these flows is therefore the same conformal line defect with which we started, but without the empty line sector. As a consequence, the defect-changing operators themselves disappear, and the previously ``stable'' line becomes genuinely stable.

The removal of the trivial defect from the Hilbert space can be understood as resulting from the condensation of infinitesimal line segments separated by holes. These configurations generate an infinitely large cosmological-constant suppression factor for the empty line.

Another consequence of this picture is that, already in the planar limit, the stable conformal line is effectively ``full of holes'' on the line. We therefore expect that the holes in the conformal flux sourced by the line, which arise at higher orders in the $1/N$ expansion, produce only small corrections to the defect CFT.

\subsection{Planar DRG Flows Triggered by Defect Changing Operators}

The planar RG flow triggered by a relevant defect-creation/annihilation operator  can be analysed on completely general grounds, without reference to any particular model. The only assumptions required are
\begin{itemize}
\item A relevant defect-creation operator.
\item Large-$N$ factorisation of the expectation value of a product of conformal defects, each stretched between a pair of defect-changing operators.
\end{itemize}

Such flows were analysed in section 4 of \cite{Nagar:2024mjz} using a differential-equation approach. Here, we rederive and extend that analysis through an explicit resummation, with emphasis on the removal of the empty line sector from the defect Hilbert space. Readers already familiar with \cite{Nagar:2024mjz} may wish to skip the remainder of this section.

We distinguish between two qualitatively different regimes. In the first, the relevant defect-changing operator has dimension $\Delta_+>1/2$. In our example, this operator lives at the end of the stable defect and has dimension $\Delta_+={1\over2}(1+\vartheta)$. In the second regime, the relevant operator has dimension $\Delta_-\equiv1-\Delta_+<1/2$. In our example, this is the operator associated with the unstable defect; see (\ref{betaeta}). We denote the corresponding defect-changing operator by $\cO_\pm$ respectively, and normalise it so that the expectation value of a straight conformal line segment of length $L$, stretched between two such defect-changing operators, is given by
\beq
M_\pm(L)\equiv\<\cO^\dagger_\pm(0)\,\cW_\theta[0,L]\,\cO_\pm(L)\>={1\over L^{2\Delta_\pm}}\,.
\eeq

To set the stage, we begin with the flow generated by a relevant defect-changing operator of dimension $\Delta_-$. This flow is somewhat simpler because it does not remove the empty line sector from the defect Hilbert space.

To trigger an RG on the direct sum of the conformal and trivial defects, we deform the defect action as
\beq\la{deformation}
S_\text{line}\quad\rightarrow\quad S_\text{line}+\sqrt b\,\rho^{1-\Delta_-}\int\dd x\begin{pmatrix}0&\cO_-(x)\\\cO_-^\dagger(x)&0\end{pmatrix}\,.
\eeq
Here, $b$ is the deformation parameter, while $\rho$ sets the mass scale of the flow. The upper component corresponds to the conformal line defect, and the lower component to the trivial defect.

The line defect deformed by (\ref{deformation}) consists of partially filled and partially empty segments, connected by insertions of $\cO_-$ and $\cO_-^\dagger$. These ordered operators are integrated along the line. Consider, for example, the integral over the insertion point of a single $\cO_-$ separating an empty segment from a filled one. This integral evaluates to
\beq\la{emptyfilled}
\int\limits_0^L{\dd s\over s^{2\Delta_-}}={1\over1-2\Delta_-}{1\over L^{2\Delta_--1}}\,,
\eeq
where we used the fact that $\Delta_-<1/2$, ensuring that the integral is finite. We denote by $\widetilde M_b(L)$ the two-point function of the empty endpoint on the trivial defect after the deformation (\ref{deformation}) is turned on. This quantity can be determined, for example, by observing that it satisfies the Schwinger--Dyson equation
\begin{align}\la{SDeq}
\widetilde M_b(L)=&1+{b\over1-2\Delta_-}\rho^{2(1-\Delta_-)}\int\limits_0^L{\dd s\over s^{2\Delta_--1}}\widetilde M_b(L-s)\,,
\end{align}
or by performing a direct brute-force resummation of all insertions of the deformation operator in (\ref{deformation}). This leads to the series expansion
\beq\la{Msum}
\widetilde M_b(L)=\sum_{n=0}^\infty\({b\,(\rho L)^{2(1-\Delta_-)}\over1-2\Delta_-}\)^nJ_n\,,
\eeq
where $J_n$ denotes a generalised free-field diagram on the line containing $2n$ insertions of the deformation operator (\ref{deformation}),
\beq
J_n=\int\limits_{0<\sigma_1<\ldots<\sigma_n<1}{\dd\sigma_1\ldots\dd\sigma_n\over[(\sigma_n-\sigma_{n-1})\ldots(\sigma_2-\sigma_1)\sigma_1]^{2\Delta_--1}}=\frac{\Gamma\(2(1-\Delta_-)\)^n}{\Gamma\(2n(1-\Delta_-)+1\)}\,,
\eeq
see \cite{Nagar:2024mjz} for further details. Solving (\ref{SDeq}), or equivalently resumming the series in (\ref{Msum}), yields the closed-form expression
\beq
\widetilde M_b(L)=E_{2(1-\Delta_-)\,,1}\(b(\rho L)^{2(1-\Delta_-)} \Gamma(1-2\Delta_-)\)\,,
\eeq
where $E_{c,d}(z)$ denotes the Mittag--Leffler function; see \cite{Tsitouras:2011aa} for a review of its relevant properties.

For $b<0$, the correlator exhibits the following large-$L\rho$ behaviour,
\beq
\widetilde M_{b<0}(L)={1\over|b|\,\Gamma(1-2\Delta_-)^2(\rho L)^{2(1-\Delta_-)}}\[1+O\({1\over\rho L}\)\]\,.
\eeq
Hence, we conclude that the deformation (\ref{deformation}) generates an RG flow from the direct sum of the unstable conformal line and the trivial defect, with trivial identity defect-changing operator of dimension $\Delta=0$, to a new conformal line with boundary dimension
\beq\la{Delastar}
\Delta_*=1-\Delta_-=\Delta_+\,.
\eeq

For $b>0$ the large $L\rho$ behaviour of this correlator is 
\beq
\widetilde M_{b>0}(L)={1\over2\Delta_+}e^{\rho L\,\(b\Gamma(2\vartheta)\)^{1\over2\Delta_+}}+{(\rho L)^{-2\Delta_+}\over b\,\Gamma(2\vartheta)^2}\(1+O\left({1\over\rho L}\right)\)\,.
\eeq
Here, the first term represents a cosmological-constant factor multiplying the trivial defect, while the second term is of the form of the one obtained for the $b<0$ case.

Next, we consider the same class of defect RG flows, triggered by an operator of dimension $\Delta_+$, with $\cO_+$ replacing $\cO_-$ in (\ref{deformation}). This flow was analysed in detail in section 4 of \cite{Nagar:2024mjz} and closely parallels the discussion above. The main difference is that, since $\Delta_+>1/2$, the integral in (\ref{emptyfilled}) develops a UV divergence from the region where $s\to0$,
\beq\la{emptyfilled2}
\int\limits_\epsilon^L{\dd s\over s^{2\Delta_+}}={1\over1-2\Delta_+}\({1\over L^{2\Delta_+-1}}-{1\over\epsilon^{2\Delta_+-1}}\)\,,
\eeq
where $\epsilon$ is a point-splitting regulator. The first term arises from the upper limit of integration, where the empty segment shrinks to zero size. Assuming $b<0$, the flow again terminates at a conformal defect whose upper component in the direct sum has dimension $\Delta_{**}=1-\Delta_+=\Delta_-$.

The second term in (\ref{emptyfilled2}) originates from the lower limit of integration, where the filled segment shrinks to zero size. This divergent contribution exponentiates into the suppression factor
\beq
\lim_{\epsilon\to0}e^{-{|b|\over(2\Delta_+-1)}\epsilon^{1-2\Delta_+}}=0\,,
\eeq
at the lower component of the direct sum. Its effect is to remove the empty line sector from the Hilbert space of the theory. As a consequence, the defect-changing operators disappear as well, and no further defect RG flows can be generated by them. What remains are defect operators of the form $\cO_-^\dagger\times\cO_-$. This operator drives a flow back to the ``stable'' conformal line, which is now genuinely stable -- the empty line sector has been removed from the Hilbert space, and with it the relevant defect-changing operators.

\section{Quark Anti-quark Fusion in ${\cal N}=4$ SYM Theory}\label{sec: QSC}

We now turn to our main example, the fusion of $1/2$-BPS Wilson lines in ${\cal N}=4$ SYM theory at large $N$. 
We show that the qualitative structure is very similar to that of the planar ladder model.
The key difference is that the gauge theory contains infinitely many conformal families, 
rather than the two families encountered in the ladder model. Each family is characterised by its own Casimir ${\bf C}(g)$ and, consequently, by its own critical value of the 't~Hooft coupling at which conformality is lost and walking behaviour sets in when ${\bf C}(g_{\rm crit})=1/4$.
It is convenient to introduce the notation
\beq\la{geffdeff}
\vartheta(g)\equiv \frac{1}{2}\sqrt{1-4{\bf C}(g)}\,,%
\eeq
which plays the same role as (\ref{varthetadef}) in the ladder model.

The $1/2$-BPS Wilson line is defined as in (\ref{MWL}) \cite{Maldacena:1998im}
\beq\la{halfBPSline}
\cW_q=\cP\,\exp\int\(\ii A_\mu\dd x^\mu+\vec n\cdot\vec\Phi\,|\dd x|\)\,,
\eeq
where $\vec\Phi$ denotes the six scalar fields of the theory and $\vec n$ is a unit vector. We consider this Wilson line in the cusped configuration (\ref{cusp2pf}); see figure \ref{cylindermap}. Rather than choosing the same scalar-coupling vector on the quark and antiquark lines forming the cusp, we take the couplings to be specified by two unit vectors, $\vec n$ and $\vec n'$. This introduces a second angle, defined by $\cos\theta=\vec n\cdot\vec n'$ \cite{Maldacena:1998im,Drukker:1999zq}. The corresponding generalised cusp anomalous dimension therefore depends on the geometric cusp angle $\delta$, the internal angle $\theta$, the choice of cusp operator $\cO_{\rm cusp}$, and the `t Hooft coupling $g={\sqrt{g_{YM}^2N}\over4\pi}$. Note that the fact that the internal angle is a good quantum number of the cusp implies that the Hamiltonian on the sphere does not mix states with different internal angles. Through the fusion limit (\ref{gammacusp}), this property is inherited by the fused defects. In particular, the fused defects are likewise characterised by the same internal angle.\footnote{We thank O.~Aharony for discussions on this point.}

The problem of computing $\Gamma_\text{cusp}(\delta,\theta,g)$ admits an integrability-based description. It began with the TBA/Y-system formulation \cite{Correa:2012hh,Drukker:2012de}, which consists of an infinite set of integral equations for an infinite set of Y-functions. The Quantum Spectral Curve (QSC) reformulation trades this infinite system of functional relations for a finite number of functions with well-controlled analyticity properties and asymptotic behaviour. It was developed for the current set-up in~\cite{Gromov:2015dfa}, used to study the quark-anti--quark potential non-perturbatively for the first time in~\cite{Gromov:2016rrp}, and further
enabled the study of excited states in \cite{Cavaglia:2018lxi,Grabner:2020nis}.
To solve the fusion problem, we study the WL QSC for excited states in the fusion limit $\delta\to0$. We find that the system simplifies dramatically in this limit. Rather than computing the full spectrum of $\Gamma_{\rm cusp}$, the QSC re-adjusts to produce $\vartheta(g)$ and ${\cal S}(g)$, with the spectrum determined by the FME equation \eq{gluing}, exactly as in the ladder model. In particular, one of its sectors reduces precisely to the Casimir equation (\ref{casimireq}) with $\Delta_1=\Delta_2$.

\subsection{QSC for the Ladder Model and the Fusion Limit}\la{QSCsec}

The central objects of the QSC are a finite collection of functions, known as Q-functions \cite{Gromov:2015dfa}. These depend on a single complex variable $u$, called the \emph{spectral parameter}. The Q-functions satisfy
 a set of functional difference relations, called QQ-relations. In the ladder limit, the QFT essentially reduces to a quantum-mechanical system; as a consequence, the QSC also simplifies considerably, while still retaining all the crucial elements of the full ${\cal N}=4$ construction.
 Furthermore, the derivation of the fusion limit of the ladder model in the QSC form will be highly instructive for the next section, where the fusion of the full quark and anti-quark lines (\ref{halfBPSline}) in ${\cal N}=4$ SYM is considered.

\subsubsection{QSC of the Ladder Model}

The ladder model 
is related to the cusp in ${\cal N}=4$ SYM 
through a double-scaling limit, in which $\theta\to +i\infty$ while keeping $\hat g=\tfrac{1}{2}g\,e^{-i\theta}$ fixed, \cite{Correa:2012nk}. In this limit, the cusp QSC simplifies dramatically \cite{Gromov:2016rrp}, so we start with it.
While in the full ${\cal N}=4$ theory, there is a large number of Q-functions whose direct QFT representation is less transparent, the ladder Q-function can be linked directly to the Feynman diagram --- more specifically, to the wave function (\ref{Gst}) arising from the Bethe--Salpeter procedure. Specifically, they are linked by
a Mellin-like integral transformation \cite{Gromov:2016rrp,Cavaglia:2018lxi}
\begin{equation}\label{eq:qToF}
\psi(\sigma)=-\ii e^{-{\sigma\over2}\Gamma_\text{cusp}}\int\limits_{c-\ii\infty}^{c+\ii\infty}\frac{du}{u} \; \mathbf{q}(u) \, e^{w_\delta(\sigma) \, u}\,,\qquad c>0\,,
\end{equation}
where $w_\delta(\sigma)$ is defined as 
\beq\la{MTfactor}
e^{\ii w_\delta(\sigma)}=-\dfrac{\sinh(\frac{\sigma+\ii\delta}{2})}{\sinh(\frac{\sigma-\ii\delta}{2})}\,.
\eeq
Under this transformation, the Bethe--Salpeter equation \eq{BSPlad} becomes a finite-difference equation for $\mathbf{q}(u)$,
with polynomial coefficients
\begin{equation}\label{Bax2p}
\left(2u^2 \cos\delta +2u\,\Gamma_\text{cusp}\sin\delta +4 \hat{g}^2\right)\mathbf{q}(u) +u^2
	\mathbf{q}(u-\ii)+u^2 \mathbf{q}(u+\ii)=0\,.
\end{equation}
We refer to this equation as the Baxter equation.
As a second-order equation, it has two independent solutions, which we denote by $\mathbf{q}_\pm(u)$, with the following asymptotic behaviour
\beq\label{qasFish}
\mathbf{q}_\pm(u)\simeq u^{\pm\Gamma_\text{cusp}}e^{\pm(\pi-\delta)u}\quad{\rm as}\quad u\to +\infty\,.
\eeq
In the Schr\"{o}dinger equation approach, the spectrum of $\Gamma_\text{cusp}$ was determined by imposing certain boundary conditions on the wave function.
In the Q-function language, the same discrete spectrum is obtained instead from the following condition at $u=0$ \cite{Gromov:2016rrp,Cavaglia:2018lxi}\footnote{For a complete derivation, see Appendix B of \cite{Cavaglia:2018lxi}.}
\begin{equation}\label{qquant}
    \Gamma_{\text{cusp}}=-\frac{2\hat g^2}{\sin\delta}\frac{\mathbf{q}_+(0)\,\overline{\mathbf{q}}_+'(0)+\overline{\mathbf{q}}_+(0)\,\mathbf{q}'_+(0)}{\mathbf{q}_+(0)\,\overline{\mathbf{q}}_+(0)}\,.
\end{equation}
This condition is implicitly non-linear, as the right-hand side depends on $\mathbf{q}(u)$, which in turn depends on $\Gamma_\text{cusp}$ through the Baxter equation \eqref{Bax2p}. Solving \eqref{Bax2p} and \eqref{qquant} simultaneously yields a discrete spectrum of cusp dimensions identical to that found earlier.

\subsubsection{Fusion Limit of the Ladder QSC}
In the preceding analysis, we found that the $\delta\to 0$ limit requires considering and gluing together two distinct regimes, ``far'' and ``near''. In the Q-function language, these correspond, respectively, to large spectral parameters $u\sim 1/\delta$ and to $u\sim 1$. We now verify this correspondence and show how the previous results emerge in this new language.

Indeed, the parameter $\delta$ enters the Baxter equation (\ref{Bax2p}) only through the combination $\Gamma_\text{cusp}\cdot u\,\delta$. For finite $\Gamma_\text{cusp}\sim 1$, this naturally leads to the two asymptotic $u$-regimes described above. They are the counterparts of the $\sigma\sim1$ and $\sigma\sim\delta$ regions studied in Section~\ref{sec:ladders}. To see this, note that the variable conjugate to $u$ in the integral transform \eqref{eq:qToF} 
in the fusion limit takes the form 
\beq\la{wsmalldelta}
w_\delta(\sigma)\ \text{mod}\ 2\pi=\delta\coth{\sigma\over2}-\pi+O(\delta^3)\,,
\eeq
so when $\sigma\sim 1$ the variable $U=u\,\delta$ is natural, whereas when $\sigma\sim\delta$ the factor $\coth(\sigma/2)$ cancels the $\delta$ dependence, making $u\sim 1$ the appropriate scaling.
We shall now analyse each regime separately and then match the two solutions, in close parallel with the analysis of section~\ref{sec:ladders}.

\paragraph{Far Region.} 
For $u\sim1/\delta$ we introduce the rescaled spectral parameter $U=u\,\delta$ and define
\beq\la{qfardef}
q_\pm(U)\equiv e^{\mp \pi U/\delta}\mathbf{q}_\pm(U/\delta)\,.%
\eeq
Taking $\delta$ to be small while keeping $U\sim 1$, we expand the Baxter equation \eqref{Bax2p} in powers of $\delta$. In this limit, shifts of $u$ by $\pm i$ turn into derivatives of $q(U)$, and the Baxter equation reduces to the following second-order differential equation 
\begin{equation}\la{farDif}
 \left(4 \hat{g}^2-U^2+2 \Gamma_{\text{cusp}}  U\right)q(U)+U^2 q''(U)=0\,.
\end{equation}
The Mellin-type transform (\ref{eq:qToF}) in this regime takes the form%
\beq\la{Mellinfar}
\psi_\text{far}(\sigma) =-\ii e^{-{\sigma\over2}\Gamma_\text{cusp}} 
\int\limits_{c-i\infty}^{c+i\infty}\dfrac{dU}{U} e^{U\coth{\sigma\over2}} q_+(U)\,,
\eeq  
and similarly for $q_-(U)$. Under this transform, the Baxter equation (\ref{farDif}) is mapped to the far-region Bethe--Salpeter equation \eqref{largex}.

At large $U$, the two independent solutions of (\ref{farDif}) behave as $q_\pm\simeq e^{\mp U}U^{\pm\Gamma_\text{cusp}}$, in accordance with (\ref{qasFish}). Equation~\eq{farDif} can be solved analytically in terms of the Whittaker functions $M$ and $W$; the solutions with pure asymptotic behaviour at large $U$ are specific linear combinations of these functions. Since the quantisation condition \eq{qquant} requires only $q_+$, we give the properly normalised expression for that solution only
\beqa
    q_+(U)&=&2^{-\Gamma_\text{cusp} } W_{\Gamma_\text{cusp} ,\vartheta }(2 U)\,.
\eeqa
To match with the near regime we expand $q_+(U)$ at small $U$. The two leading powers, whose coefficients are needed for the matching, are
\begin{equation}\label{q1Lad}
    q_+(U)=\frac{2^{-\Gamma_\text{cusp} -\vartheta +\frac{1}{2}} \Gamma (2 \vartheta )}{\Gamma
   \left(-\Gamma_\text{cusp} +\vartheta +\frac{1}{2}\right)}U^{\frac{1}{2}-\vartheta }+\frac{2^{-\Gamma_\text{cusp} +\vartheta +\frac{1}{2}} \Gamma (-2 \vartheta )}{\Gamma \left(-\Gamma_\text{cusp} -\vartheta +\frac{1}{2}\right)}U^{\frac{1}{2}+\vartheta}+\ldots\,.
\end{equation}

\paragraph{Near Region.} 
In the regime $u\sim1$, the limit $\delta\to0$ can be taken directly in the Baxter equation (\ref{Bax2p}), yielding
\begin{equation}\label{BaxLads}
2 \left(u^2+2 \hat{g}^2\right) \mathbbm{q}(u)-u^2 (\mathbbm{q}(u-\ii)+\mathbbm{q}(u+\ii))=0\,,
\end{equation}
where we removed the anti-periodic factor ${\mathbbm q}(u)\equiv e^{\mp \pi u}{\bf q}(u)$. 
This equation admits two independent solutions with large-$u$ asymptotics (we use the labels $1$ and $2$ here to match the notation of the ${\cal N}=4$ SYM QSC introduced below):
\begin{equation}\la{largeubq}
    \bq_1\simeq u^{\frac12+\vartheta}\,,\quad\bq_2\simeq u^{\frac12-\vartheta}\,,\qquad u\to\infty\,.
\end{equation}
We choose this basis so that they have a regular large $u$ power expansion. Note that in analogy with the standard QSC asymptotic e.g. (\ref{qasFish}) the $\tfrac{1}{2}+{\vartheta}$ plays the role of the energy 
for the ``near'' sub-system. 
These solutions can be expressed analytically in terms of ${}_3F_2$ hypergeometric functions; explicit formulae are given in the Appendix as eq.~\eq{p14}.
To complete the re-derivation of the FME for the ladder model, we match the two regimes to determine the combination of $\bq_1$ and $\bq_2$ that corresponds to ${\bf q}_+$ in the $\delta\to 0$ limit.

\paragraph{Gluing.} The far and near solutions have a common domain of validity in the region $1\ll u\ll1/\delta$. Matching the small-$U$ expansion \eq{q1Lad} of $q_+$ with the large-$u$ asymptotics \eq{largeubq}, we find that as $\delta\to 0$, up to an overall coefficient,
\beq\la{qp}
{\bf q}_+(u)\simeq 
e^{\pi u}\(\bq_1 {\mathbb S}
+\,\bq_2\)\,,
\eeq
where we introduce the notation that will prove useful in the full QSC case below
\begin{equation}\la{SandA}
{\mathbb S}=2^{+8\vartheta}\frac{\Gamma(-2\vartheta)}{\Gamma(+2\vartheta)}\frac{1}{\cal S}
\qquad{\rm with}\qquad
    {{\cal S}}=\frac{ \Gamma \left(\frac{1}{2} -\vartheta-\Gamma_\text{cusp} \right)}{\Gamma \left(\frac{1}{2} +\vartheta-\Gamma_\text{cusp} \right)}\; \(\frac{\delta}{8}\) ^{-2 \vartheta }\,.%
\end{equation}
We recognise \eq{SandA} as the FME \eq{gluing}. It remains to fix ${\cal S}$ by imposing the quantisation condition \eqref{qquant}. Since the right-hand side of \eqref{qquant} contains a divergent factor of $1/\sin\delta$, the numerator multiplying it must vanish in the fusion limit. Using the explicit form of ${\bq}_1$ and ${\bq}_2$ from (\ref{p14}) and (\ref{qp}), the quantisation condition reduces to the following quadratic equation for ${\cal S}$
\beq
\mathcal{S}^2 - 2\sec(\pi\vartheta)\,\mathcal{R}\,\mathcal{S} + \mathcal{R}^2 = 0\,,\qquad\mathcal{R} = \frac{\pi^2\,\Gamma\!\left(\vartheta+\tfrac{3}{2}\right)}{\vartheta^2\sin^2(\pi\vartheta)\,\Gamma\!\left(\tfrac{3}{2}-\vartheta\right)\,\Gamma(\vartheta)^4}\,.
\eeq
The two solutions to this quadratic equation precisely match the two solutions ${\cal S}_{\pm}$ for even and odd states \eq{SpmSh}, thus accomplishing the re-derivation of the FME (\ref{gluingprod}).

We now show how all these key steps translate to the full ${\cal N}=4$ cusp case.

\subsection{Cusp QSC in ${\cal N}=4$ SYM Theory}\label{sec:QSCSolve}

We now generalise our QSC analysis of the fusion limit to ${\cal N}=4$ SYM theory. 
We begin with a brief review of the cusp QSC following~\cite{Gromov:2015dfa}. For a pedagogical introduction to the QSC formalism, see~\cite{Gromov:2017blm}.

A quantum field theory, and ${\cal N}=4$ SYM in particular, possesses infinitely many degrees of freedom, in contrast to the ladder limit, where the problem reduces to one-dimensional quantum mechanics. Correspondingly, the generalisation to the full theory involves an infinite set of parameters to be fixed by the quantisation conditions, rather than just the single parameter $\Gamma_{\rm cusp}$ appearing in the ladder model. Apart from this, the conceptual steps remain very similar, as we explain in this and the following sections.

Let us make a systematic comparison between the full QSC and its ladder limit
\begin{enumerate}
\item There are more fundamental Q-functions. Besides the four $\bQ_i$ associated with the $AdS_5$ sector, which reduce to $\sqrt{u}\,{\bf q}(u)$ in the ladder (or fishnet) limit, the full theory contains four additional Q-functions $\bP_a$ associated with the $S^5$ sector. These functions encode the infinitely many additional degrees of freedom absent in the ladder limit. More precisely, the $\bP_a$ admit a simple Laurent expansion in the Zhukovsky variable $x$ (see Appendix~\ref{app: QSCGeneral}), with the expansion coefficients providing the corresponding infinite set of unknown parameters.
\item The Q-functions encode $\Gamma_{\rm cusp}$ and the deformation angle $\delta$ through their large-$u$ asymptotics, in the same way as in the ladder limit \eq{qasFish}, (up to the relative factor of $\sqrt{u}$ from above) 
\begin{align}\label{qasymp}
    \bQ_1 \simeq&\,\mathbf{B}_1\, u^{1/2+\Gamma_\text{cusp}} e^{u(\pi-\delta)}\,, \qquad
\bQ_2 \simeq \mathbf{B}_2\, u^{1/2+\Gamma_\text{cusp}} e^{-u(\pi-\delta)}\,,\\
\bQ_3 \simeq&\,\mathbf{B}_3\, u^{1/2-\Gamma_\text{cusp}} e^{u(\pi-\delta)}\,, \qquad
\bQ_4 \simeq \mathbf{B}_4\, u^{1/2-\Gamma_\text{cusp}} e^{-u(\pi-\delta)}\,.
\end{align}
In a similar fashion, the new functions $\bP_a(u)$ encode the $\theta$-angle in the sphere and the R-charge of the cusp (see \eq{Pthph}), which we will set to zero here.

\item The Q-functions, in general, are no longer meromorphic. Instead, they possess an infinite sequence of branch cuts in the complex $u$-plane at $[-2g+in,2g+in]$ where $n\in{\mathbb Z}$. This is how the 't~Hooft coupling enters the construction.
\item As we now have an infinite set of parameters to fix, the single quantisation condition \eq{qquant} is replaced by a functional constraint.
More precisely, the analytic continuation of $\bQ_i$ across the cut $[-2g,2g]$ (denoted by the index $\gamma$) is related to a certain linear combination of $\bQ_i$ themselves as
\begin{equation}\label{GlueShort}
    \left(\frac{\bQ_i}{\sqrt{u}}\right)^\gamma=\Omega_i^{~j}(u)\left.\frac{\bQ_j(u)}{\sqrt{u}}\right|_{u\to-u}\,,\quad\Omega=\begin{pmatrix}1&0&0&0\\0&1&0&0\\0&\mathbf{a}_1\sinh(2\pi u)&1&0\\\mathbf{a}_2\sinh(2\pi u)&0&0&1\end{pmatrix}\,.
\end{equation}
The constants ${\bf a}_1$ and ${\bf a}_2$, appearing in the gluing matrix, are additional parameters which have to be tuned. As was noticed in the previous QSC studies \cite{Gromov:2015vua}, only the first two conditions (which do not involve any additional parameters) are sufficient to fix all parameters, and the other two are not independent.\footnote{At present, there is no analytic proof of this statement; it is based on numerical evidence and weak-coupling studies.} 
\item
The simple 2nd-order Baxter equation with fixed polynomial coefficients \eq{Bax2p} is replaced by a 4th-order finite-difference equation for $\bQ_i$ with coefficients depending on $\bP_a$
\begin{align}
    &\bQ_i^{[+4]}D_0-\bQ_i^{[+2]}\left[D_1-\bP^{a[+2]}\bP_a^{[+4]}D_0\right]+\bQ_i\left[D_2-\bP_a\bP^{a[+2]}D_1+\bP_a\bP^{a[+4]}D_0\right]\nn\\
    +&\bQ_i^{[-4]}\bar{D}_0-\bQ_i^{[-2]}\left[\bar{D}_1-\bP_a^{[-2]}\bP^{a[-4]}\bar{D}_0\right]=0\,,\label{baxter}
\end{align}
where $f^{[\pm m]}(u)\equiv f(u\pm{\ii m\over2})$. The matrices $D_n$ are defined in Appendix~\ref{app: QSCGeneral} in terms of $\bP_a$.
\end{enumerate}
This completes our summary of the key differences with respect to the ladder model. For a full exposition, we refer the reader to~\cite{Gromov:2015dfa}. Further details, as well as the technical definitions used throughout our calculations, are collected in Appendix~\ref{app: QSCGeneral}.

In the next section, we proceed with the fusion limit $\delta\to0$, which can again be done in parallel with the ladder-limit case.

\subsection{Fusion Limit of the ${\cal N}=4$ Cusp QSC}

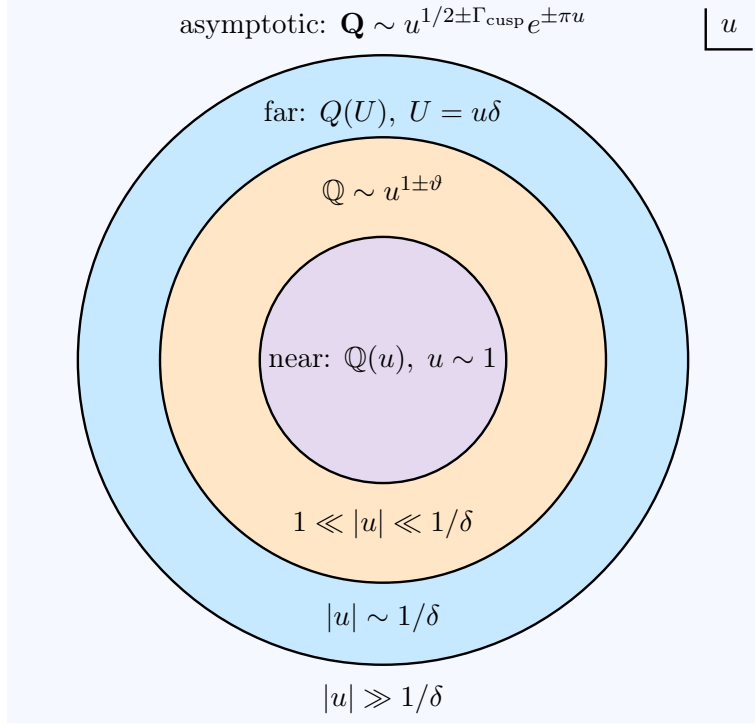
\begin{figure}
\begin{center}
\begin{tikzpicture}[line width=0.9pt, scale=1.55]

  \definecolor{outercol}{RGB}{0,153,255}
  \definecolor{midcol}{RGB}{255,140,0}
  \definecolor{innercol}{RGB}{111,45,168}
  \definecolor{bgcol}{RGB}{245,248,255}
   \fill[bgcol] (-3.2,-3.1) rectangle (3.2,3.1);

  \begin{scope}

    \path[fill=outercol!22, even odd rule]
      (0,0) circle (2.60) (0,0) circle (1.90);

    \path[fill=midcol!22, even odd rule]
      (0,0) circle (1.90) (0,0) circle (1.05);

    \fill[innercol!18] (0,0) circle (1.05);

  \end{scope}

  \draw (0,0) circle (2.60);
  \draw (0,0) circle (1.90);
  \draw (0,0) circle (1.05);

  \draw (2.75,3.00) -- (2.75,2.65) -- (3.15,2.65);
  \node at (2.95,2.85) {$u$};

  \node at (0, 2.88)
    {asymptotic: ${\bf Q} \sim u^{1/2\pm \Gamma_\text{cusp}} e^{\pm \pi u}$};

  \node at (0, 2.1)
    {far: ${Q}(U),\;U=u\delta$};

  \node at (0, 1.45)
    {${\mathbb Q} \sim u^{1\pm \vartheta}$};

  \node at (0, 0)
    {near: ${\mathbb Q}(u),\;u\sim 1$};

  \node at (0,-1.4)
    {$1\ll |u| \ll 1/\delta$};

  \node at (0,-2.20)
    {$|u|\sim 1/\delta$};

  \node at (0,-2.89)
    {$|u|\gg 1/\delta$};

\end{tikzpicture}
\end{center}

\caption{Schematic representation of the different regimes in the 
anti-parallel limit $\delta\to0$. The near region corresponds to $u\sim 1$, 
where the Q-function ${\mathbb Q}(u)$ satisfies a $\Gamma_\text{cusp}$-independent Baxter equation. 
The far region corresponds to $U=u\delta\sim 1$, 
where the rescaled Q-function $Q(U)$ satisfies a differential equation. 
The asymptotic region $|u|\gg 1/\delta$ is described by the large-$u$ 
behaviour of the original Q-function $\mathbf{Q}(u)$ up to an antiperiodic factor $e^{\pm \pi u}$.}

\label{fig:Dwarf}
\end{figure}

To take the fusion limit, we follow the same blueprint as in the ladder model, splitting the problem into two regimes: a far regime, where $U=u\delta\sim1$, and a near regime, where $u\sim1$. 

To avoid confusion between the various QSC limits, we introduce separate symbols for the Q-functions in each regime. For a summary of our notation, see Fig.~\ref{fig:Dwarf} and Table~\ref{tab:Qs}.

\paragraph{Far Regime of the Full QSC.}
As in the ladder model, the full fourth-order finite-difference equation \eq{baxter} reduces, in this regime, to a fourth-order differential equation
\begin{equation}\label{4Bax_far}
    \left(\frac{1-4 \Gamma_\text{cusp} ^2-\text{N}_1}{U^2}+\frac{\text{N}_2}{U^4}+1\right)Q(U)-\frac{2
   \text{N}_1}{U^3} Q'(U)+\left(\frac{\text{N}_1}{U^2}-2\right) Q''(U)+Q^{(4)}(U)=0\,,
\end{equation}
where, to take the limit, we have separated the rapidly oscillating exponent $e^{\pm \pi u}$ from the smooth piece $Q_i(U) = e^{\pm \pi U/\delta}{\bf Q}_i(U/\delta)$, in analogy with \eq{qfardef}.
The coefficients $N_1, N_2$ depend on the parameters stored in ${\bf P}$, see Appendix~\ref{app: QSCGeneral}, and are unknown at this stage.
For what follows, it is important that the relative error in $\delta$ to the equation \eq{4Bax_far} is of order $\delta^2$.

While we were not able to find an analytic solution to \eq{4Bax_far}, we can proceed to the next step and analyse the 
$U\to0$ asymptotics dictated by \eqref{4Bax_far}, which produce the following four power-law solutions
\begin{equation}\label{Qsmall}
    \{U^{1+\vartheta_1}~,~\;U^{1-\vartheta_2}~,~U^{2+\vartheta_2}~,~\;U^{2-\vartheta_1}\}\,,
\end{equation}
with the powers determined by the coefficients $N_1$ and $N_2$ via
\begin{equation}\label{thetann}
    \text{N}_1=-(\text{$\vartheta_1 $}-1) \text{$\vartheta_1 $}-\text{$\vartheta_2 $}^2-\text{$\vartheta_2 $}+2~~,~~\text{N}_2=(\text{$\vartheta_1 $}-2)
   (\text{$\vartheta_1 $}+1) (\text{$\vartheta_2 $}-1) (\text{$\vartheta_2 $}+2)\,.
\end{equation}
This differs slightly from the ladder model, where $\vartheta_1=\vartheta_2=\vartheta$.
Appealing to physical intuition --- as seen in the classical limit, $\vartheta$ is the effective energy of the solution at the scale where $\delta$ is negligible.
Having two independent exponents would imply a (potentially non-integer) effective spin in the directions transverse to the defect. While conceivable in general, for the simplest non-trivial state we expect the effective spin to vanish, giving $\vartheta_1=\vartheta_2$, or equivalently

\begin{equation}\label{constr}
4\text{N}_2-\text{N}_1(\text{N}_1+6)=0\,.
\end{equation}

While it would be interesting to find a more appealing analytic argument for this equality, instead we intensively tested the relation between the parameters $N_1$ and $N_2$ at finite small $\delta$'s numerically. We reached values of $\delta$ as small as $0.008\pi$, and observe that the numerical value of the left-hand side of \eqref{constr} decreases steadily, reaching $2\times10^{-2}$ at $\delta=0.008\pi$. We furthermore fitted the left-hand side of \eqref{constr} using powers of $\delta^k$, with $k=0,1,2,\dots$, and found the constant term in the fit to be $1.7\times10^{-5}$. The most stable fit we obtained starts at order $\delta^2$. 
We are therefore convinced that this relation is indeed correct in the $\delta\to0$ limit, and from now on we set
\begin{equation}\label{fullconstr}
\vartheta=\vartheta_1=\vartheta_2\,,\quad N_1=2(1-\vartheta^2)\,,\quad N_2=4-5\vartheta^2+\vartheta^4\,.
\end{equation}

\begin{table}[t]
    \centering
    \begin{tabular}{{|C{1.5cm}|C{7.3cm}|C{2.2cm}|C{1.4cm}|}}\hline
       {\small \bfseries Notation} &{\small \bfseries Definition} & {\small\bfseries Domain} & {\small\bfseries \shortstack{{}\vspace{1mm}\\Ladder\\limit}} \\\hline
        $\bQ_i$ & $\begin{matrix} \text{Default cusped Wilson Line QSC basis.}\\  \text{ Finite $\delta$. Pure asymptotics.}\end{matrix}$ & $u$ and $\delta$ not restricted & ${\mathbf{q}_i}{\sqrt{u}}$ \\ \hline
        $\mathbb{Q}_i$ & $\begin{matrix} \text{Near-region basis.}\\\text{ Pure asymptotics with exponents removed.}\end{matrix}$& $|u|\ll\frac1\delta$& ${\bq_i}{\sqrt{u}}$ \\\hline
        $Q_i$ & $\begin{matrix} \text{Far-region basis.}\\ Q_i(U)=e^{\pm \pi u}\bQ_i(U/\delta)\\\text{ Pure asymptotics with exponents removed.}\end{matrix}$ & $\frac1\delta<|u|$ & ${q_i}{\sqrt{u}}$\\\hline
        $\mathcal{Q}_i$ &$\begin{matrix} \text{Near-region basis with convenient gluing.}& \\\begin{matrix}
    \mathcal{Q}_1&=&\mathbb{Q}_1 \mathbb{S}+\mathbb{Q}_2=\mathcal{A}_n^{-1}Q_1~~~~~~~~~\\\mathcal{Q}_2&=&\mathbb{Q}_1 \mathbb{S}e^{-2\pi i \vartheta}+\mathbb{Q}_2=\mathcal{A}_n^{-1}Q_2\\
    \mathcal{Q}_3&=&\mathbb{Q}_3 \mathbb{S}+\mathbb{Q}_4~~~~~~~~~~~~~~~~~~~~~~\\\mathcal{Q}_4&=&\mathbb{Q}_3 \mathbb{S}e^{-2\pi i \vartheta}+\mathbb{Q}_4~~~~~~~~~~~~~
\end{matrix}\end{matrix}$ & $|u|\ll\frac1\delta$&--\\\hline
    \end{tabular}
    \caption{Conventions for different Q-function bases and their relation to the ladder limit notations.}
    \label{tab:Qs}
\end{table}
\noindent The above restriction simplifies the ODE $\eqref{4Bax_far}$ tremendously, as we found that this leads to the factorisation of the 4th order equation into two 2nd order ODEs
\begin{equation}\label{Baxterdecomp}
    \mathcal{L}_2\mathcal{L}_1Q(U)=0\,,
\end{equation}
where $\mathcal{L}_1$ and $\mathcal{L}_2$ are two second order differential operators, see Appendix \ref{app:BaxFactor}. In particular, \eq{Baxterdecomp} implies that two solutions of the initial equations are given by $\mathcal{L}_1Q=0$, or more explicitly
\begin{equation}\la{difff}
    \left(4g_\text{eff}^2-U^2+2 \Gamma_{\text{cusp}}  U\right)\left(\frac{Q(U)}{\sqrt{U}}\right)+U^2 \left(\frac{Q(U)}{\sqrt{U}}\right)''=0\,,
\end{equation}
where we have used \eqref{fullconstr} and $\vartheta(g)\equiv \frac{1}{2}\sqrt{1-16g_\text{eff}^2(g)}$
to express $N_1$ and $N_2$ in terms of $g_\text{eff}$. Remarkably, this equation coincides exactly with its ladder-limit counterpart \eqref{farDif}, up to the replacement $\hat g\leftrightarrow g_\text{eff}(g)$. As a result, the remainder of the far-regime analysis proceeds exactly as in the ladder limit. In particular, solutions to \eqref{difff} are given by $Q/\sqrt{U}\simeq\mathbf{q}_\pm$, with $\hat g$ replaced by $g_\text{eff}$. Since their asymptotics \eqref{qasFish} match those of the first two Baxter Q-functions of the Wilson line QSC, given in \eqref{qasymp}, we conclude that \eqref{difff} is solved by $Q_1(U)$ and $Q_2(U)$. See Table~\ref{tab:Qs} for their exact definition and Appendix~\ref{app:BaxFactor} for their analytic expressions.

\paragraph{Near regime.}
In the near regime we introduce, following the ladder model, ${\mathbb Q}_i$, which solves the Baxter equation \eq{baxter} (adjusted to remove the $e^{\pi u}$ factor) in the limit $\delta\to 0$ and $u\sim 1$, while having pure asymptotics when $1\ll u\ll 1/\delta$ similar to \eq{largeubq}. More precisely
\begin{equation}\label{qasnear}
\mathbb{Q}_i\simeq\left\{ u^{1+\vartheta},\  u^{1-\vartheta},\ \mathbb{B}_3 u^{2+\vartheta},\ \mathbb{B}_4 u^{2-\vartheta}\right\}_{i}\,,
\end{equation}
where the coefficients $\mathbb{B}_i$ are given in \eqref{BBBs}. Unlike in the ladder limit, the Baxter equation can no longer be solved analytically; we must resort to numerics or to the small-$g$ expansion described in Section~\ref{sec: QSCWeak}. To complete the fusion reduction, we must still derive the gluing condition.

When we take the limit $\delta\to 0$, we are not guaranteed that ${\bf Q}_i\to \mathbb{Q}_i$ rather than to some linear combination, as was the case in the ladder limit. At the same time, the gluing condition \eq{GlueShort} is sensitive to such linear transformations, so we must determine the precise mixing of $\mathbb{Q}_i$, which we do next.

\paragraph{Gluing.} The far and near Q-functions can be related to each other in their common domain. We do so by expanding the far-region Q-functions in the $U\to0$ limit and identifying within this expansion the large-$u$ asymptotics of $\mathbb{Q}_i$. The coefficients multiplying the $\mathbb{Q}_i$ asymptotics determine the transformation matrix relating the two Q-function bases. We summarise this result as the matrix relation
\begin{equation}
   Q_i(u\delta)=M_{i}^{~j}\mathbb{Q}_j(u)\,,\quad 1\ll u \ll\frac{1}{\delta}\,.
\end{equation}
Further details of this calculation, together with the explicit form of the matrix $M$, are provided in Appendix~\ref{app:BaxFactor}. Below we present this relation to leading order in $\delta$, up to an overall normalisation factor.
\begin{equation}\label{QfarQnear1delta}
    \mathbf{Q}_1\simeq e^{\pi u}(\mathbb{Q}_1\mathbb{S}+\mathbb{Q}_2)\,,\quad \mathbf{Q}_2\simeq e^{-\pi u}(\mathbb{Q}_1\mathbb{S}e^{-2\pi i \vartheta}+\mathbb{Q}_2)\,,\quad\mathbf{Q}_3\simeq\mathbf{Q}_4\simeq\mathbb{Q}_2\,,
\end{equation}
where the factor $\mathbb{S}$ is
given in terms of ${\Gamma}_{\rm cusp}$ and $\vartheta$ as before \eqref{SandA}. 

Note that since the equation \eq{difff} is identical to its ladder model counterpart, for ${\bf Q}_1$ the relation above is identical to that of the latter model. However, unlike in the ladder limit, due to the lack of analytic expressions for $\mathbb{Q}_i$, we cannot write an explicit expression for ${\mathbb Q}$. Instead, it must be found by solving \eq{baxter} either perturbatively or numerically.

As we noticed previously, it is sufficient to only impose the gluing for ${\bf Q}_1$ and ${\bf Q}_2$, which becomes
\beqa\la{gluQ1Q2}
e^{+\pi u}\(\frac{\mathbb{Q}^\gamma_1\mathbb{S}+\mathbb{Q}^\gamma_2}{\sqrt u}\) &=& 
e^{-\pi u}
\left.
\(\frac{\mathbb{Q}_1\mathbb{S}+\mathbb{Q}_2}{\sqrt u}\)\right|_{u\to -u}\,,\\
e^{-\pi u}\(\frac{\mathbb{Q}_1^\gamma\mathbb{S}e^{-2\pi i \vartheta}+\mathbb{Q}^\gamma_2}{\sqrt u}\) &=& 
e^{+\pi u}
\left.
\(\frac{\mathbb{Q}_1\mathbb{S}e^{-2\pi i \vartheta}+\mathbb{Q}_2}{\sqrt u}\)\right|_{u\to -u}\,.
\eeqa
Note that these two equations completely determine $\mathbb{Q}^\gamma_1$ and $\mathbb{Q}^\gamma_2$. Then one could be worried that the gluing condition for ${\bf Q}_3$, which, up to normalisation, coincides with ${\mathbb Q}_2$ at the leading order, could lead to a contradiction with \eq{gluQ1Q2}. In fact, however, the structure of $e^{\pm \pi u}$ \eq{gluQ1Q2} is very delicate, and the contradiction is avoided by fixing the unknown gluing matrix coefficients ${\bf a}_1$ and ${\bf a}_2$ as derived in Appendix \ref{app:BaxFactor}, eq. \eqref{bolda1a2}.

\paragraph{Curious Simplification of the Gluing Condition in the Fusion Limit.}

Even though, for the purpose of closing the equations for numerics or weak-coupling analytics, \eq{gluQ1Q2} is sufficient, we would like to highlight the hidden elegance in the fused QSC gluing conditions.

To reveal this hidden structure, we must push \eq{QfarQnear1delta} to the next order in $\delta$. Interestingly, the first correction to \eqref{4Bax_far} vanishes, so we simply retain terms of order $\delta$ in \eqref{QfarQnear1delta},
which are modified as follows:
\beqa
\label{QfarQnear}
    e^{-\pi u}\mathbf{Q}_1&\simeq& \mathbb{Q}_1 \;\mathbb{S}+\mathbb{Q}_2+\delta\;\mathcal{D}_n\left(\mathbb{Q}_3 \;\mathbb{S}+\mathbb{Q}_4\right)\,,\\
    e^{+\pi u}\mathbf{Q}_2&\simeq&\mathbb{Q}_1 \;\mathbb{S} e^{-2\pi i\vartheta}+\mathbb{Q}_2-\delta\;\mathcal{D}_n\left(\mathbb{Q}_3\;\mathbb{S} e^{-2\pi i\vartheta}+\mathbb{Q}_4\right)\,,
\eeqa
where $\mathcal{D}_n$ is some known but irrelevant constant factor, which can be found in Appendix \ref{app:BaxFactor}. Finally, to obtain the gluing conditions on the functions $\mathbb{Q}_i$, we substitute the above relations into \eqref{GlueShort}. We find that the functions $\mathbb{Q}_i$ naturally combine into pairs obeying a remarkably simple set of gluing conditions
\begin{equation}\label{THEGLUE}
\tilde{\mathcal{Q}}_i(u)=\mathcal{Q}_i(-u)\,\,,\qquad i=1,\dots,4\,,
\end{equation}
where we have defined combinations that naturally generalise \eq{gluQ1Q2}
\beqa\la{fullguingnear}
    &&\mathcal{Q}_1=\mathbb{Q}_1 \mathbb{S}+\mathbb{Q}_2\,,\quad\mathcal{Q}_2=\mathbb{Q}_1 \mathbb{S}e^{-2\pi i \vartheta}+\mathbb{Q}_2\,,
\\
\nn&&\mathcal{Q}_3=\mathbb{Q}_3 \mathbb{S}+\mathbb{Q}_4\,,\quad\mathcal{Q}_4=\mathbb{Q}_3 \mathbb{S}e^{-2\pi i \vartheta}+\mathbb{Q}_4\,.
\eeqa
Curiously, all $\sinh$-terms are absorbed into the twist factors $e^{\pm\pi u}$, giving way to the natural and elegant gluing expressions \eq{fullguingnear}.
They are reminiscent of the gluing conditions for local operators, for which the gluing matrix likewise lacks the $u$-dependent periodic exponentials. This simplification may hint at a hidden structure yet to be fully revealed.

\subsection{Summary of the Numerical Results}\label{sec: num}

Let us summarise the status thus far --- we have reduced the QSC to a new closed set of equations for ${\mathbb Q}_i$, free of any reference to $\delta$ or $\Gamma_{\rm cusp}$, which yield the pure quantities $\vartheta(g)$ and ${\cal S}(g)$. From these, the full spectrum of the conformal family $\Gamma_{\rm cusp,n}$ at arbitrarily small $\delta$ can be reconstructed via the FME equation \eq{gluing}.

In this section, we present a numerical analysis of the reduced QSC system and generate data for ${\vartheta}$ and ${\mathbb S}$ for the two simplest conformal families, corresponding to the insertion of a single scalar operator at the cusp. The precise form of these operator insertions is determined in the next section. At the level of the QSC, however, they are characterised solely by their quantum numbers, and the formalism is agnostic to their perturbative field content. To specify a particular state, one may, for example, use perturbative data as the starting point for the numerical solver, as was done in~\cite{Gromov:2023hzc}.
In our case, we initialise the QSC solver using a combination of random starting points and data from the ladder limit, where the solution is known analytically.

We use the well-established numerical procedure of \cite{Gromov:2015wca} described in detail in a recent paper \cite{Gromov:2023hzc}. We only have to specify a different type of asymptotics of $\bP$ and $\bQ$ and change the gluing conditions, but otherwise we use the method of \cite{Gromov:2015wca}. All the relevant properties of the Q-functions required to implement the algorithm, as well as its main steps, are summarised in Appendix~\ref{app: QSCGeneral}.

\begin{figure}[H]
    \centering
    \includegraphics[width=0.9\linewidth]{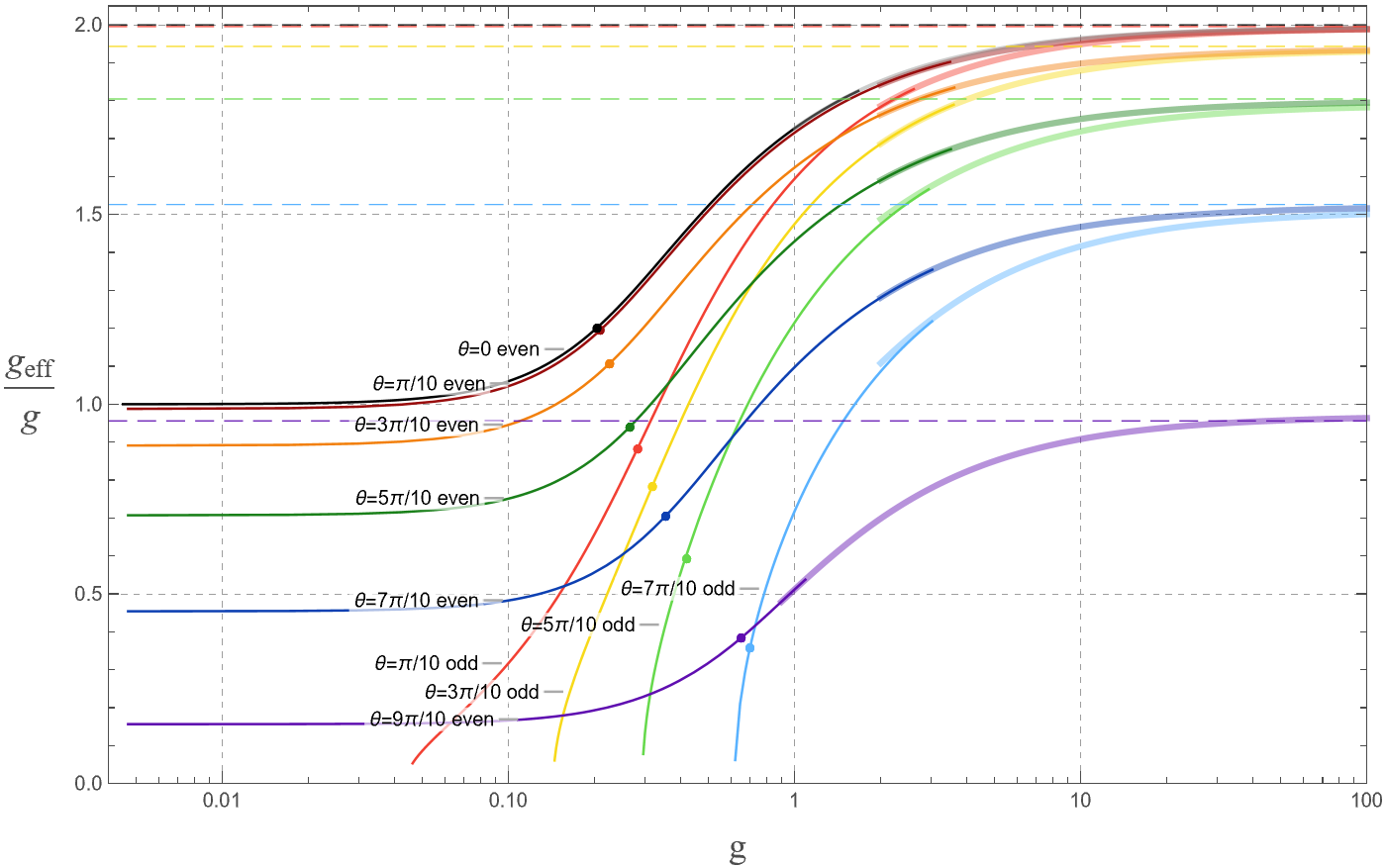}
    \caption{The ratio $g_{\text{eff}}/g$ (where $g_{\rm eff}(g)\equiv \frac{1}{4}\sqrt{1-4\vartheta^2(g)}$) as a function of coupling in six cases of angle $\theta\in\{0,\frac{\pi}{10},\frac{3\pi}{10},\frac{5\pi}{10},\frac{7\pi}{10},\frac{9\pi}{10}\}$. Each curve is labelled with the corresponding angle $\theta$ as well as a label ``even'' or ``odd'' indicating the type of state it belongs to. Each curve is also split into two parts: The solid line represents data obtained by a numerical QSC solver, while the transparent part of each plot represents a fit in inverse powers of coupling for numerical data starting from $g>1.5$. Dots on the lines mark the critical points for each case, which are written out in Table \ref{tab:gcrit} below. Horizontal dashed lines represent the String Theory prediction 
 for the $g_{\text{eff}}/g$ ratio given in Table \ref{tab:strong-coupling}. We see that at strong coupling both types of lines collapse to the same values predicted by the string theory description.}
    \label{fig:geffoverg}
\end{figure}

We now describe the main numerical results. We have solved the QSC for both even and odd states for a range of values of the angle $\theta$,
\begin{equation}\label{angles}
    \theta=\left\{0,\frac{\pi}{10},\frac{3\pi}{10},\frac{5\pi}{10},\frac{7\pi}{10},\frac{9\pi}{10}\right\}\,,
\end{equation}
and for coupling values in the range $0.004 \lesssim g \lesssim 3$ for most even states, and $0.09 \lesssim g \lesssim 3$ for most odd states. The difference in the lower bound on $g$ arises because $\mathbb{S}$ diverges as $g^{-4}$ in the $g\to0$ limit; see \eqref{SandA} together with the weak-coupling expansion \eqref{oddS}. As a consequence, significantly higher numerical precision is required to solve the system for odd states at weak coupling. For comparison, in the even-state case it is sufficient to retain about 14 terms in the Laurent expansion of $\bP_a$ in \eqref{Pthph}, whereas the odd states typically require roughly twice as many terms.

As mentioned, the highest value of the coupling we managed to reach is around 3 for most states (with the actual maximum being $g=3.61$ for $\theta=3\pi/10$). As $g$ increases, so does the number of terms that must be retained in the expansion of $\bP_a$, since the series converges more efficiently at weak coupling. As a result, the numerical algorithm becomes significantly slower at larger values of $g$. For $g\sim3$, we typically need to retain around 80 terms on average in the expansion \eqref{Pthph}.

\paragraph{Special Cases.} Both states at $\theta=9\pi/10$ could be computed only up to significantly smaller coupling values ($g\sim1$ for the even state and $g\sim0.6$ for the odd state). As before, we lose numerical precision rapidly at this special value of $\theta$, which approaches the limit $\theta=\pi-\delta$, where one expects new phenomena to emerge. For example, the quark--antiquark potential vanishes in this limit. Additionally, ${\mathbb S}$ for the odd state diverges as $(\pi-\theta)^{-4}$ as $\theta\to\pi$. This is evident from its weak-coupling expansion (see Appendix~\ref{app: weak}) and, in turn, significantly slows down the numerics. By contrast, the weak-coupling expansion of the even state indicates that $\vartheta$ approaches $1/2$ from below at every loop order. Through \eqref{geffdeff}, this implies that the associated phase transition, occurring at $g_{\text{eff}}=1/4$, is pushed to infinite coupling as $\theta\to\pi$; see Table~\ref{tab:gcrit}. Owing to these additional numerical challenges, the $\theta=9\pi/10$ case is not included in all of our tables.
\begin{table}[H]
    \centering
    \small
    \setlength{\tabcolsep}{9pt}
    \renewcommand{\arraystretch}{1.15}

    \textbf{Critical coupling
    $\boldsymbol{g_{\mathrm{crit}}}$ at which
    $\boldsymbol{\vartheta=0}$}

    \vspace{0.4em}

    \begin{tabular}{lcccccc}
        \toprule
        & \multicolumn{6}{c}{$\theta$} \\
        \cmidrule(lr){2-7}
        \textbf{State}
        & $0$
        & $\frac{\pi}{10}$
        & $\frac{3\pi}{10}$
        & $\frac{5\pi}{10}$
        & $\frac{7\pi}{10}$
        & $\frac{9\pi}{10}$ \\
        \midrule

        Odd state
        & --
        & $0.282853$
        & $0.318667$
        & $0.417350$
        & $0.698638$
        & -- \\

        Even state
        & $0.207242$
        & $0.209179$
        & $0.225735$
        & $0.265788$
        & $0.354421$
        & $0.650836$ \\

        \bottomrule
    \end{tabular}

        \caption{
        Critical values of the coupling $g_{\mathrm{crit}}$ at which $g_{\mathrm{eff}}=1/4$ and $\vartheta=0$. Across this point, $\vartheta$ changes from real to purely imaginary. The rows correspond to the odd and even states, while the columns correspond to different values of the angle $\theta$. Dashes indicate cases for which no critical value was obtained. The same results are listed to much higher numerical precision in Appendix~\ref{app: weak}, Table~\ref{tab:gcritFull}.
    }
    \label{tab:gcrit}

\end{table}

For all values of the angle $\theta$, the numerical algorithm is initialised at a small coupling value $g\sim0.1$ to ensure faster convergence. It is then gradually increased to probe the strong-coupling regime $g\gg1$.
An additional difficulty is that near the critical value of $g$ at which $\vartheta$ vanishes, the convergence of the Newton method is affected because two different solutions of the QSC with $\vartheta\leftrightarrow-\vartheta$ collide at this point, making convergence linear. In practice, this issue can be circumvented by introducing a small imaginary shift of the coupling in the vicinity of $g_{\text{crit}}$. We have determined these critical values $g_{\text{crit}}$ with at least five-digit precision for all cases considered and list them in Table~\ref{tab:gcrit}. In both even and odd states, $g_{\text{crit}}$ increases with $\theta$.

Another special case is $\theta\to0$, where some of the functions $\bP_a$ develop divergences. This can be seen by inspecting their prefactors $\mathbf{A}_i$ in \eqref{AAAA} (in the $\delta\to0$ limit), and is also reflected in the weak-coupling expansions of the coefficients $a_k$ and $b_k$. This issue can be addressed by making certain linear combinations of $\bP$'s or modifying the ansatz \eqref{Pthph} to work in the $\theta=0$ case specifically. We adopt the modified ansatz from \cite{Grabner:2020nis} and adjust the QSC.  Using this method, we found the numerical values of $g_{\text{eff}}/g$ for $\theta=0$. Comparing this value with $\theta=\pi/10$
, we can see that they are in good agreement with each other for the even state, as can be seen in Fig.~\ref{fig:geffoverg}.

We would also like to note that an estimate for the $g_\text{crit}$ value of the even state $\theta=0$ case was previously obtained in \cite{Alday:2025pmg}. In that work, the numerical QSC algorithm for arbitrary cusp angle was applied at small but finite values of $\delta$ in order to approximate the fusion limit. The value listed in Table~\ref{tab:gcrit} improves upon that earlier result in terms of numerical accuracy.   

\begin{table}[H]
    \centering
    \small
    \setlength{\tabcolsep}{9pt}
    \renewcommand{\arraystretch}{1.15}

    \textbf{(a) Effective-coupling ratio
    $\boldsymbol{\beta=g_{\mathrm{eff}}/g}$}

    \vspace{0.4em}

    \begin{tabular}{lcccccc}
        \toprule
        & \multicolumn{6}{c}{$\theta$} \\
        \cmidrule(lr){2-7}
        \textbf{State/source}
        & $0$
        & $\frac{\pi}{10}$
        & $\frac{3\pi}{10}$
        & $\frac{5\pi}{10}$
        & $\frac{7\pi}{10}$
        & $\frac{9\pi}{10}$ \\
        \midrule
        Odd state
        & -- & $1.9903$ & $1.9427$ & $1.803{\color{white}0}$ & $1.524{\color{white}0}$ & -- \\

        Even state
        & $1.999{\color{white}0}$ & $1.994{\color{white}0}$ & $1.9428$
        & $1.8047$ & $1.5265$ & $0.97{\color{white}00}$ \\

        Analytics
        & $2{\color{white}.0000}$ & $1.9954$ & $1.9434$
        & $1.8048$ & $1.5266$ & $0.9561$ \\
        \bottomrule
    \end{tabular}

    \vspace{1.2em}

    \textbf{(b) Constant term
    $\boldsymbol{\mathcal C}$ in
    $\boldsymbol{\log(\mathbb S)/(2ig)
      =-2\beta\log g+\mathcal C+O(g^{-1})}$}

    \vspace{0.4em}

    \begin{tabular}{lccccc}
        \toprule
        & \multicolumn{5}{c}{$\theta$} \\
        \cmidrule(lr){2-6}
        \textbf{State/source}
        & $0$
        & $\frac{\pi}{10}$
        & $\frac{3\pi}{10}$
        & $\frac{5\pi}{10}$
        & $\frac{7\pi}{10}$ \\
        \midrule
        Odd state
        & -- & $-1.578~$ & $-1.822~$ & $-2.241~$ & $-2.682~$ \\

        Even state
        & $-1.541~$ & $-1.5797$ & $-1.8326$
        & $-2.2449$ & $-2.6884$ \\

        Analytics
        & $-1.5452$ & $-1.5818$ & $-1.8331$
        & $-2.2452$ & $-2.6886$ \\
        \bottomrule
    \end{tabular}
        \caption{
       Strong-coupling comparison between numerical fits (the first two rows in each table) and analytical predictions (the last row). Panel (a) shows the limiting effective-coupling ratio $\beta=g_{\mathrm{eff}}/g$, with the analytical prediction obtained from \eqref{QL}. Panel (b) shows the constant term $\mathcal C$ in $
\log(\mathbb S)/(2ig)
=
-2\beta\log g+\mathcal C+O(g^{-1})\,,
$
with the analytical prediction following from \eqref{STSexp}. Dashes indicate cases for which no reliable numerical fit could be obtained.
    }
    \label{tab:strong-coupling}

\end{table}

\paragraph{Strong Coupling.} Having solved the QSC for coupling values as large as $g\sim 3$, we can compare our results with string theory predictions. As we show in later sections, from the string theory side, we have identified only a single solution for each value of $\theta$, with no distinction between even and odd states. This appears to be consistent with our numerical results, as the ratio $\beta=g_{\text{eff}}/g$ converges to the same value for both even and odd states at sufficiently large $g$. We verify this by fitting our numerical data in the range $1.5<g\lesssim 3$ for both states and for most values of $\theta$ in \eqref{angles}. We use an expansion in inverse powers of $g$, and present the results in Fig.~\ref{fig:geffoverg}. The leading coefficients of these fits are determined with three- to four-digit accuracy and agree between even and odd states to approximately two-digit precision (see Table~\ref{tab:strong-coupling}). Moreover, string theory provides an analytic prediction for the large-$g$ limit of $\alpha$ via \eqref{QL}, and our numerical results for both states are consistent with this prediction at two-digit accuracy, see panel (a) of Table~\ref{tab:strong-coupling}, last row.

We have also obtained an analytical prediction for the first two terms in the large-$g$ expansion of the quantity $\mathbb{S}$ (appearing in gluing \eqref{fullguingnear}). It takes the form
\begin{equation}
    \frac{\log(\mathbb{S})}{2 i g}=-\beta\log(g^2)+\text{Const}+O(g^{-1})\,,
\end{equation}
with the exact expression given in \eqref{STSexp}. We are able to reproduce both coefficients numerically. The coefficient in front of the $\log(g^2)$ term is in agreement with the predicted value $-\beta=-\sqrt{{\bf C}(g)}/(2g)$, although this essentially repeats the comparison already shown in panel (a) of Table~\ref{tab:strong-coupling}. The constant term provides a more independent check and is listed in panel (b) of Table~\ref{tab:strong-coupling}.
The entries in each column agree to two-digit precision. This once again indicates the absence of a distinction between even and odd states in the string-theory limit and demonstrates consistency between our numerical results and the string-theory prediction.

\paragraph{Recovering Finite $\delta$ Spectrum.} Once we have $\vartheta(g)$ and ${\mathbb S}(g)$ computed we can use the FME equation \eq{gluing} to determine the whole family spectrum at small $\delta$, recovering infinitely many states. We assembled this data into the figure~\ref{fig:families}.

\section{Perturbative Checks and Definition of the Fused Line}\la{pertsec}

In this section, we give a perturbative definition of the fused line defect and perform several perturbative checks of our finite-coupling results at one- and two-loop order. A natural strategy might be to start from the perturbative expansion of $\Gamma_{\text{cusp}}$ at finite $\delta$ and only then take the limit $\delta\to0$. However, the fusion limit $\delta\to0$ does not commute with perturbation theory \cite{Appelquist:1977es,Erickson:1999qv,Pineda:2007kz,Correa:2012nk}. In flat space, this non-commutativity manifests itself through infrared divergences in perturbation theory. On the sphere, by contrast, there are no IR divergences; instead, one finds logarithms of $\delta$ multiplying power divergences. For example, up to three loops one obtains
\beq\la{Gammapert}
\Gamma_\text{cusp}=-g^2\[{4\pi\over\delta}+\dots\]-g^4\[16\pi{\log\delta\over\delta}+\dots\]-g^6\[{64\pi^4\over3}{1\over\delta^2}+\dots\]+O(g^8)\,,
\eeq
which does not take the form of a perturbative expansion of (\ref{gammacusp}). However, once $\delta$ is interpreted as a UV regulator for the defect theory, the structure of (\ref{Gammapert}) becomes completely natural. At one-loop order, we see the need for a counterterm associated with a defect operator of classical dimension zero. At two loops, this operator, as well as the defect-changing operator at the endpoint of the line, acquires an anomalous dimension. At three loops, we encounter further corrections to the counterterm together with wavefunction renormalisation effects, and so on.

Extracting perturbative data for the fused defect from the expansion of $\Gamma_\text{cusp}$ in this way involves several subtleties. First, the quantity $\Gamma_\text{cusp}$ in (\ref{Gammapert}) is obtained after regulating the UV divergences associated with the cusp. One therefore deals with two distinct regulators, the cusp regulator at the endpoints of the fused line, and the parameter $\delta$, which regulates the bulk of the defect. Second, the perturbative expansion in (\ref{Gammapert}) is computed prior to the introduction of the necessary counterterms. As a result, the perturbative data must first be reorganised by including loop corrections arising from counterterm insertions. Third, the existing perturbative results obtained from Feynman-diagram computations must be generalised to excited cusp operators, rather than being restricted to the ground-state empty cusp.

Here, instead, we adopt a direct perturbative definition of the fused line defect that does not rely on the cusp fusion construction and involves only a single UV regulator. In this approach, the parameter $\delta$ is absent, and the quark and antiquark lines are placed directly on top of one another from the outset. Once the defect is defined in this way, one is free to use any convenient regularisation scheme.

The fused line defect lives in the tensor-product space ${\bf F}\otimes\bar{\bf F}$. The gauge field of ${\cal N}=4$ SYM acts on this space and is therefore embedded into ${\bf A}\otimes{\bf A}$ as
\beq
{\cal A}_\mu\equiv A^a_\mu\(T^a_1\otimes\One_2-\One_1\otimes(T^a_2)^{\mathsf T}\)\,,
\eeq
where the $T^a$ are the generators of $SU(N)$ in the fundamental representation. The first factor acts on the quark line, while the second acts on the antiquark line. Similarly, the scalar fields are embedded into ${\bf A}\otimes{\bf A}$ as
\beq\la{Phitheta}
\Phi_\theta\equiv\vec\phi^a\(\hat {\bf n}_1T^a_1\otimes\One_2+\hat {\bf n}_2\One_1\otimes (T^a_2)^{\mathsf T}\)\,,
\eeq
where $\vec{\bf n}_1\cdot\vec{\bf n}_2=\cos\theta$. 

The fused line is defined as
\beq\la{fusedline}
\cW_\theta[0,L]=\left.\cP\,\exp\int\limits_0^L\(\ii\cA\cdot\dd x+\Phi_\theta(x)|\dd x|+{\rm c.t.}\)\right|_\text{reg}\,,%
\eeq
where ``c.t.'' denotes local, scheme-dependent counterterms. They are required to cancel UV power divergences and are tuned to render the defect conformal. Here, $\cP$ denotes the path ordering, which is the same for fields acting on the fundamental and anti-fundamental factors due to the transposition. The resulting line defect mixes the adjoint and singlet sectors appearing in the decomposition ${\bf F}\otimes\bar{\bf F}={\bf A}\oplus{\bf S}$.\footnote{An exception is provided by the supersymmetric scalar coupling for $\theta=\pi$
\beq
\Phi^\text{susy}=\Phi_0=\phi^a\(T^a_1\otimes\One_2-\One_1\otimes(T^a_2)^{\mathsf T}\)\,,
\eeq
which preserves the adjoint sector and prevents mixing with the singlet.}

Note that there is another possible scalar coupling in the line action, one that interpolates between the adjoint and singlet sectors. As we will see, however, no such term is generated at two-loop order in the regularisation scheme that we employ.

Next, we discuss the defect-changing operators on which the line (\ref{fusedline}) can terminate. Since the endpoint of the line carries one fundamental and one anti-fundamental index, it can couple either to an adjoint operator or to a singlet operator. We classify these defect-changing operators according to their representation in the free theory.

The lowest-dimension operator at weak coupling is the identity operator, $\cO_\One$. It corresponds to contracting the fundamental and anti-fundamental indices into a singlet, without any additional field insertion. In the free theory, it is a primary operator with vanishing conformal dimension.

The first two non-trivial excited states were studied perturbatively in \cite{Cavaglia:2018lxi} in the ladder limit. The first operator with bare dimension one is the first conformal descendant of $\cO_\One$. It is obtained by acting with a longitudinal derivative on $\cO_\One\cW_\theta$,
\beq\la{descendant}
\dd\cO_\One=\vec\phi^a\,T^a\cdot\(\vec{\bf n}_1+\vec{\bf n}_2\)\,.
\eeq

Another operator with classical dimension one is the antisymmetric combination of the scalar fields to which the line couples
\beq\la{operators}
\cO_\text{odd}=\vec\phi^a\,T^a\cdot\(\vec{\bf n}_1-\vec{\bf n}_2\)\,.
\eeq
All other operators either have higher tree-level dimensions or involve the remaining scalar fields of the theory.

To study this defect-changing operator and the corresponding line defect, we consider the expectation value
\beq\la{Mop}
M_\cO(L)\equiv\<\cO^\dagger(0)\,\cW_\theta[0,L]\,\cO(L)\>\,.
\eeq
We will use this observable to extract the conformal dimensions of $\cO_\One$ and $\cO_\text{odd}$ at two-loop and one-loop orders, respectively. As a warm-up, we first compute the dimensions of the corresponding operators in the ladder model, following~ \cite{Cavaglia:2018lxi} closely but paying attention to the specifics of the $\delta\to0$ limit, and only then turn to the full ${\cal N}=4$ SYM theory.

\subsection{Ladder Limit}
The ladder limit is obtained by taking $\theta\to i\infty$ and $g\to0$ while keeping the combination $\hat g^2=g^2 e^{-i\theta}/4$ fixed \cite{Erickson:2000af,Correa:2012nk}. In this limit, all diagrams are suppressed except for the ladder diagrams built from exchanges of the two scalar fields appearing in (\ref{Phitheta}).

To perform the computation, we must first choose a regularisation scheme. We will make use of two different schemes. For the operator $\cO_\text{odd}$, we employ point-splitting regularisation, in which any two points on the line connected by a propagator are required to be separated by at least a UV cutoff distance $\epsilon\ll L$. For the operator $\cO_\One$, we use dimensional regularisation instead.

\begin{figure}[t]
\centering
\includegraphics[width = \textwidth]{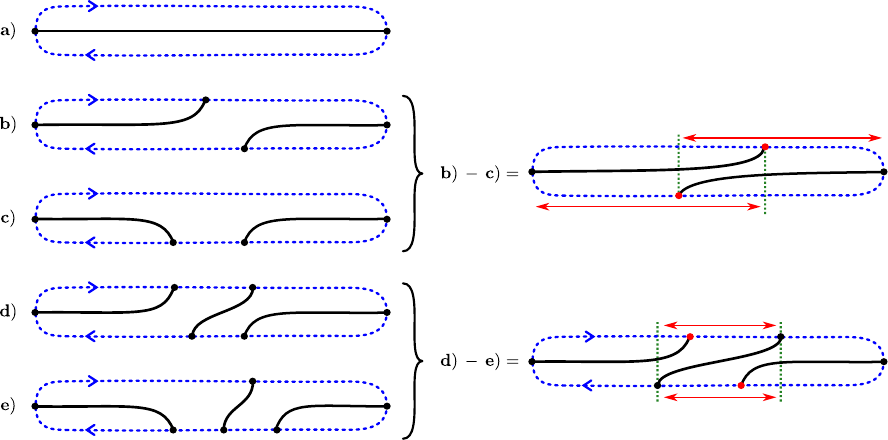}
\caption{Diagrams contributing to the expectation value $M_\text{odd}=\<\cO^\dagger_\text{odd}(0)\,\cW_\theta[0,L]\,\cO_\text{odd}(L)\>$. The dashed blue lines represent the fundamental and anti-fundamental sectors in (\ref{fusedline}), which are localised on top of one another. ${\bf a})$ At tree level, the correlator is given by free scalar propagators connecting the $\cO_\text{odd}$ defect-changing operators. ${\bf b}),{\bf c})$ At one-loop order, scalar propagators can connect $\cO_\text{odd}$ to the line. Subtracting diagram ${\bf c})$ from diagram ${\bf b})$ is equivalent to restricting the integration points on the fundamental and anti-fundamental lines to configurations in which they cross. ${\bf d}),{\bf e})$ At two-loop order (and beyond), the diagrams in ${\bf b})$ and ${\bf c})$ are dressed by ladder corrections. Taking their difference is equivalent to imposing that the integration points of the first and last propagators in ${\bf d})$ cross the ladder insertions on the fundamental and anti-fundamental lines, respectively.}\label{diagrams}
\end{figure}

\subsubsection{The Odd Defect Changing Operator}

The operator $\cO_\text{odd}$ is simpler to analyse, and we therefore begin with it. The reason is that the local power divergences generated on the line are even under the exchange of the two scalar fields appearing in (\ref{operators}), and are therefore orthogonal to $\cO_\text{odd}$.\footnote{For this simplification to hold, it is important that the regularisation scheme respects the symmetry exchanging the two scalars.} As a result, the counterterms in (\ref{fusedline}) decouple.

The operator $\cO_\text{odd}$ reduces, in the ladder limit, to the odd state described by $\psi_0^\text{odd}$ in (\ref{psievenodd}). To see this, note that at tree level $\psi_{0,\text{tree}}^\text{odd}(\sigma)=\sinh{\sigma\over2}$. The two exponentials appearing in the hyperbolic sine arise from free propagators connecting each of the two scalar fields in $\cO_\text{odd}$ to points on the quark and antiquark lines, integrated up to $s$ and $t$ as in (\ref{Gst}). We therefore expect its conformal dimension to be given by $\Delta_0={1\over2}(1+\sqrt{1-16\hat g^2})$.

The diagrams contributing to $M_\text{odd}$ up to two-loop order are shown in figure \ref{diagrams}. At tree level, only diagram ${\bf a})$ contributes,
\beq
M_\text{odd}^\text{tree}(L)=-2(a)=-2\times {4\hat g^2\over L^2}\,,
\eeq
where the factor of $-2$ arises from the two possible contractions between $\vec{\bf n}_1\cdot\vec\phi$ in one $\cO_\text{odd}$ insertion and $\vec{\bf n}_2\cdot\vec\phi$ in the other. At finite coupling, we expect the correlator \eq{Mop} to take the form
\beq\la{Mfinitg}
M_\text{odd}(L)={\cal N}_\text{odd}(\hat g)\({2\epsilon\over L}\)^{2\gamma_\text{odd}(\hat g)}\times  M_\text{odd}^\text{tree}(L)\,.
\eeq
In what follows, we extract the anomalous dimension $\gamma_\text{odd}$ and the wavefunction renormalisation factor ${\cal N}_\text{odd}(\hat g)$ through two-loop order.

At one loop, there are two contributing diagrams, ${\bf b})$ and ${\bf c})$ in figure \ref{diagrams}. Since $\cO_\text{odd}$ is antisymmetric under the exchange of the two scalar fields, these diagrams enter with opposite signs. This symmetry is preserved by our regularisation scheme. Consequently, taking their difference is equivalent to restricting the integration points on the fundamental and anti-fundamental lines to configurations in which they cross; see figure \ref{diagrams}. Using this simplification, we find
\beq
M_\text{odd}^\text{1-loop}=2({\bf c})-2({\bf b}
)=-4\hat g^2\(2\log{2\epsilon\over L}+1\)M_\text{odd}^\text{tree}(L)\,,
\eeq
As anticipated, the power divergences cancel automatically, or equivalently, are absent once the integration points on the line are restricted to cross. Comparing with (\ref{Mfinitg}), we find
\beq\la{gamma1l}
\gamma_\text{odd}^\text{1-loop}=-4\hat g^2\,,\qquad{\cal N}_\text{odd}^\text{1-loop}=-4\hat g^2\,.
\eeq

At two loops, the contributing diagrams are ${\bf d})$ and ${\bf e})$ in figure \ref{diagrams}, which enter with opposite signs. As before, taking their difference is equivalent to restricting the integration points of the first and last propagators in diagram ${\bf d})$ so that they cross the ladder insertion points on the fundamental and anti-fundamental lines, respectively. We find
\beq
M_\text{odd}^\text{2-loop}=2({\bf e})-2({\bf d})=2(4\hat g^2)^2\(\log^2{\epsilon\over L}-1%
\)M_\text{odd}^\text{tree}(L)\,.
\eeq
From the coefficient of $\log^2\epsilon$, we observe the exponentiation of the one-loop anomalous dimension (\ref{gamma1l}). The absence of a single-logarithmic divergence implies that
\beq
\gamma_\text{odd}^\text{2-loop}=-{\cal N}_\text{odd}^\text{1-loop}\gamma_\text{odd}^\text{1-loop}=-(4\hat g^2)^2\,.
\eeq

In total, we have
\beq
\Delta_\text{odd}=1-4\hat g^2-(4\hat g^2)^2+O(\hat g^6)\,,
\eeq
in agreement with the perturbative expansion of $\Delta_0={1\over2}(1+\sqrt{1-16 \hat{g}^2})$. Extending the computation of $\cO_\text{odd}$ to higher loop orders is relatively straightforward. At any loop order, there are only two contributing diagrams, one in which the first and last ladder propagators end on the same scalar field in $\cO_\text{odd}$, and another in which they end on different scalar fields. These diagrams enter with opposite signs, and taking their difference is equivalent to imposing the same restriction on the line integration points as above. As a result, power divergences are absent at every order, and no counterterms are required.

\subsubsection{The Identity Defect Changing Operator}
At tree level, the identity defect-changing operator $\cO_\One$ has vanishing conformal dimension. We therefore expect it to correspond to the primary operator on the unstable defect, with dimension $\Delta_\One=\Delta_-={1\over2}(1-\sqrt{1-16\hat g^2})$. This operator is related, through the defect RG flow discussed in section \ref{DCORGsec}, to the even operator on the stable defect, whose dimension satisfies $\Delta_+=1-\Delta_\One$.

To reach the unstable fixed point, one must tune the coupling of the relevant operator $\cO_\One^\dagger\times\cO_\One$ on the line. In addition, there is a marginally irrelevant operator of the form $\cO_\One^\dagger\times\dd\cO_\One$. These are the operators whose couplings we collectively denoted by the counterterms in (\ref{fusedline}). Tuning these couplings in perturbation theory amounts to cancelling the local divergences generated on the line. Since both operators are symmetric under the exchange of the two scalar fields, the defect-changing operator $\cO_\One$ appearing in $M_\One$ has a nonzero overlap with them. As a result, the simplifications that occurred for $\cO_\text{odd}$ no longer apply.

Another manifestation of the power divergences is the non-commutativity of the $\hat g\to0$ and $\delta\to0$ limits. For the operator of dimension $\Delta_+$ on the stable defect, the situation is analogous to that of the lowest bound state discussed below (\ref{Gammapert}). The main difference is that the small-$\delta$ expansion of the perturbative result for $\Gamma_\text{cusp}(\delta)$ is regular and does not contain logarithmic terms of the type appearing in (\ref{Gammapert}).

To extract the anomalous dimension of $\cO_\One$ beyond one-loop order, we must keep track not only of the power-divergent and logarithmically divergent terms but also of the finite terms. This is because at higher loop orders, a finite correction to the wavefunction renormalisation factor can dress a correction to the anomalous dimension. Keeping track of such finite terms against a background of power-divergent contributions in point-splitting regularisation is a delicate matter. To avoid this issue, we employ a dimensional regularisation scheme. That is, we keep the line one-dimensional and analytically continue the power in the scalar propagator as
\beq
{1\over\sigma^2}\quad\rightarrow\quad{(\mu/e)^\epsilon\over\sigma^{2-\varepsilon}}\,,
\eeq
where $\mu$ is the dimensional-regularisation scale. We begin in the regime $\varepsilon=4-d>1$, where UV power divergences are absent, and analytically continue to $\varepsilon\to0$ at the end of the computation. Note that this regularisation scheme would be problematic for an infinite straight line, as it would introduce infrared divergences. In the present case, however, we consider the finite line segment (\ref{Mop}), and no such IR divergences arise.

\begin{figure}[t]
\centering
\includegraphics[width = 7cm]{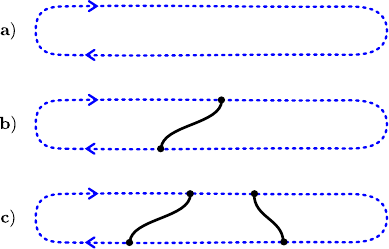}
\caption{Diagrams contributing to the expectation value $M_\One$ through two-loop order.}\label{diagrams2}
\end{figure}
The diagrams contributing to $M_\One$ at tree level and at one- and two-loop order are shown in figure \ref{diagrams2}. At tree level, $M_\One$ is given by the trivial diagram ${\bf a})$,
\beq
M_\One^\text{tree}(L)=(a)=1\,.
\eeq
Beyond tree level, we renormalise both the operator and the line defect so that
\beq\la{Mevenfinitgev}
M_\One(L)^\text{reg}={1\over(\mu L)^{2\gamma_\One(\hat g)}}\,,\qquad \cO_\One^\text{reg}={\cal N}_\One(\hat g)\times\cO_\One\,,
\eeq
where ${\cal N}_\One(\hat g)$ is a wavefunction renormalisation factor.

At one loop, there is a single contributing diagram, namely diagram ${\bf b})$,
\beq
M_\One^\text{1-loop}(L)=({\bf b})+2{\cal N}_\One^\text{1-loop}\times 1\,,
\eeq
with
\beq
({\bf b})=-2(4\hat g^2)\({1\over\varepsilon}+\log\mu+{\varepsilon\over2}\[\log^2\mu+1\]\)\,.
\eeq
We conclude that
\beq
\gamma_\One^\text{1-loop}=4\hat g^2\,,\qquad {\cal N}_\One^\text{1-loop}={4\hat g^2\over\varepsilon}\,.
\eeq
Note that the terms of order $\varepsilon$ in diagram ${\bf b})$ must be kept, as they contribute to the result at the next loop order.

At two-loop order, the contributions come from diagram ${\bf c})$ as well as from the one-loop wavefunction renormalisation, multiplied by the lower-order diagrams,
\beq\la{twoloopeven}
M_\One^\text{2-loop}(L)=({\bf c})+2{\cal N}_\One^\text{1-loop}\times({\bf b})+2{\cal N}_\One^\text{2-loop}\times({\bf a})\,,
\eeq
with
\beq
({\bf c})=2(4\hat g^2)^2\({1\over\varepsilon^2}+{2\over\varepsilon}\log\mu-{1\over2\varepsilon}+2\log^2\mu-\log\mu-{\pi^2\over6}\)\,.
\eeq
Substituting diagram ${\bf c})$ into (\ref{twoloopeven}), we obtain
\beq
M_\One^\text{2-loop}(L)=-2(4\hat g^2)^2\({1\over\varepsilon^2}+{1\over2\varepsilon}-\log^2\mu+\log\mu+1+{\pi^2\over6}\)+2{\cal N}_\One^\text{2-loop}\,.
\eeq
We see that the $\log\mu/\varepsilon$ term has cancelled. The $\log^2\mu$ term reproduces the exponentiation of the one-loop anomalous dimension, while the $1/\varepsilon^2$, $1/\varepsilon$, and constant terms are cancelled by ${\cal N}_\One^\text{2-loop}$. The coefficient of the single logarithm then yields the two-loop anomalous dimension,
 
\beq
\gamma_\One^\text{2-loop}=(4\hat g^2)^2\,.
\eeq
We conclude that
\beq
\Delta_\One=0+4\hat g^2+(4\hat g^2)^2+O(\hat g^6)\,,
\eeq
in agreement with the perturbative expansion of $\Delta_-={1\over2}(1-\sqrt{1-16\hat g^2})$. Note that, in the dimensional regularisation scheme employed here, no counterterms localised on the line are generated at this order. 

Note that, in the presence of a small cusp angle $\delta>0$, the identity cusp operator acquires a divergent conformal dimension of order $1/\delta$. As discussed at the beginning of this section and in section \ref{noncommutativity}, the limits $\delta\to0$ and $\hat g\to0$ do not commute. Since we work directly at $\delta=0$, the identity operator should not be identified with the orphan state. The FME neatly captures this structure.

\subsection{In ${\cal N}=4$ SYM Theory}

We now extend the ladder computations to ${\cal N}=4$ SYM at finite $\theta$. The diagrams contributing to $M(L)$ include the ladder diagrams considered above, together with additional ones. At one- and two-loop order, the new diagrams can be extracted from existing computations in the literature \cite{Erickson:2000af,Makeenko:2006ds,Drukker:2011za,Bykov:2012sc} by taking the $\delta\to0$ limit of the corresponding cusp diagrams. The reason is that these diagrams do not contribute to the Casimir energy and therefore are not affected by the strong infrared effects associated with the fusion limit. At three loops, however, non-ladder diagrams begin to contribute to the Casimir energy itself. We therefore expect this simplification to break down beyond two-loop order.

\subsubsection{The Odd Defect Changing Operator}

It may seem that the operator $\cO_\text{odd}$ in (\ref{operators}) vanishes at $\theta=0$ and therefore only exists for $\theta\neq0$. This cannot be the case, however, since the number of defect-changing operators, or equivalently, the number of states on the sphere, cannot jump as $\theta$ is varied. Instead, we note that the two-point function of $\cO_\text{odd}$ already carries an overall factor of $(1-\cos\theta)$ at tree level; see diagram ${\bf a})$ in figure \ref{diagrams},
\beq
M_\text{odd}^\text{tree}(L)=(1-\cos\theta){4g^2\over L^2}\,.
\eeq
Hence, after properly normalising $\cO_\text{odd}$, its tree-level two-point function becomes independent of $\theta$,
\beq\la{Onormalization}
\widehat M_\text{odd}^\text{tree}(L)={1\over L^2}\,,\qquad M_\text{odd}^\text{tree}=4g^2(1-\cos\theta)\widehat M_\text{odd}^\text{tree}\,.
\eeq

\begin{figure}[h]
\centering
\includegraphics[width = 12cm]{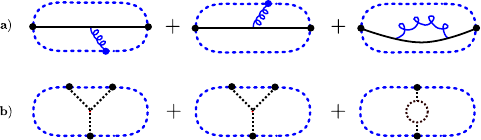}
\caption{${\bf a})$ Non-ladder diagrams contributing to $M_\text{odd}(L)=\<\cO^\dagger_\text{odd}(0)\,\cW_\theta[0,L]\,\cO_\text{odd}(L)\>$. ${\bf b})$ Non-ladder diagrams contributing to $M_\One(L)=\<\cO^\dagger_\One(0)\,\cW_\theta[0,L]\,\cO_\One(L)\>$.}
\label{non_ladder}
\end{figure}

At one-loop order, the contributing diagrams consist of the ladder diagrams ${\bf b})$ and ${\bf c})$ in figure \ref{diagrams}, together with the gluon-exchange diagrams shown in figure \ref{non_ladder}.a.
Every scalar leg ending on $\cO_\text{odd}$ or $\cO_\text{odd}^\dagger$ can contract with either of the two scalar fields appearing in the operator, yielding a factor of $(1-\cos\theta)$. Consequently, all ladder-type diagrams, such as those in figure \ref{diagrams}.b, carry an overall factor of $(1-\cos\theta)^2$. After properly normalising $\cO_\text{odd}$ as in (\ref{Onormalization}), these contributions vanish as $\theta\to0$.

The diagrams in figure \ref{non_ladder}.a, on the other hand, carry only a single factor of $(1-\cos\theta)$. Therefore, after properly normalising $\cO_\text{odd}$ as in (\ref{Onormalization}), they survive the limit $\theta\to0$. More generally, at any loop order, the only diagrams that survive this limit are those proportional to a single power of $(1-\cos\theta)$. These are precisely the diagrams in which a single scalar propagates all the way between the two boundary operators $\cO_\text{odd}$ and $\cO_\text{odd}^\dagger$.

The gluon-exchange diagrams shown in Figure \ref{non_ladder} are identical to those that arise at one-loop order for a cusp operator with a scalar insertion orthogonal to the scalar fields coupled to the Wilson lines. Such diagrams do not generate an anomalous dimension. Consequently, they do not contribute to $\Delta_\text{odd}$, and the only contributions that remain are the ladder-type diagrams shown in figure \ref{diagrams}.

Computing the one-loop ladder diagrams in ${\cal N}=4$ SYM and in the ladder model is essentially the same problem. The ladder limit is obtained by taking $\theta\to\ii\infty$ and $g\to0$ while keeping the combination $\hat g^2=g^2 e^{-\ii\theta}/4$ fixed \cite{Correa:2012nk}. It follows that the one-loop anomalous dimension of $\cO_\text{odd}$ can be computed in the ladder model and then promoted to the full theory by replacing $\hat g^2\to{g^2\over2}(\cos\theta-1)$. Using (\ref{gamma1l}), we find that
\beq\la{gammaodd1l}
\gamma_\text{odd}^\text{1-loop}=2(1-\cos\theta) g^2\,.
\eeq

\subsubsection{The Identity Defect Changing Operator}

At one-loop order, the result for $\cO_\One$ in ${\cal N}=4$ SYM theory is the same as in the ladder limit, upon making the replacement
\beq\la{rep}
\hat g^2\quad\rightarrow\quad {1\over2}g^2\(1+\cos\theta\)\,.
\eeq
Here, the term proportional to $\cos\theta$ originates from scalar exchange, whereas the term proportional to $1$ originates from gluon exchange. Hence,
\beq\la{gamma1loop}
\gamma_\One^\text{1-loop}=2g^2(1+\cos\theta)\,.
\eeq
This result was previously obtained in \cite{Cuomo:2024psk} by studying the $\delta\to0$ limit of the cusp dimension; see equation (3.35) therein. As discussed above, the cusp dimension receives strong infrared contributions at higher orders, and extracting the anomalous dimension from it requires the resummation of infinitely many diagrams.

At two-loop order, there can be contributions proportional to $(1+\cos\theta)^2$ as well as contributions proportional to $(1+\cos\theta)$. The former must coincide with the two-loop ladder result upon making the replacement (\ref{rep}). The latter is proportional to the $O(\hat g^2 g^2)$ term in the expansion around the ladder limit. It arises from the sum of a $Y$-shaped diagram and a ladder self-energy diagram; see figure \ref{non_ladder}.b. The contribution of these diagrams can be extracted from existing computations of the cusp dimension \cite{Erickson:2000af,Makeenko:2006ds,Drukker:2011za,Bykov:2012sc}, since they possess a smooth $\delta\to0$ limit. The $Y$-shaped diagram is a total derivative along the line and therefore reduces to a sum of two boundary terms. One of these boundary terms cancels against the self-energy diagram, while the other produces the following contribution to the cusp dimension\footnote{See equation (3.2) in \cite{Drukker:2011za}.}
\beq
g^4 V_\text{int}^{(2)}={4g^4\over3}{\cos\theta+\cos\delta\over\sin\delta}(2\pi-\delta)(\pi-\delta)\delta\,,\qquad\lim_{\delta\to0}V_\text{int}^{(2)}={8\pi^2\over3}(1+\cos\theta)\,.
\eeq
Hence,
\beq\la{gammaOne2}
\gamma_\One^\text{2-loop}=4g^4\Big[{2\pi^2\over3}(1+\cos\theta)+(1+\cos\theta)^2\Big]\,.
\eeq

We conclude that
\beq\la{deltaone}
\Delta_\One=0+2g^2(1+\cos\theta)+4g^4\Big[{2\pi^2\over3}(1+\cos\theta)+(1+\cos\theta)^2\Big]+O(g^6)\,.
\eeq
To relate this result to our predictions for the stable line, we must apply the defect RG flow described in section \ref{DCORGsec}. Since this flow maps $\Delta_\One$ to $\Delta_+=1-\Delta_\One$, we obtain
\beq\la{deltaplus}
\Delta_+=1-2g^2(1+\cos\theta)-4g^4\Big[{2\pi^2\over3}(1+\cos\theta)+(1+\cos\theta)^2\Big]+O(g^6)\,.
\eeq
The perturbative analysis is already rather intricate at two loops. In the next section, we use integrability to extend the calculation to much higher orders. The resulting perfect agreement provides a non-trivial check of both the fused QSC construction and of the perturbative framework developed above.

\subsection{Comparison with QSC}\label{sec: QSCWeak}
 
In this section, we explain how the above results can be reproduced using the fused QSC derived in Section~\ref{sec: QSC}. At the level of the QSC, all quantities are already finite, so there is no need to choose a particular regularisation or renormalisation scheme. This avoids many of the technical complications encountered in a conventional Feynman-diagram calculation.

The perturbative solution of the QSC was pioneered in~\cite{Marboe:2014gma} and later developed into a general iterative procedure in~\cite{Gromov:2015vua}, which we adopt here. The resulting algorithm is sufficiently simple to be implemented efficiently in programming languages such as Wolfram Mathematica.

The technical details of the full perturbative procedure are given in Appendix~\ref{app: QSCWeak}, where we largely follow \cite{Gromov:2015vua}. Since the algorithm is structurally similar to other known perturbative QSC constructions, we outline its main steps below and mainly focus on the aspects specific to the present case.

When considering the weak-coupling expansion in QSC, it is often convenient to start from the Algebraic Bethe Ansatz equations, which can, in principle, be derived analytically from the QQ-system; see, for example, \cite{Gromov:2014caa} or from the S-matrix bootstrap~\cite{Beisert:2005tm}. Solving these algebraic equations in the $g\to0$ limit, even at leading order, provides a reliable starting point for the perturbative iterative procedure. In the context of the present work, this would require studying the cusped Wilson line with an arbitrary number of scalar insertions in the cusp. Unfortunately, the corresponding ABA equations are not known yet, and therefore, we have to adopt an alternative approach to initialise the weak-coupling expansion.

Much like in the numerical approach, the perturbative solution of the QSC is based on the following two QQ-relations
\begin{equation}\la{fm}
    Q_{a|i}^+-Q_{a|i}^-=-\mathbf{P}_a\mathbf{P}^bQ_{b|i}^{+}\,,\qquad e^{\pm \pi u}\mathbb{Q}_{i}=-\mathbf{P}^{a}Q_{a|i}^{\pm}\,.
\end{equation}
Here we restored the exponential factor removed previously from $\mathbb{Q}_i$ for convenience. First, the ansatz for $\mathbf{P}_a$ in \eqref{Pthph} is expanded at weak coupling. This requires the knowledge of how the coefficients $a_k$ and $b_k$ scale in the $g\to 0 $ limit: in general, they behave as $\sim g^k$, but there are exceptions for some of the first few coefficients. 
The exact scaling in $g$ was extracted from the numerical solution by probing sufficiently small values of coupling and fitting the resulting data in powers of $g$. It is precisely at this stage that the difference between even and odd states becomes apparent, as the corresponding initial data exhibit different scaling behaviour. 

Afterwards, the leftmost difference equation in \eq{fm} for $Q_{a|i}$ can be solved in terms of a finite number of unknown coefficients, and then $\mathbb{Q}_i$ is reconstructed through the rightmost relation in \eq{fm}. For Q-functions without exponential asymptotics (sometimes referred to as a {\it twist}), the general solution to these equations can be written in the basis of spectral parameter powers $u^{\pm k}$, $k=0,1,2,\dots$, together with the so-called $\eta$-functions \cite{Marboe:2014gma}
\begin{equation}\label{etas}
    \eta_{a_1,a_2,...a_k}(u)=\sum_{n_1>n_2...>n_k\geq0}^{\infty}\dfrac{1}{(u+i n_1)^{a_1}(u+i n_2)^{a_2}...(u+i n_k)^{a_k}}\,.
\end{equation}
For generic twist angles, there is, in principle, no reason for the solution space to remain this simple; see, for example, the case of the ABJM model \cite{Anselmetti:2015mda}, where antiperiodic $e^{\pm \pi u}$ factors are also naturally present. A more general set of functions has to be used. Nevertheless, fortunately, we found that the standard basis of $\eta$'s is sufficient for both even and odd states, which greatly reduced the complexity of the calculation. At leading order in the $g\to0$ expansion, the functions $\mathbb{Q}_i$ depend only on regular $\eta$-functions; the explicit expressions for both even and odd states are listed in Appendix~\ref{app: QSCWeak}. The appendix also contains a more detailed discussion of the full perturbative algorithm, in particular an explanation of why twisted generalisations of $\eta$-functions do not appear at higher orders.

\paragraph{Results.} We have successfully implemented the weak-coupling perturbative algorithm and obtained the series expansion for $\vartheta(g)$ up to six loops for the even state and up to five loops for the odd state. We have also determined the first six and five terms in the expansion of $\mathbb{S}$ for the even and odd states, respectively. The full expansions are listed in Appendix~\ref{app: weak}, as well as in the \textit{Mathematica} notebook \texttt{EffQSCWeakResults.nb} included in the arXiv submission. Below, we present only the first two non-trivial terms to ease the comparison with the results of the previous sections:
\paragraph{Even state}
\begin{equation}\la{sympert}
    \vartheta_\text{even}(g)=\frac{1}{2}-2 g^2 (1+\cos\theta)-16 g^4 \left(\frac{1}{6} \pi ^2 (1+\cos\theta)+\frac{1}{4} (1+\cos\theta)^2\right)+O(g^6)\,,
\end{equation}
\begin{flalign}\label{evenS}
~~~~&\mathcal{S}_\text{even}(g)=\frac{4}{g^2 \pi  (\cos \theta +1)}-\frac{16 (-1+\gamma +4 \log2)}{\pi }+O(g^2)\,.&
\end{flalign}
\paragraph{Odd state}
\begin{equation}\la{antisympert}
\vartheta_\text{odd}(g)=\frac{1}{2}+2 g^2 (1-\cos\theta)-g^4 \left(4 (1-\cos\theta)^2+\frac{8}{3} \pi ^2 (1+\cos\theta)\right)+O(g^6)\,,
\end{equation}
 \begin{flalign}\label{oddS}
&&\mathcal{S}_\text{odd}(g)=\frac{4 \pi  (\cos \theta +1)^2 g^2}{\cos \theta -1}
&+\frac{16}{3} \pi  (\cos \theta +1)^2 g^4\times\notag&\\&&&\times\left(-12 \log2-3 \gamma +6-\
\frac{2 \pi ^2}{(\cos \theta -1)^2}\right)+O(g^6)\,.&
\end{flalign}
Through the relation $\Delta={1\over2}+\vartheta(g)$, we find perfect agreement with the Feynman-diagram approach of the previous section. Namely, between (\ref{sympert}) and (\ref{deltaplus}) for the even state, and between (\ref{antisympert}) and (\ref{gammaodd1l}) for the odd state.

We also find that both \eqref{evenS} and \eqref{oddS} reduce in the ladder limit to the corresponding expressions $\mathcal{S}_\pm$ in \eqref{SpmSh}, which holds for all orders we computed. 

Notably, in the limit $\theta\to\pi$, the even-state expression \eqref{sympert} reduces to $\vartheta=1/2$ or $\Delta_{\rm even}=1$, see its expansion in $\epsilon=\pi-\theta$ below
\begin{align}
    \vartheta_\text{even}&=\frac{1}{2}-g^2 \left(\epsilon ^2-\dfrac{\epsilon^4}{12}\right)-g^4\left(\frac{4}{3} \pi ^2 \epsilon ^2+\left(1-\frac{\pi^2}{9}\right)\epsilon^4\right)+\notag\\
    &+g^6 \left(6 \zeta (3)-\frac{5 \pi ^2}{3}\right) \epsilon ^4+g^8 \left(\frac{32 \pi ^2 \zeta (3)}{3}-90 \zeta (5)-\frac{16 \pi ^4}{9}\right) \epsilon ^4+\notag\\&+g^{10} \left(\frac{64 \pi ^4 \zeta (3)}{15}-176 \pi ^2 \zeta (5)+1204 \zeta (7)+\frac{16 \pi
   ^6}{45}\right) \epsilon ^4-\notag\\
   &- g^{12}\frac{4}{135} \left(32 \pi ^6 \zeta (3)+2280 \pi ^4 \zeta (5)-83790 \pi ^2 \zeta (7)+535815 \zeta (9)+16 \pi ^8\right) \epsilon ^4\,.
\end{align}
At the same time, the function $\mathcal{S}_+$ clearly diverges due to the leading term in \eqref{evenS}, which is the main manifestation of the non-commutativity of $\delta\to 0$ and $g\to 0$ limits for this state. We also noticed that $\mathbb{S}_{\rm even}$ at $\theta=\pi$ can be guessed in all orders to be
\beq
\left.\mathbb{S}_\text{even}\right|_{\theta=\pi}=-\dfrac{\pi}{1+\frac{(2\pi g)^2}{3}}\,.
\eeq
The odd state also exhibits peculiar behaviour: the quantity $\mathcal{S}_-$ vanishes in the $\theta\to\pi$ limit, as follows from \eqref{oddS}, and we find this property to persist through the next three orders of the expansion.

The leading order in the $\theta\to \pi$ expansion of $\vartheta$ can usually be reached analytically, see for example \cite{Gromov:2015dfa}. It would be interesting to do so in the present case as well.

\section{Semi-Classical Falling String in Global $AdS$ with Internal Angle}\la{AdSsec}

At strong coupling, the cusp configuration admits a dual description through the AdS/CFT correspondence, which can be analysed directly in the fusion limit $\delta\to0$. In this regime, the system lies above criticality, and it is therefore natural to focus on the density of states. For vanishing internal angle, $\theta=0$, the relevant classical solution is a folded string in $AdS$ ending on the fused line at the $AdS$ boundary; see figure \ref{fallingstring}.a. In section \ref{foldedstringsec}, we analysed this solution and showed that it reproduces the density (\ref{dsdensity}) with ${\bf C}=8g$. We now generalise this string solution to a nonzero internal cusp angle, $\theta>0$, and use it to determine the corresponding Casimir ${\bf C}(\theta)$.

Turning on the scalar angle modifies the boundary conditions of the string, so that its two endpoints are attached to points on the $S^5$ separated by an angle $\theta$. The corresponding classical solution extends inside an $AdS_2$ slice of $AdS_5$ and along an equator of $S^5$; see figure \ref{fallingstring}.b. It is therefore convenient to use the following parametrisation of the $AdS_5\times S^5$ metric,
\beq\la{AdS5xS5}
\dd s^2_{AdS_5\times S^1}=R^2\[\dd\rho^2+\cosh^2\!\rho\,\dd s_{AdS_2}^2+\sinh^2\!\rho\,\dd\Omega_2^2+\dd\Theta^2\]\,,
\eeq
where $\Theta$ parametrises an equator of $S^5$, while the solution is localised at $\rho=0$. Restricting the metric to $\rho=0$ yields the global $AdS_2$ metric used in section \ref{foldedstringsec}, namely
\beq\la{LAdS2}
\dd s_{AdS_2}^2={\dd\chi^2-\dd t^2\over\sin^2\chi}\,.
\eeq

The shape of this solution in $AdS$ is shown in figure \ref{fallingstring}.b. The extension of the string along the sphere endows it with an effective mass density in $AdS$, causing the tip to move at a speed lower than the speed of light. 

The solution we seek should preserve the element of the line $SL(2,{\mathbb R})$ symmetry whose fixed points are the two cusp points before the fusion limit $\delta\to0$ is taken. This generator is conjugate to translations along the Rindler-$AdS$ time direction. The $AdS_2$ metric in Rindler-$AdS$ coordinates takes the form
\beq\la{RAdS2}
\dd s_{AdS_2}^2={\dd\xi^2-\dd t_R^2\over\sinh^2\xi}\,.
\eeq
In these coordinates, we make an ansatz in which the accelerating tip of the string is stationary and evolves along the Rindler-$AdS$ time direction. 

These Rindler coordinates cover only a patch of global $AdS$ described by (\ref{LAdS2}). The two coordinate systems are related by
\beq\la{RindlertoGlobal}
\tan\chi=\frac{\sinh \xi}{\cosh\eta\,\cosh t_R-\sinh\eta\,\cosh \xi}\,,\quad\tan t=\frac{\sinh t_R}{\cosh\eta\, \cosh\xi-\sinh\eta\,\cosh t_R}\,,
\eeq
where $\eta$ is a free boost parameter that determines the maximal radial position of the string, and hence the height from which it falls towards the boundary.

If we were below criticality, we would be considering a Euclidean $AdS_2$ solution, and the corresponding Euclidean Rindler-$AdS$ energy should be identified with the dimension of the highest-weight state, $\lim_{g\to\infty}\Delta_+=\ii\sqrt{\bf C}$. Since at strong coupling we are above criticality, however, the solution under consideration is timelike, and the corresponding Rindler-$AdS$ energy should instead be identified with $E_R=-\sqrt{\bf C}$.

It is instructive to first describe the solution in static gauge (section \ref{staticgauge}) and then reanalyse it using the standard strong-coupling integrability framework in conformal gauge (section \ref{sec:coset}).

\subsection{Solution in Static Gauge}\la{staticgauge}

In static gauge, we choose the worldsheet coordinates to coincide with $\tau=t_R$ and $\sigma=\Theta$. The Nambu-Goto action then takes the form
\beq
S_{NG}=2g\int\dd\tau\,\dd\sigma{\sqrt{1-\dot\xi^2+\xi'^2/\sinh^2\xi}\over\sinh\xi}\,,
\eeq
where $g={R^2\over4\pi\alpha'}$. Here, the dot and prime denote derivatives with respect to $\tau$ and $\sigma$, respectively. We make an ansatz for a solution that is static in Rindler time, $\xi(\tau,\sigma)=\xi(\sigma)$. Since the Lagrangian does not depend explicitly on $\Theta$, we have
\beq
\d_\sigma{\cal J}=0\,,\qquad{\cal J}={1\over\sqrt{\xi'(\sigma)^2+\sinh^2\xi(\sigma)}}\,.
\eeq

We look for a solution $\xi(\sigma)$ that extends into the bulk from the boundary lines at $\xi(0)=\xi(\theta)=0$ to a maximal radial position $\xi(\theta/2)=\xi_m$. At the turning point, we have $\xi'(\theta/2)=0$, which implies that $\cJ=1/\sinh\xi_m$. It follows that
\beq\la{constJ}
\xi'(\sigma)=\sqrt{\sinh^2\xi_m-\sinh^2\xi(\sigma)}\,.
\eeq
The solution of this first-order equation that starts at the boundary, $\xi(0)=0$, is given by
\beq\la{Rindlersol}
\xi(\sigma)=-i\,\text{am}\left(i\sigma\sinh\xi_m\left|-\text{csch}^2\xi_m\right.\right)\,.
\eeq
where $\text{am}$ is the \verb|JacobiAmplitude| function. The corresponding opening angle is given in parametric form by
\beq\la{Deltatheta}
\theta=2\int\limits_0^{\xi_m}{\dd\xi\over\xi'}=-{2\ii\over\sinh\xi_m}F\left(\ii\xi _m|-1/\sinh^2\xi_m\right)\,,
\eeq
where $F$ is the \verb|EllipticF| function.

Since the Lagrangian does not depend explicitly on the Rindler time, the solution (\ref{Rindlersol}) is associated with a conserved Rindler energy. To regulate this energy, we truncate the solution at a large radial position $\xi_M$ near the boundary and subtract a boundary counterterm $1/\xi_M$ for each fold. In this way, we obtain 
\begin{align}\la{ERindler}
E_R=&4g\lim_{\xi_M\to0}\Big(\int\limits_{\xi_M}^{\xi_m}{\dd\xi\over\sinh^2\xi}{\sinh\xi_m\over\sqrt{\sinh^2\xi_m-\sinh^2\xi}}-{1\over\xi_M}\Big)\\
=&4g\,  \ii\[F\left(i\xi_m|-1/\sinh^2\xi _m\right)-E\left(i\xi_m|-1/\sinh^2\xi_m\right)\]\,,\nn
\end{align}
where $E$ is the \verb|EllipticE| function. Note that this Rindler energy differs from the energy in global $AdS$ and should therefore not be identified with the dimension of the defect-changing operator. Instead, it is the continuation of the highest weight energy, and we have ${\bf C}(\theta)=E_R^2$. Note also that the absence of a logarithmic divergence means that no UV RG scale is needed to define the solution. This sits nicely with the fact that the density is defined at $\delta\to0$.

As the maximal radial position of the tip approaches the $AdS$ boundary, the Rindler energy approaches
\beq
\lim_{\xi_m\to0}(\theta,E_R)=(\pi,0)\,.
\eeq
In this limit, the Rindler--$AdS$ and global-$AdS$ energies coincide. It is also the supersymmetric limit of the cusp and, consequently, of the fused defect.\footnote{By expanding (\ref{Rindlersol}) at small $\xi_m$ we find that $\lim_{\xi_m\to0}\xi(\sigma)=\xi_m\sin\sigma+O(\xi_m^2)$. This solution should capture the Bremsstrahlung function at strong coupling, \cite{Correa:2012at}.}

In the opposite limit, the tip of the string approaches the centre of $AdS$, and
\beq
\lim_{\xi_m\to\infty}(\theta,E_R)=(0,-4g)\,.%
\eeq
The solution covers the entire Rindler $AdS_2$ twice.\footnote{If we analytically continue $t_R\to\ii t_R$ then we find two sheets covering the Euclidean $AdS_2$ that are connected through a thin tube, along which $\Theta$ ranges from $0$ to $\pi$. Consequently, its regularised area approaches twice that of the circular Wilson loop solution.} For all intermediate values $0<\xi_m<\infty$, one has $\pi>\theta>0$ and $0>E_R(\theta)>-4g$.%

Applying the coordinate transformation (\ref{RindlertoGlobal}), we obtain a family of solutions in global $AdS$ parametrised by $\eta$. These solutions take the form
\beq
\cos\chi_\eta(t,\sigma)=\frac{\sqrt{\left(\sin^2t\,\sinh ^2\eta-\cos^2t\right) \tanh ^2\xi(\sigma)+1}-\cos t\,\cosh\eta\,\sinh\eta\,\tanh ^2\xi(\sigma)}{\sinh^2\eta\, \tanh^2\xi(\sigma)+1}\,,
\eeq
where $\xi(\sigma)$ is given by (\ref{Rindlersol}). Evaluating (\ref{RindlertoGlobal}) at the tip of the string, $\xi=\xi_m$, $t=t_R=0$, we find that the maximal radial position is
\beq\la{chimax}
\cos\chi_m=\frac{\cosh\xi _m-\cosh\eta\,\sinh\eta\,\sinh^2\xi _m}{\cosh^2\eta\,\sinh^2\xi _m+1}\,.
\eeq

To reveal the integrability structure underlying this solution, we shall embed it in the $\mathrm{AdS}_3\times S^1$ coset sigma-model framework of \cite{Kazakov:2004nh} in the next section.

\subsection{Coset Construction of the Solution in Conformal Gauge}
\la{sec:coset}

The goal of this section is to re-derive the solution found in the previous section within the coset formulation of the worldsheet $\sigma$-model, which is better suited for integrability analysis.
We will consider the $\mathrm{AdS}_3\times S^1$ coset sigma-model framework of \cite{Kazakov:2004nh}, to which we refer for all generalities on the $\mathbb{Z}_2$-graded construction and the algebraic curve. We will identify the solution in this formalism, compute its conserved charges, and discuss the deformation induced by the small angle $\delta$. As a result, we will rederive the strong coupling limit of the quadratic Casimir, $\lim_{g\to\infty}C(g,\theta)=E_R(g,\theta)^2$ 
and compare it to our numerical non-perturbative solution. Furthermore, we will derive an analytic expression for the leading strong-coupling asymptotics of ${\cal S}(g)$ using the quasiclassical Bohr--Sommerfeld approach.

\subsubsection{Coset Formulation}
\paragraph{Setup} In order to benefit from the integrability formulation, we start by re-deriving the $AdS_2$ solution found in the previous section in terms of the $\mathrm{SL}(2,\mathbb{R})$ group element, which we parametrise as follows%
\beq\la{gparam}
{\mathbf g}=\begin{pmatrix}\cosh t_R\,\operatorname{csch}\xi +\coth\xi  & \mathrm{e}^{-\psi }\,\sinh t_R\,\operatorname{csch}\xi \\-\mathrm{e}^{\psi }\,\sinh t_R\,\operatorname{csch}\xi  & \coth\xi -\cosh t_R\,\operatorname{csch}\xi \end{pmatrix}\,.
\eeq
The AdS${}_3$ metric coincides with the Haar measure 
\beq
ds^2=-\frac{1}{2}\tr({\mathbf g}^{-1}\dd{\mathbf g})^2=
\operatorname{csch}^2\xi\,\left({dt_R}^2-d\xi ^2-d\psi ^2\,\sinh^2t_R\right)\,,
\eeq
which is left and right invariant, making the $\mathrm{SL}_L(2,\mathbb{R})\times\mathrm{SL}_R(2,\mathbb{R})$ isometry manifest.
Next we write the action of the string on $AdS_3\times S^1$ as follows%
\beq\la{Spoly}
S=%
g\int\dd\sigma\,\dd\tau
\left[
\frac{1}{2}\operatorname{tr}{j}_a^{\,2}
+(\partial_a\Theta)^2
\right]\,.
\eeq
The corresponding left and right Noether currents are given by%
\beq
j_a={\mathbf g}^{-1}\partial_a{\mathbf g}\,,
\qquad
l_a={\mathbf g}\,j_a\,{\mathbf g}^{-1}=\partial_a {\mathbf g}\,{\mathbf g}^{-1}\,,
\eeq
so that the properly normalised conserved charges are
\beq\la{Qrl}
Q_L=%
2g\int\dd\sigma\, j_0 \,,\qquad Q_R=%
2g\int\dd\sigma\, l_0\,.
\eeq
The equations of motion are identical to the current conservation condition $d* j =0$. More explicitly, in terms of the parametrisation \eq{gparam}
\beqa
\nn&&\d_\sigma^2\chi-
\d_\tau^2
\chi-\left[(\d_\sigma{\chi})^2-(\d_\tau{\chi})^2
+(\d_\sigma t)^2-(\d_\tau t)^2\right]\coth\chi=0\,,\\
\la{eomp}&&\d_\sigma^2 t-\d_\tau^2 t-\left[2\d_\sigma\chi 
\,\d_\sigma t
-2\d_\tau\chi \,\d_\tau t\right]\coth\chi=0\,,\\
\nn&&\d_\sigma^2\Theta-\d_\tau^2\Theta=0\,.
\eeqa
where we set $\psi=0$ for simplicity.
Since \eq{Spoly} is the Polyakov string action in conformal gauge, we must also impose the Virasoro constraints
\beqa\la{virasoro}
\frac{1}{2}\tr j_\pm^2+\d_\pm\Theta^2=0\,.
\eeqa

\paragraph{Solution} In order to find the solution discussed in the previous section, we consider the following ansatz
\beq\la{ansatz2}
t_R=\kappa\,\tau\,,\qquad \xi=f(\sigma)\,,\qquad \Theta= m\,\sigma\,,\qquad\psi=0\,.
\eeq
Since we are considering an open-string solution, it is convenient to choose $\sigma\in[0,\pi]$, in which case $m=\theta/\pi$. We further require the solution to extend from the boundary at $f(0)=0$ into the bulk and then back to the boundary at $f(\pi)=0$. These boundary conditions break the $SL(2,{\mathbb R})_L\times SL(2,{\mathbb R})_R$ down to the diagonal subgroup ${\mathbf g}\to h\cdot{\mathbf g}\cdot h^{-1}$, $h\in SL(2,{\mathbb R})_\text{diag}$. This symmetry is further broken by the solution to a single element of $SL(2,{\mathbb R})_\text{diag}$ that is conjugate to the Rindler time $t_R$. It acts on ${\mathbf g}$ as
\beq\la{tRshift}
{\mathbf g}(t_R+a,\xi,0)=e^{-{a\over2}\sigma_1}\cdot{\mathbf g}(t_R,\xi,0)\cdot e^{{a\over2}\sigma_1}\,.
\eeq

The equations of motion \eq{eomp} evaluated on the ansatz (\ref{ansatz2}) reduce to
\beq\la{eompf}
f''\left(\sigma \right)= \left[f'\left(\sigma \right)^2-\kappa ^2\right]\coth f\left(\sigma \right)\,,
\eeq
whereas the Virasoro condition \eq{virasoro} gives
\beq
\left[f'\left(\sigma \right)^2-\kappa ^2\right]\operatorname{csch}^2f\left(\sigma \right)+m^2=0\,.
\eeq
The Virasoro constraint is a first-order equation that implies \eq{eompf}. Direct integration then yields
\beq
f(\sigma)=-\mathrm{i}\,\mathrm{am}\left(\mathrm{i}\,\kappa \,\sigma \mid -\frac{\theta ^2}{\pi ^2\,\kappa ^2}\right)\,.
\eeq
Finally, in order to fix the remaining parameter $\kappa$, we require that at $\sigma=\pi$ the string returns to the boundary $f(\pi)=f(0)=0$. It is convenient to express this solution in parametric form as
\beq\la{param}
\theta = 2\,\sqrt{1-\zeta}\,K\left(\zeta\right)\,,\qquad \kappa = \frac{2}{\pi }\,\sqrt{\zeta} \,K\left(\zeta\right)\,,
\eeq
where $K$ is the \verb|EllipticK| function. Equation~\eq{param} should be understood as an implicit expression for $\kappa(\theta)$.
Note that $\zeta\in [0,1]$ with $\left.\theta\right|_{\zeta=0}=\pi$ and $\left.\theta\right|_{\zeta=1}=0$.

\paragraph{Conserved Charge} Having found the solution, let us compute the corresponding conserved charge. The charge corresponding to (\ref{tRshift}) is power-divergent near the boundary. Since it does not have a logarithmic divergence, its finite part is well-defined. It is given by
\beq\la{QL}
E_R={1\over4}\tr\!\([Q_R-Q_L]\sigma_1\)_\text{reg}=%
-4g\frac{\left(\zeta-1\right)\,K\left(\zeta\right)+E\left(\zeta\right)}{\sqrt{\zeta}}\,.
\eeq
Note that $j_0$ itself has non-trivial time dependence, but the time-dependent part is a total derivative and vanishes under the regularisation.

There are two important differences between an infinite straight-line configuration and a pair of infinite parallel lines separated by a small $\delta>0$. First, the latter breaks the geometrical $SL(2,{\mathbb R})$ symmetry of the straight line. Second, the corresponding string worldsheet is no longer confined to $AdS_2$; for $\delta>0$ it must extend into $AdS_3$.
As a consequence, the $AdS_3$ multiplets that share the same ${\bf C}$ but differ in $\Delta$ are split. We will discuss this effect more explicitly below, where we reproduce the FME equation in the classical limit.

\subsubsection{Finite-gap Method}
To place our specific solution in a broader context, we briefly review the general finite-gap construction for strings in $AdS_3$, following~\cite{Kazakov:2004nh}. Since we shall need to introduce the deformation parameter $\delta$, which takes the string outside the $AdS_2$ subsector, it is convenient to work in this more general setting from the outset. Another important difference is that we study a double cover of the open-string solution, obtained through a reflection trick from a quasi-closed string, namely a string that is closed only up to the action of a global symmetry.

The integrability of the model is encoded in the flat Lax connection $J(x)$, defined by
\beq
J_\sigma = \frac{1}{2}\left(
\frac{j_+}{1-x}-
\frac{j_-}{1+x}
\right)\,,\quad
J_\tau = \frac{1}{2}\left(
\frac{j_+}{1-x}+
\frac{j_-}{1+x}
\right)\,,\qquad j_\pm = j_\tau\pm j_\sigma\,.
\eeq
The conservation of $j$ together with the Bianchi identity imply
the flatness condition of $J$ for any $x\in {\mathbb C}$
\beq
\d_\tau J_\sigma - \d_\sigma J_\tau - [J_\sigma, J_\tau]=0\,.
\eeq
Under the full symmetry group, $J$ transforms non-trivially only under one of the two $\mathrm{SL}(2,\mathbb{R})$ factors.

We impose the twisted boundary condition $J(\tau,\sigma+2\pi;x)=G_{R}^{-1}\cdot J(\tau,\sigma;x)\cdot G_{R}$. As a consequence, the monodromy matrix
\beq\la{monodromy}
\Omega(\tau,x) = G_R\cdot\overleftarrow{\mathrm{P}}\exp\int\limits_{0}^{2\pi}\dd\sigma\, J_\sigma(\tau,\sigma; x)\,,
\eeq
transforms by a similarity transformation under the time translation, $\tau\to\tau+{\rm const}$. In our case a rotation by an angle $\delta$ is generated by ${\mathbf g}\to \exp\({\ii\over2}\sigma_2\delta\)\cdot{\mathbf g}\cdot \exp\({\ii\over2}\sigma_2\delta\)$, which parametrises a spacelike compact direction. Correspondingly, we have $G_R=\exp\(-{\ii\over2}\delta\,\sigma_2\)$ in (\ref{monodromy}).
The key observation is that the spectral data of $\Omega$ encode an infinite set of integrals of motion. In the present case these are fully described by the \textit{quasimomentum} $p(x)\equiv\frac{1}{i}\log\omega$, where $\pm\omega$ are the eigenvalues of $\Omega$. Equivalently, one defines $p(x)$ implicitly via
\beq\la{pdef}
\tr \Omega(x) = 2 \cos p(x).
\eeq
The quasimomentum $p(x)$ is a meromorphic function on a hyperelliptic Riemann surface, with simple poles at $x = \pm 1$ of the form
\beq\la{ppoles}
p(x) \simeq \pi\frac{\cJ/\sqrt{\lambda}\pm m}{x \mp 1}\,, \qquad x \to \pm 1\,.
\eeq
This is a manifestation of the Virasoro constraint,
where $\cJ$ and $m$ are parameters of the $S^1$ part of the solution $\Theta= \frac{\cJ}{\sqrt\lambda}\tau+m\,\sigma$. From \eq{monodromy}, the large- and small-$x$ asymptotics of $p(x)$ encode the global conserved charges, with the twist $\pi-\delta$ appearing as a constant term
\beq\la{xlargesmall}
p(x)\simeq \frac{\pi-\delta}{2}+\frac{1}{x}\frac{\Delta+S}{2g}+{\cal O}\(\frac{1}{x^2}\)
\,,\qquad
p(x)\simeq \frac{\pi-\delta}{2}-x\frac{\Delta-S}{2g}+{\cal O}\({x^2}\)\,.
\eeq
More precisely, we have 
\beq\la{charges2}
(\Delta+S)^2 =%
-\frac{1}{2}\tr Q_L^2\,,
\eeq
where $Q_L$ is given in (\ref{Qrl}), see \cite{Kazakov:2004nh} for more details.\footnote{The sign in (\ref{charges2}) will prove crucial in what follows.} 

The analytic structure of $p(x)$ is characterised by a collection of branch cuts $C_k$ in the complex $x$-plane, across which the quasimomentum obeys the discontinuity condition
\beq\la{pdiscont}
p(x+i0) + p(x-i0) = 2\pi n_k\,, \qquad x \in C_k\,,
\eeq
where $n_k \in \mathbb{Z}$ are mode numbers.

Finally, the quasimomentum provides a simple and natural representation of all the action variables $\mathcal{N}_k$ of this classically integrable system via the contour integrals
\beq\la{Sk}
\cN_k = \frac{1}{2}\oint_{\mathcal{C}_k}\dd u\, p = 
\frac{g}{2\pi i} \oint_{\mathcal{C}_k}\dd x  \left( 1- \frac{1}{x^2} \right) p(x)= 
-\frac{g}{2\pi i} \oint_{\mathcal{C}_k}\dd x  \left( x+ \frac{1}{x} \right) p'(x)\,,
\eeq
where $\mathcal{C}_k$ encircles the cut $C_k$, $u=g\,(x+1/x)$ is the conjugate variable to $p$ \cite{Dorey:2006zj}, {and $\cN_k$ is the filling fraction}.

Let us now describe how the above construction extends to open strings. The key ingredient is a boundary reflection matrix, which allows one to glue two copies of the open string into a single closed string. This doubling has two consequences. First, each branch cut $C_k$ acquires a mirror image under $x\to -1/x$. Second, the right-hand sides of the asymptotic relations \eq{xlargesmall} are doubled, since one now traverses two copies of the same open string. Moreover, the spin $S$ is generally not preserved by the reflection, and one therefore sets $S=0$. Combining \eq{xlargesmall}, \eq{charges2}, and \eq{QL}, we obtain
\beq\la{plargex}
p(x)\simeq(\pi-\delta)\pm\frac{\ii E_R}{g\,x}\,.
\eeq
Note that the coefficient of $1/x$ in \eq{xlargesmall} is purely imaginary. This reflects the fact that $E_R$ is conjugate to a timelike direction and is consistent with \eq{charges2}.

\subsubsection{Classical Spectral Curve for the Fused Line Defect}

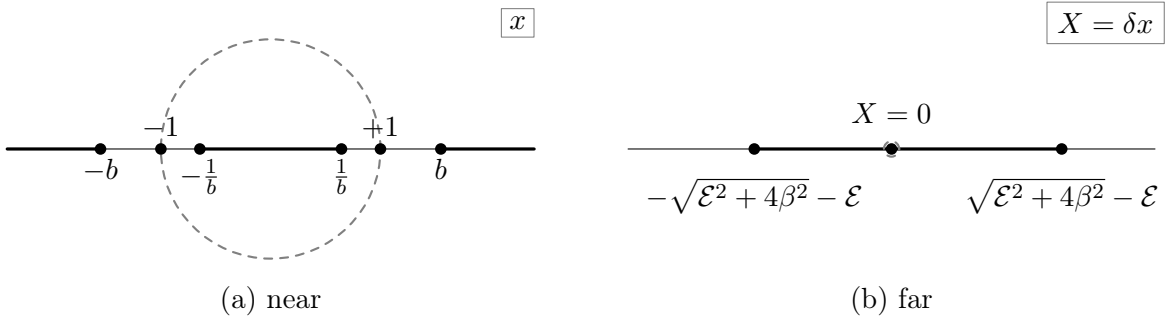
\begin{figure}[t]
\centering
\begin{tikzpicture}[line cap=round, line join=round]

\begin{scope}[scale=1.45]

  \def\b{1.55}
  \def\ib{0.645}
  \def\xL{-2.4}
  \def\xR{ 2.4}

  \draw[thin] (\xL,0) -- (\xR,0);

  \draw[gray, dashed, thick] (0,0) circle (1);

  \draw[very thick] (-\ib,0) -- (\ib,0);
  \draw[very thick] (\xL,0) -- (-\b,0);
  \draw[very thick] (\b,0) -- (\xR,0);

  \foreach \x in {-\b,-1,-\ib,\ib,1,\b}
    \fill (\x,0) circle (1.5pt);

  \node[below] at (-\b,0) {$-b$};
  \node[above] at (-1,0) {$-1$};
  \node[below] at (-\ib,0) {$-\frac{1}{b}$};
  \node[below] at (\ib,0) {$\frac{1}{b}$};
  \node[above] at (1,0) {$+1$};
  \node[below] at (\b,0) {$b$};

  \node[draw=gray, inner sep=3pt] at (2.25,1.15) {$x$};

  \node at (0,-1.38) {(a) near};

\end{scope}

\begin{scope}[xshift=8.2cm, scale=1.45]

  \def\xL{-2.4}
  \def\xR{ 2.4}
  \def\xa{-1.25}
  \def\xc{ 0.0}
  \def\xb{ 1.55}
  \def\eps{0.07}

  \draw[thin] (\xL,0) -- (\xR,0);

  \draw[very thick] (\xa,0) -- (\xb,0);

  \fill (\xa,0) circle (1.5pt);
  \fill (\xc,0) circle (1.5pt);
  \fill (\xb,0) circle (1.5pt);

  \draw[gray, dashed, thick] (\xc,0) circle (\eps);

  \node[below=7pt] at (\xa,0)
    {$-\sqrt{{{\cal E}}^{2}+4\beta^{2}}-{{\cal E}}$};

  \node[below=7pt] at (\xb,0)
    {$\sqrt{{{\cal E}}^{2}+4\beta^{2}}-{{\cal E}}$};

  \node[above=6pt] at (\xc,0) {$X=0$};

  \node[draw=gray, inner sep=4pt] at (1.95,1.15) {$X=\delta x$};

  \node at (0,-1.38) {(b) far};

\end{scope}

\end{tikzpicture}

\caption{\la{fig:cuts}Analytic structure of the spectral curve for the fused line. Two complementary regimes are relevant: the near'' regime, described by the variable $x$, and the far'' regime, described by the rescaled variable $X=\delta x$. In the near regime, the curve is symmetric under both $x\to -x$ and $x\to 1/x$, and possesses two cuts extending through the origin and infinity. In the far regime, obtained by zooming out, the structure near the unit circle in the $x$-plane collapses to the single point $X=0$. The cuts that previously extended to infinity now terminate at two branch points, which are no longer symmetric under $X\to -X$, located at $-\sqrt{{{\cal E}}^{2}+4\beta^{2}}-{{\cal E}}$ and $\sqrt{{{\cal E}}^{2}+4\beta^{2}}-{{\cal E}}$, where $\beta=-E_R/(2g)$.}
\label{fig:x-X-planes}
\end{figure} 

To construct the classical spectral curve $p(x)$, one must diagonalise the monodromy matrix $\Omega(x)$, as discussed in the next section. However, determining the monodromy matrix is a difficult problem even when the flat connection is known explicitly, since one must still solve a second-order differential equation with non-trivial coefficients. In the present case, an additional complication arises from the need to regularise the solution near the boundary, rendering such a brute-force approach technically cumbersome. In practice, it is far more efficient to determine $p(x)$ by bootstrapping its analytic properties.

Since we are dealing with the simplest solution of this type—namely, one whose profile is described by elliptic functions—the natural ansatz is a two-cut curve possessing both the parity symmetry $x\to -x$ and the additional $\mathbb{Z}_2$ symmetry $x\to 1/x$. The most convenient starting point is an ansatz for $p'(x)$, which lives on a simpler two-sheeted Riemann surface. The most general function with these symmetries, having only double poles at $x=\pm1$ and decaying as $1/x^2$ at large $x$, is~\footnote{
A very similar curve was considered, for example, in \cite{Sizov:2013joa}. Our case can be obtained formally by setting $\psi=\pi/2$ and $r=ib$ in that work. We therefore refer the reader there for further details.
}
\beq\la{pprime}
p'(x)=
\frac{b\,\left(A_1\,x^4+A_1+A_3\,x^2\right)}{{\left(x^2-1\right)}^2\,\sqrt{b-x}\,\sqrt{b+x}\,\sqrt{b\,x-1}\,\sqrt{b\,x+1}}\,,
\eeq
leaving only three free parameters: $b$, $A_1$, and $A_3$. We choose the branch cuts as shown in figure ~\ref{fig:cuts}.
The final set of conditions requires $p(x)$ itself, which we can define via
\beq
p(x)=\pi+\int\limits_b^x\dd x\,  p'(x)\,.
\eeq
We require that
\beq
p(x)= \pi+{\cal O}\(\frac{1}{x}\)\,,\qquad
\left.p(x)\right|_{x\to 1}\simeq \frac{\theta}{x-1}\,,\qquad p(1/b)=-\pi\,,
\eeq
which fixes all remaining parameters. It is convenient to parametrise the branch point $b$ as
$
b=\frac{\sqrt{1-\zeta }+1}{\sqrt{\zeta }}
$,
which then gives
\beq\la{factors}
\theta=2\sqrt{1-\zeta }K\left(\zeta \right)\,,\quad 
A_1=-\frac{4\left(\left(\zeta -1\right)K\left(\zeta \right)+E\left(\zeta \right)\right)}{\sqrt{\zeta }}\,,\quad
A_3=\frac{8\left((1-\zeta) K\left(\zeta \right)+E\left(\zeta \right)\right)}{\sqrt{\zeta }}\,.
\eeq
As a check, one can ``measure'' the charge by expanding $p(x)$ at large $x$, which gives
\beq
p(x)\simeq \pi \pm 4i\frac{\left(\zeta -1\right)\,K\left(\zeta \right)+E\left(\zeta \right)}{x \sqrt{\zeta }}\,,
\eeq
where the sign depends on which side of the cut $p$ is evaluated, i.e.\ above or below the real axis. Comparing with \eq{plargex}, we recognise $E_R$ in perfect agreement with \eq{QL}, confirming that this is the correct curve. 

\paragraph{Spectral Curve in the Far Regime.}
So far, we have considered the $AdS_2$ solution with $\delta$ set strictly to zero. We now turn on a small $\delta$ and reconstruct $p(x)$ in the far regime. Although the full solution at finite $\delta$ is not known, the large-$x$ behaviour of $\tr\Omega(x)$ can be fixed from the asymptotic formula \eq{xlargesmall} using symmetry considerations alone. 
For example, when $x\to\infty$ at the leading order the flat connection is zero and only the twist matrix survives. At the next order, as the boundary condition preserves the energy, the first contribution of the flat connection enters as the only conserved global charge ${\cal E}=\Delta/g$, so that one finds
\beqa
\tr\Omega(x) &=&
-2\,\cos\delta +\frac{2\,{{\cal E}} \,\sin\delta}{x}-\frac{\beta}{x^2}+\mathcal{O}\left(\frac{1}{x^3}\right)\,,\qquad{{\cal E}}\equiv\frac{\Delta}{g}\,,
\eeqa
where $\Delta$ is conjugate to global $AdS$ time, $t$ in (\ref{ds2}). The $1/x^2$ term in general is not given by a Noether charge. Instead, $\beta$ is some higher conserved charge.

Next, we expand to second order in $\delta=\pi-\phi$ and introduce the rescaled variable $x=X/\delta$, assuming that all coefficients in the $1/x$ expansion remain regular as $\delta\to0$. We find
\beqa
\tr\Omega &=& -2+\delta ^2\,\left(\frac{\beta}{X^2}+\frac{2\,{{\cal E}}}{X}+1\right)+\mathcal{O}\left(\delta ^3\right)\,,
\eeqa
using that $\tr\Omega=2\cos p= -2+ (p-\pi)^2+O((p-\pi)^4)$ we conclude that
\beq\la{pfar}
p^{\rm far} = \pi+\delta\frac{
\sqrt{\beta+2\,{\cal E} X+X^2}}{X} +{\cal O}(\delta^2)\,.
\eeq
This gives the quasimomentum in the far regime totally fixed by the symmetries (except for one free coefficient $\beta$). We see that $p^{\rm far}$ has two branch points at
\beq
X_\pm=\pm\sqrt{{{\cal E}}^2+4\,\beta ^2}-{{\cal E}}\,.
\eeq
One may ask how this is consistent with the near regime that we have studied in section \ref{sec:QSCSolve}, where we found four branch points. In fact, $X_\pm$ are the endpoints of the cuts that originate at $x=\pm b$ in the near regime, as shown in Fig.~\ref{fig:cuts}. For finite $\delta$ we therefore expect eight branch points in total: two finite-size cuts outside the unit circle and two inside. As $\delta\to 0$, the two external branch points recede to infinity and become invisible in the near regime. Finally, in order to fix $\beta$ we can compare the two expressions for $p(x)$ in the overlapping regime $1\ll x\ll1/\delta$ i.e. $x$ is large while $X\to 0$, which gives $\beta=-E_R/(2g)$.

\subsubsection{Extracting the Density in the Classical Regime}
Now that we have good control over the quasimomentum, we can apply the quasiclassical Bohr--Sommerfeld quantisation condition. Since the action variables are given by the contour integrals \eq{Sk} of the quasimomentum, the quantisation condition simply requires an integer filling fraction, $\cN_k\in\mathbb{Z}$.

Since the density is insensitive to ${{\cal E}}$-independent shifts of the filling fraction $\cN_k$, it suffices to use only $p^{\rm far}(x)$ \eq{pfar}. We impose the quantisation condition \eq{Sk} on the cut along the negative real axis
\beqa\la{nfar}
\frac{\pi}{g}\cN^{\,\rm far}&=&\frac{1}{i\delta}\int\limits_{-\sqrt{{{\cal E}}^{2}+4\beta^{2}}-{{\cal E}}}^{-R\delta}\!\!\!\!\!\!\dd X\(p^{\rm far}-\pi\)\\
\nn&=&
\beta \,\left(\log\frac{64\,\beta ^4}{(R\delta)^2\,\left({{\cal E}}^2+4\,\beta ^2\right)}-2\right)+
{{\cal E}}\,\left(\pi-\frac{\ii}{2}\,\log\frac{{{\cal E}}-2\,\mathrm{i}\,\beta }{{{\cal E}}+2\,\mathrm{i}\,\beta }\right)\,.
\eeqa
The result depends on the small-$X$ cut-off $R\delta$, which cancels when the near-regime contribution is included. For the purpose of computing the density, however, all that matters is that the near contribution is independent of ${{\cal E}}$. We therefore have
\beq
\rho = \frac{1}{g}\frac{d\cN}{d{{\cal E}}}
= \frac{1}{g}\frac{d\cN^{\rm far}}{d{{\cal E}}}
=\frac{1}{2}-\frac{\mathrm{i}}{2\pi}\,\log\frac{2\,\mathrm{i}\,\beta-{{\cal E}} }{2\,\mathrm{i}\,\beta+{{\cal E}} }\,,
\eeq
which is exactly the result of \eq{dsdensity} (recalling that ${\bf C}=E_R^2=(2g\beta)^2$). Note also that from \eq{QL}, $\beta=2$ at $\theta=0$, in agreement with the analysis of section~\ref{foldedstringsec}.

\subsubsection{Computing ${\cal S}$ in the Classical Regime}
Let us now perform a more careful analysis of the Bohr-Sommerfeld quantisation by including the near regime contribution into the action variable $\cN$. We again use \eq{Sk}. The near contribution to the filling fraction is 
\beq\la{nneardef}
\frac{\pi}{g}\cN^{\,\rm near} = \frac{1}{i}
\int\limits_{-R}^{-b}(p(x)-\pi)\,\dd\(x+\frac{1}{x}\)
= - (p(R)-\pi)R-\frac{1}{i}
\int\limits_{-R}^{-b}\dd x\(x+\frac{1}{x}\)p'(x)\,.
\eeq
Recall that $p'(x)$ is given explicitly by the simple expression (\ref{pprime}), with the coefficients specified in (\ref{factors}) and $b=(\sqrt{1-\zeta }+1)/\sqrt{\zeta }$. The corresponding integral can be evaluated analytically, yielding
\beq\la{nnear}
\frac{\pi}{g}\cN^{\,\rm near}=\beta\,\left(\log\frac{\zeta \,R^2}{1-\zeta }+2\right)-4\,\sqrt{\zeta }\,K\left(\zeta \right)\,.
\eeq
Now combining with \eq{nfar}, we get the full expression for Bohr-Sommerfeld quantised $n$
\beq\nn
\frac{\pi}{g}\cN^{\,\rm BS}=
\beta \,\log\frac{64\,\zeta\,\beta ^4}{\delta^2\,(1-\zeta)\,\left({{\cal E}}^2+4\,\beta ^2\right)}+
{{\cal E}}\,\left(\pi-\frac{\ii}{2}\,\log\frac{{{\cal E}}-2\,\mathrm{i}\,\beta }{{{\cal E}}+2\,\mathrm{i}\,\beta }\right)-4\,\sqrt{\zeta }\,K\left(\zeta \right)\,.
\eeq
The above equation is nothing but the classical limit of the FME equation \eq{gluing2}. Indeed, by using that $\tilde\vartheta = 2g\beta$ and expanding at large $g$ from 
\eq{gluing2} we obtain
\beq
\frac{\pi}{g}n^{\rm FME}=
\beta \,\left(\log\frac{64}{\delta^2\left({{\cal E}}^2+4\,\beta ^2\right)g^2}+2\right)+
{{\cal E}}\,
\left(\pi-\frac{\mathrm{i}}{2}\,\log\frac{{{\cal E}}-2\,\mathrm{i}\,\beta }{{{\cal E}}+2\,\mathrm{i}\,\beta } \right)
-\frac{1}{2ig}\log {\cal S}\,.
\eeq
By comparing the two we see that the dependence on ${{\cal E}}$ is identical and, furthermore, we can identify ${\cal S}$ with
\beqa\la{calSres}
\frac{1}{2ig}\log {\cal S}=
\beta \,\left(2-\log\frac{\zeta\,g^2\,\beta ^4}{(1-\zeta)}\right)+4\,\sqrt{\zeta }\,K\left(\zeta \right)\,.
\eeqa
We see that this formula predicts a non-trivial $\log(g)$ dependence in addition to the linear scaling. We shall compare it with our non-perturbative QSC data below. Before doing so, let us address a conceptual question. Since ${\cal S}$ is, by construction, an intrinsically near-region quantity, it is natural to ask how the Bohr--Sommerfeld quantisation condition arises from the QSC at strong coupling.

\paragraph{${\cal S}$ from the Near Regime Only.}
Let us step back and see how ${\cal S}$ appears in the QSC formalism of section \ref{sec: QSC}. It emerged, e.g. in the limit of ${\bf Q}_2(u)$ at $\delta\to 0$, where ${\bf Q}_2(u)$ becomes a combination of ${\mathbb Q}_1(u)$ and
${\mathbb Q}_2(u)$ -- two solutions which have pure asymptotic $u^{1\pm \vartheta}$ at large $u$, or more precisely for  $1\ll u\ll 1/\delta$\footnote{Up to a $\sim 1$ irrelevant factor at large $g$.}
\beq\la{Q2S2}
e^{-\pi u}{\bf Q}_2(u)= {\mathbb Q}_1(u)\,{\mathbb S}+{\mathbb Q}_2(u)\,,
\eeq
where ${\mathbb S}$ is basically $1/\cal S$ up to a simple factor, see \eq{SandA}. In order to understand how to fix the factor ${\mathbb S}$ in the classical limit, let us recall the relation between $\bQ_i$ and quasimomentum \cite{Gromov:2014caa}
\beq
\bQ_2\simeq {\cal C}\exp{\int\limits_i^u\dd u\, p(u)}\,,\qquad p(u)=\d_u\log {\bf Q}_2\,,
\eeq
which is essentially a WKB approximation.

From \eq{Q2S2} we can now clearly see how the branch cut develops at $g\to\infty$. As $\vartheta(g)\simeq 2i\beta g$ becomes purely imaginary, both terms in (\ref{Q2S2}) oscillate along the real axis -- it is exactly like in the usual quantum mechanics in the classically allowed domain, where two WKB oscillating wavefunctions co-exist. Now, in order to estimate the factor $S$, we can use the fact that at the branch point $b$ (turning point), both oscillating factors have to synchronise their phases in order to merge with the near turning point approximation. For that, we have to require that
\beq
\int\limits_{i}^b\dd u\, p(u+i0)\simeq \int\limits_{i}^b\dd u\, p(u-i0)\,.
\eeq
At the same time, at large $u$ we have to require the normalisation of ${\mathbb Q}_1(u)$ and ${\mathbb Q}_2(u)$ to be just $e^{\pm \vartheta\log u}$. Thus, we should have
\beqa
\log {\mathbb S}{-}\log {\cal C}+\int_{i}^{R_u} p(u+i0) du&\simeq& -\vartheta \log R_u\,,\\
{-}\log {\cal C}+\int_{i}^{R_u} p(u-i0) du&\simeq& +\vartheta \log R_u\,.
\eeqa
After subtracting these two expressions and converting cut-off in $u$ to cut-off in $x$ as $R_u=g\, R$, we get
\beq
\log {\mathbb S}=\subset\hspace{-4.5mm}\int\limits_{b}^{R_u} p(u) du  +2\vartheta \log R_u=
2\pi\ii\,\cN^{\,\rm near}-2\vartheta \log (g\,R)\,,
\eeq
where we used \eq{nneardef}. Using (\ref{nnear}) along with $\vartheta\simeq2i\beta g$ we find
\beq\label{STSexp}
\frac{\log {\mathbb S}}{2\ii g}=
\beta\,\left(\log\frac{\zeta}{g^2(1-\zeta) }+2\right)-4\,\sqrt{\zeta }\,K\left(\zeta \right)\,,
\eeq
which agrees precisely with \eq{calSres}.

To summarise, we have determined the strong-coupling behaviour of the fusion data, $\vartheta(g)$ and ${\mathbb S}$, together with the strong-coupling density $\rho(\Delta)$, for arbitrary internal angle $\theta$. An interesting next step would be to quantise the fluctuations around the falling folded-string solution and derive the subleading strong-coupling corrections to these quantities.

\section{Future Directions}\la{sec:future}

At a conceptual level, the central new observation of this work is that the
quadratic Casimir continues to commute with the Hamiltonian on the sphere
above the critical coupling, despite the loss of conformality on the fused
defect. It would be important to understand the general conditions under
which this phenomenon occurs. 
In particular, one may ask whether an analogous structure can emerge in the RG dynamics of an ordinary quantum field theory, rather than being restricted to defects.

A natural next step is to study correlation functions in the walking regime. In particular, one would like to understand how four-point functions on the cylinder decompose into the conformal families identified in this work. Since the individual states drift along the RG flow, a
particularly interesting regime is reached when the radius of the sphere is
much larger than the characteristic RG scale. The dynamics at energies of
order the inverse sphere radius should then be controlled by the universal
density $\rho(\Delta)$ in (\ref{density}). It is tempting to conjecture that
the usual discrete sum over intermediate states is replaced in this regime by
a spectral integral weighted by $\rho(\Delta)$, possibly involving an
appropriate analytic continuation of the conformal partial waves.

Developing such a representation may provide a starting point for bootstrap
methods in non-conformal walking theories. Recent endpoint-bootstrap
constructions explicitly incorporate the defect-creation operators at the
ends of an endable conformal line and derive crossing equations that mix
bulk and defect data while retaining positivity
\cite{Lanzetta:2025xfw}. The defect-changing operators considered here
are precisely of this type. Below criticality, our exact QSC spectrum could
therefore provide nontrivial input and a benchmark for this bootstrap.
Above criticality, the replacement of discrete spectrum by the
density $\rho(\Delta)$ suggests a possible extension of the endpoint
bootstrap beyond conformal fixed points.

This programme may also offer an indirect route to complex CFTs. Standard
conformal-bootstrap techniques rely strongly on unitarity and positivity
and are difficult to apply directly to complex fixed points. In the setting
studied here, one may instead attempt to bootstrap the intermediate theory,
which is unitary but non-conformal. Since its drifting states carry the same
Casimir data as the conformal families of the associated complex fixed
points, its correlation functions may retain nontrivial information about
those theories.

There are several natural questions from the integrability perspective. The
closed system of QSC equations obtained in the fusion limit exhibits an
unexpected degree of simplification and, in several respects, resembles the
QSC for local operators. It would be important to determine whether this
resemblance reflects a deeper integrable structure intrinsic to the fused
line. A first step would be to extend the present construction to general
field insertions at the cusp and derive the corresponding asymptotic Bethe
ansatz directly for the fused system. Such a construction should reveal
more transparently how the different conformal families are organised and
may provide a more efficient analytic description of their critical
couplings.

A new ingredient in the fusion problem is the function ${\mathbb S}(g)$.
Together with $\vartheta(g)$, which is related to the Casimir of the conformal family, it contains the near-region information and
determines how the states in a conformal family are corrected at small but
finite $\delta$. Since ${\mathbb S}$ is related to a defect structure
constant, it would be particularly interesting to study its asymptotic
Bethe-ansatz limit for general states. Previous studies of structure
constants involving defect-changing operators and local operators on
Wilson lines suggest that this limit may admit an interpretation in terms
of boundary or hexagon-like building blocks
\cite{Basso:2015zoa,Kim:2017sju}. Establishing such a
relation would connect the spectral problem of fused defects directly to
integrability-based methods for correlation functions. It could also
supply new non-protected structure-constant data for the bootstrability
programme \cite{Cavaglia:2024dkk}.

The operators analysed in this work are defect-creation and annihilation operators.  
A complete description
of the fused defect must also include local operators living directly on
top of the fused line.\footnote{Only a subset of these operators factorise into products of defect-creation and defect-annihilation operators.} Optimistically, their spectrum and correlation
functions may themselves be governed by integrability. It is not clear,
however, how these states can be accessed using the existing
cusped-Wilson-line or QSC constructions, since they do not naturally arise
as endpoint states. Finding an integrable description of this sector is
therefore an important open problem.

Finally, there are two particularly interesting limits that merit further investigation. One is the limit $\theta\to\pi$ of the fusion problem. In this limit, one expects the supersymmetry of the fused $1/2$-BPS Wilson lines to be restored. At the same time, this limit is distinct from the near-supersymmetric regime, in which one first takes $\theta\to(\pi-\delta)$ and only afterwards the fusion limit $\delta\to0$ may be taken. Another interesting regime is the limit where both $\delta\to0$ and $g\to0$. As we have seen, these two limits do not commute. It would therefore be interesting to study a double-scaling limit in which both parameters are taken to zero together, as such a limit may reveal a novel structure.

\section*{Acknowledgments}

We thank O.~Aharony, E.~Armanini, S.~Itzhaki, Z.~Komargodski, S.~Komatsu, P.~Kravchuk and A.~Zhiboedov for useful discussions. We thank O.~Aharony, S.~Ekhammar, Z.~Komargodski, P.~Kravchuk, F.~Levkovich-Maslyuk, and M.~Preti for comments on the manuscript. AS is supported by the Israel Science Foundation, grant number 1099/24. The work of FC is funded by the STFC Research Council grant ST/Y509279/1. NG and AS are grateful to CERN for hospitality during the preparation of this work. The work of NG was partially supported by the European Research Council (ERC) under the European Union's Horizon 2020 research and innovation programme (grant agreement No.~865075, EXACTC), and by the Science and Technology Facilities Council under grants ST/P000258/1 and ST/X000753/1.
Some computations in this work were facilitated by the \href{https://github.com/vanbaalon/wolfbook}{Wolfbook} VS~Code extension.

\newpage

\appendix

\section{QSC Generalities}\label{app: QSCGeneral}
\subsection{Details of the Fusion of the ladder model in QSC language}

The ladder model Baxter equation takes the following form in the near regime
\begin{equation}
2 \left(u^2+2 \hat{g}^2\right) \mathbbm{q}(u)-u^2 (\mathbbm{q}(u-\ii)+\mathbbm{q}(u+\ii))=0\,.
\end{equation}
It has two linearly independent solutions that can be written in terms of the hypergeometric function $_{3}F_2$

\begin{eqnarray}\la{p14}
\mathbbm{q}_1(u)
&=&
\frac{u\,\Gamma\!\left(\frac12+\vartheta\right)^{2}\Gamma\!\left(\frac32+\vartheta\right)}
     {2i\,\Gamma(2\vartheta)\,\sin\!\left(\pi\vartheta\right)}\,
e^{\,i\pi\vartheta/2}
\Bigg[
e^{\,\ii\pi/4}\,{}_3F_{2}\!\bigg(\!\begin{matrix}
1-\ii u,\ \tfrac{1}{2}-\vartheta,\ \tfrac{1}{2}+\vartheta
\\[-2pt]
1,\ 2
\end{matrix}\!;1\bigg)
\\[-1pt]
&&\hspace{9.5em}
-\ e^{-\ii\pi/4}\,
\frac{\cosh\!\big(\pi(u-\ii \vartheta\big)}{\sinh(\pi u)}\,
{}_3F_{2}\!\bigg(\!\begin{matrix}
1+\ii u,\ \tfrac{1}{2}-\vartheta,\ \tfrac{1}{2}+\vartheta
\\[-2pt]
1,\ 2
\end{matrix}\!;1\bigg)
\Bigg]\,,\nn
\\[6pt]
\mathbbm{q}_2(u)
&=&
\frac{2^{\,2\vartheta+1}\pi^{3/2}u\,\Gamma\!\left(\tfrac{1}{2}-\vartheta\right)}
     {\ii\,\sin(2\pi\vartheta)\,\Gamma\!\left(\vartheta-\tfrac{1}{2}\right)\Gamma\!\left(-\vartheta\right)}\,
e^{\,\ii\pi\vartheta/2}
\Bigg[
e^{\,\ii\pi/4}\,{}_3F_{2}\!\bigg(\!\begin{matrix}
1-\ii u,\ \tfrac{1}{2}-\vartheta,\ \tfrac{1}{2}+\vartheta
\\[-2pt]
1,\ 2
\end{matrix}\!;1\bigg)\nn
\\[-1pt]
&&\hspace{9.5em}
+\ e^{\,3\ii\pi/4}\,
\frac{\cosh\!\big(\pi(u+\ii \vartheta)\big)}{\sinh(\pi u)}\,
{}_3F_{2}\!\bigg(\!\begin{matrix}
1+\ii u,\ \tfrac{1}{2}-\vartheta,\ \tfrac{1}{2}+\vartheta
\\[-2pt]
1,\ 2
\end{matrix}\!;1\bigg)
\Bigg]\,.\nn
\end{eqnarray}
We choose these specific linear combinations so that each $\mathbbm{q}_i$ satisfies the pure asymptotics condition, with their large-$u$ expansion given by
\begin{equation}
    \bq_1\simeq u^{\frac12+\vartheta}\,,\quad\bq_2\simeq u^{\frac12-\vartheta}\,,\qquad u\to\infty\,.
\end{equation}
Having the above expressions for $\mathbbm{q}_i$ is essential for deriving the analytic expression for $\mathcal{S}_\pm$ in \eqref{SpmSh}.

\subsection{Cusped Wilson line QSC details} 

The QSC for the cusped Wilson line configuration has been actively studied in \cite{Gromov:2015dfa}. Since it plays a central role as the starting point for our fusion-limit analysis, we briefly recall it here.

At the heart of the QSC lies the QQ-system: a set of finite-difference equations satisfied by the Q-functions. These functions encode observables relevant to the system under consideration. For example, they determine the scaling dimensions of local operators or, in the present case, the cusp anomalous dimension $\Gamma_\text{cusp}$.

The form of QQ-system relations is dictated by the $PSU(2,2|4)$ symmetry of $\mathcal{N}=4$ SYM theory and consists of relations on $2^8$ Q-functions. They all can be expressed in terms of special $4+4$ Q-functions of a single fundamental index -- the bosonic $\mathbf{P}_a(u)$ ($a = 1, \ldots, 4$) and the fermionic $\mathbf{Q}_i(u)$ ($i = 1, \ldots, 4$).
The Wilson line breaks the $PSU(2,2|4)$ symmetry down to $OSP(4|4)$. This breaking manifests through an additional constraint of the Q-functions, relating the single-index Q-functions to their Hodge-duals (Q-functions with $7$ indices)\footnote{For the most general states $\bP^{a}(u)=\chi^{ab}\bP_{b}(-u)$, and similarly for $\bQ_i$. In the parity-symmetric sector considered here, the argument flip is irrelevant.} 
\begin{equation}\la{chiconstraint}
\bP^{a}=\chi^{ab}\bP_{b}\,,\qquad\bQ^{i}=\chi^{ij}\bQ_{j}\,.
\end{equation}
Here, the invariant tensors $\chi^{ij}$ and $\chi^{ab}$ are
\begin{equation}
    \chi^{ab}=\chi^{ij}=\begin{pmatrix}
        0&0&0&-1\\0&0&1&0\\0&-1&0&0\\1&0&0&0
    \end{pmatrix}\,.
\end{equation}

The functions $\bP_a$ describe the $S^5$ sector of the theory. Their large-$u$ exponential behaviour encodes the internal angle $\theta$, while the power-law prefactor carries information about the R-charge of the cusp operator. More precisely, consider a cusp operator containing $L$ scalar insertions orthogonal to those living on the Wilson lines. It carries $L$ units of R-charge, and both this charge and the angle $\theta$ are reflected in the asymptotic behaviour as $\bP_a\sim u^{\pm L}e^{\pm\theta u}$. In addition, the functions $\bP_a$ have a simple analytic structure: they possess a Zhukovsky cut along the interval $(-2g,2g)$ on the real axis, as well as a square-root branch point at the origin.

Together with the parity symmetry, which relates $\bP_a(u)$ and $\bP_a(-u)$ through a linear map, and the $\chi$ constraint (\ref{chiconstraint}), these properties determine the large-$u$ asymptotic expansion of $\bP_a$%
\begin{equation}\label{Pthph}
\begin{aligned}
\bP_1 &=\mathbf{A}_1\,\dfrac{\sqrt{u}}{gx}(gx)^{-L}\,e^{+\theta u}\,f(x)\,,
&\qquad
f(x) &= 1 + \frac{a_1}{x} + \frac{a_2}{x^2} + \frac{a_3}{x^3} + \cdots\,, \\[6pt]
\bP_2 &= \mathbf{A}_2\,\dfrac{\sqrt{u}}{gx}(gx)^{-L}\,e^{-\theta u}\,f(-x)\,,\\[8pt]
\bP_3 &= \mathbf{A}_3\,\dfrac{\sqrt{u}}{gx}(gx)^{2+L}\,e^{+\theta u}\,g(x)\,,
&\qquad
g(x) &= 1 + \frac{b_1}{x} + \frac{b_2}{x^2} + \frac{b_3}{x^3} + \cdots\,, \\[6pt]
\bP_4 &= \mathbf{A}_4\,\dfrac{\sqrt{u}}{gx}(gx)^{2+L}\,e^{-\theta u}\,g(-x)\,,
\end{aligned}
\end{equation}
where $x(u)=\frac{u}{2g}+\sqrt{\frac{u}{2g}+1}\sqrt{\frac{u}{2g}-1}$ is the Zhukovsky map.

Similarly, the $\bQ_i(u)$ functions characterise the string dynamics in the $AdS_5$ factor. The cusp dimension and the cusp angle are encoded in their large $u$ asymptotics as 

\begin{align}
    \bQ_1 \simeq&\mathbf{B}_1\, u^{1/2+\Gamma_\text{cusp}} e^{u(\pi-\delta)}\,, \qquad
\bQ_2 \simeq \mathbf{B}_2\, u^{1/2+\Gamma_\text{cusp}} e^{-u(\pi-\delta)}\,,\\
\bQ_3 \simeq&\mathbf{B}_3\, u^{1/2-\Gamma_\text{cusp}} e^{u(\pi-\delta)}\,, \qquad
\bQ_4 \simeq \mathbf{B}_4\, u^{1/2-\Gamma_\text{cusp}} e^{-u(\pi-\delta)}\,,
\end{align}
where $\simeq$ denotes the leading term as $u\to +i\infty$. 

As with the $\bP_a$, the $\bQ_i$ carry an extra $\sqrt{u}$ factor specific to the cusp geometry; note that the combinations $\bP_a/\sqrt{u}$ and $\bQ_i/\sqrt{u}$ contain only the Zhukovsky-type cuts. The non-compact $AdS$ geometry is reflected in the choice of Riemann sheet on which $\mathbf{Q}_i/\sqrt{u}$ has a single long cut $(-\infty,-2g)\cup(+2g,+\infty)$ rather than the short one.

The strategy for solving the QQ-system exploits the asymmetry between $\bP_a$ and $\bQ_i$. The functions $\bP_a$ admit a simple closed-form parametrisation \eqref{Pthph}, making them a natural choice for the solution basis. In contrast, the analytic structure of $\bQ_i$ is far more intricate and gives rise to a set of highly nontrivial consistency relations known as the gluing conditions. The procedure therefore proceeds in two steps. First, the $\bQ_i$ are reconstructed in terms of $\bP_a$ by the methods outlined below. Then, the gluing conditions on $\bQ_i$ are imposed in order to isolate the physical solution. We now describe this construction in more detail.

There is a direct way of expressing $\bQ_i$ in terms of $\bP_a$, namely the so-called Baxter relation.
\begin{equation*}
    \bQ_i^{[+4]}D_0-\bQ_i^{[+2]}\left[D_1-\bP^{a[+2]}\bP_a^{[+4]}D_0\right]+\bQ_i\left[D_2-\bP^a\bP^{a[+2]}D_1+\bP_a\bP^{a[+4]}D_0\right]\end{equation*}
    \begin{equation*}-\bQ_i^{[-2]}\left[\bar{D}_1\bP_a^{[-2]}\bP^{a[-4]}\bar{D}_0\right]+\bQ_i^{[-4]}\bar{D}_0=0\,,
    \end{equation*}
where coefficients $D_n$ are defined as follows:
\begin{align}
D_0=&{\rm det}
\(
\bea{llll}
\bP^{1[+2]}&\bP^{2[+2]}&\bP^{3[+2]}&\bP^{4[+2]}\\
\bP^{1}&\bP^{2}&\bP^{3}&\bP^{4}\\
\bP^{1[-2]}&\bP^{2[-2]}&\bP^{3[-2]}&\bP^{4[-2]}\\
\bP^{1[-4]}&\bP^{2[-4]}&\bP^{3[-4]}&\bP^{4[-4]}
\eea
\)\,,\quad
D_1={\rm det}
\(
\bea{llll}
\bP^{1[+4]}&\bP^{2[+4]}&\bP^{3[+4]}&\bP^{4[+4]}\\
\bP^{1}&\bP^{2}&\bP^{3}&\bP^{4}\\
\bP^{1[-2]}&\bP^{2[-2]}&\bP^{3[-2]}&\bP^{4[-2]}\\
\bP^{1[-4]}&\bP^{2[-4]}&\bP^{3[-4]}&\bP^{4[-4]}
\eea
\)\,,\nn\\
D_2=&{\rm det}
\(
\bea{llll}
\bP^{1[+4]}&\bP^{2[+4]}&\bP^{3[+4]}&\bP^{4[+4]}\\
\bP^{1[+2]}&\bP^{2[+2]}&\bP^{3[+2]}&\bP^{4[+2]}\\
\bP^{1[-2]}&\bP^{2[-2]}&\bP^{3[-2]}&\bP^{4[-2]}\\
\bP^{1[-4]}&\bP^{2[-4]}&\bP^{3[-4]}&\bP^{4[-4]}
\eea
\)\,,
\end{align}
and $\bar{D}_n$ are their complex conjugates. This equation can be studied in various limits, often revealing nontrivial properties of the system. In particular, by considering the first few terms in its large-$u$ expansion, one finds that the expansion coefficients of $\bP_a$ are not all independent. The complete set of these constraints is presented below:
\beq\label{AAAA}
    \mathbf{A}_1\mathbf{A}_4=-\mathbf{A}_2\mathbf{A}_3=-\frac{i(\cos\theta+ \cos\delta)^2}{2(L+1)\sin^2\theta}\,,\quad
    \mathbf{B}_1\mathbf{B}_4=-\mathbf{B}_2\mathbf{B}_3=\dfrac{i(\cos\theta+\cos\delta)^2}{2\Gamma_{\text{cusp}} \sin^2\delta}\,,
\eeq
\begin{equation}\label{thphiconstr}
a_1 - b_1
=
-\,\frac{(L+1)\bigl(2\cos\theta\cos\phi + \cos 2\theta - 3\bigr)}
{2\sin\theta(\cos\theta - \cos\phi)}\,,
\end{equation}
\begin{equation}
\resizebox{0.8\linewidth}{!}{$
\begin{aligned}\label{gcuspapp}
\Gamma_{\text{cusp}}^{2} =\;&
a_1 \Bigg(
  -\frac{a_2 g^3 \csc\theta \csc^2\phi (\cos\theta-\cos\phi)^3}{L+1}
  +\frac{b_2 g^3 \csc\theta \csc^2\phi (\cos\theta-\cos\phi)^3}{L+1} \\[4pt]
&\qquad
  -\frac{1}{4} g \csc^3\theta \csc^2\phi (\cos\theta-\cos\phi)
  \Big[
    -2(5L+4)\cos\theta\cos\phi  \\[2pt]
&\qquad\qquad
    +\cos(2\theta)\bigl(
      2L\cos\theta\cos\phi
      +L\cos(2\phi)
      -L+2
    \bigr) \\[2pt]
&\qquad\qquad
    +(L+2)\cos(2\phi)
    +7L+4
  \Big]
\Bigg) \\[6pt]
&+\frac{a_3 g^3 \csc\theta \csc^2\phi (\cos\theta-\cos\phi)^3}{L+1}
-a_1^2 g^2 \csc^2\theta \csc^2\phi
(\cos\theta\cos\phi-1)(\cos\theta-\cos\phi)^2 \\[6pt]
&+\frac{a_2 g^2 \csc^2\theta \csc^2\phi (\cos\theta-\cos\phi)^2
\bigl(2\cos\theta\cos\phi-(L+1)\cos(2\theta)+L-1\bigr)}{2(L+1)} \\[6pt]
&-\frac{b_3 g^3 \csc\theta \csc^2\phi (\cos\theta-\cos\phi)^3}{L+1}
+\frac{b_2 g^2 L \csc^2\theta \csc^2\phi
(\cos\theta\cos\phi-1)(\cos\theta-\cos\phi)^2}{L+1} \\[6pt]
&+\frac{1}{48}\Bigg(
  24 g^2 \cos(2\theta)\csc^2\phi
  -8(18 g^2+2L^2+L)\cos\theta\cot\phi\csc\phi \\[2pt]
&\qquad
  -16(9 g^2-L^2+L)
  +216 g^2 \csc^2\phi \\[2pt]
&\qquad
  -2 \sec^2\!\left(\frac{\theta}{2}\right)\!
   \cot^2\!\left(\frac{\phi}{2}\right)
   \Big[(6 g^2+2L^2+L)\cos\phi+6 g^2-2L(2L+1)\Big] \\[2pt]
&\qquad
  +2 \csc^2\!\left(\frac{\theta}{2}\right)\!
   \tan^2\!\left(\frac{\phi}{2}\right)
   \Big[(6 g^2+2L^2+L)\cos\phi-6 g^2+2L(2L+1)\Big]
\Bigg)\,.
\end{aligned}
$}
\end{equation}
In particular, the last equation is what determines the cusp dimension $\Gamma_{\text{cusp}}$ after the QQ-system has been solved. We stress that all of the above constraints must be imposed manually when solving the QSC numerically; otherwise, the algorithm fails to converge. Note that there is some freedom in the choice of $\mathbf{A}_i$ and $\mathbf{B}_i$; for later use we fix $\mathbf{B}_i$ to be
\begin{equation}\label{Qbs}
    \mathbf{B}_1=\mathbf{B}_2=\frac{i(\cos\theta +\cos\delta)}{\sqrt{2\Gamma_{\text{cusp}} }\sin\delta}\,,\qquad \mathbf{B}_3=-\mathbf{B}_4=-\frac{\cos\theta +\cos\delta}{\sqrt{2\Gamma_{\text{cusp}} }\sin\delta}\,.
\end{equation}
The physical solution is singled out by imposing restrictions \eqref{AAAA}-\eqref{gcuspapp} along with
the gluing conditions
\begin{align}\label{gluefar1}
\left(\dfrac{\mathbf{Q}_1(u)}{\sqrt{u}}\right)^{\gamma} &= \dfrac{\mathbf{Q}_1(-u)}{\sqrt{-u}}\,, \\\label{gluefar2}
\left(\dfrac{\mathbf{Q}_2(u)}{\sqrt{u}}\right)^{\gamma} &= \dfrac{\mathbf{Q}_2(-u)}{\sqrt{-u}}\,,  \\\label{gluefar3}
\left(\dfrac{\mathbf{Q}_3(u)}{\sqrt{u}}\right)^\gamma &= \mathbf{a}_1 \sinh(2 \pi u) \dfrac{\mathbf{Q}_2(-u)}{\sqrt{-u}} + \dfrac{\mathbf{Q}_3(-u)}{\sqrt{-u}}\,, \\\label{gluefar4}
\left(\dfrac{\mathbf{Q}_4(u)}{\sqrt{u}}\right)^{\gamma}&= \mathbf{a}_2 \sinh(2 \pi u) \dfrac{\mathbf{Q}_1(-u)}{\sqrt{-u}} + \dfrac{\mathbf{Q}_4(-u)}{\sqrt{-u}} \,.
\end{align}
Here $\gamma$ denotes analytic continuation of Q-functions through the short cut $(-2g,2g)$; for a derivation of these relations see \cite{Gromov:2015dfa}. 

In practice, both $\mathbf{Q}_i$ and its analytic continuation $\mathbf{Q}^{\gamma}_i$ are often reconstructed from $\bP_a$ in two steps, avoiding the more complicated Baxter-equation approach. First, the functions $Q_{a|i}$ are determined from $\bP_a$ through
\begin{equation}
    Q_{a|i}^+-Q_{a|i}^-=-\bP_a\bP^bQ_{b|i}^{\pm}\,.
\end{equation}
One can then construct $\bQ_i$ along with its analytic continuation as follows
\begin{equation}\label{qpqpt}
\dfrac{\mathbf{Q}_i(x)}{\sqrt{u}} = -\,\dfrac{\mathbf{P}_a(x)}{\sqrt{u}}\,Q_{a|i}^{+}(x)\,,
\qquad
\left(\dfrac{\mathbf{Q}_i(x)}{\sqrt{u}}\right)^{\gamma} = -\,\dfrac{\mathbf{P}_a(1/x)}{\sqrt{u}}\,Q_{a|i}^{+}(x)\,.
\end{equation}
Substituting these expressions into \eqref{gluefar1}--\eqref{gluefar4} turns the gluing conditions into a set of equations for the coefficients $a_k$ and $b_k$. We would like to point out that the first two conditions, \eqref{gluefar1} and \eqref{gluefar2}, are already sufficient to solve the system this way.

\subsection{Fusion Limit QSC details}\label{app:FLQSC}
Here we collect some technical details of the cusped Wilson line QSC in the fusion limit $\delta\to0$. The QSC largely remains the same as in the general cusp-angle case, so we focus on highlighting the modifications specific to this limit.

The ansatz for $\bP_a$ is kept the same as in the general Wilson line setup \eqref{Pthph}; we will keep the same notation for $\bP_a$, implying that $\delta$ should be set to zero whenever the fusion limit QSC is discussed. We also focus our attention on the $L=0$ case only in this paper. With these details in mind, we choose normalisation coefficients $\mathbf{A}_a$ to be 
\begin{equation}
    \mathbf{A}_1=\mathbf{A}_2=\mathbf{A}_3=-\mathbf{A}_4=\left(\frac12+\frac{i}{2}\right)\cot\left(\frac\theta2\right)\,,
\end{equation}
making sure the restriction \eqref{AAAA} is satisfied. 

The asymptotics of the $\bQ_i$ functions are different from those in the general setup. Namely, we take $\mathbb{Q}_i$ asymptotics \eqref{Qsmall} (with $\vartheta_1=\vartheta_2$ constraint) and remember to return their exponential scaling 
\begin{equation}\label{qasnear}
e^{\pm \pi u}\mathbb{Q}_i=\left\{\mathbb{B}_1 e^{+\pi u}u^{1+\vartheta},\ \mathbb{B}_2 e^{+\pi u}u^{1-\vartheta},\ \mathbb{B}_3 e^{-\pi u}u^{2+\vartheta},\ \mathbb{B}_4 e^{-\pi u}u^{2-\vartheta}\right\}\,,
\end{equation}
where $\mathbb{B}_i$ are so far arbitrary coefficients. After truncating the series expansion of $\mathbf{P}_a$, the set of unknown parameters becomes finite and consists of $\{\vartheta,\mathcal{S},a_n,b_n,\mathbf{A}_a,\mathbb{B}_i\}$. However, these parameters are not independent, and all relations among them must be eliminated before proceeding to the numerical stage. To achieve this, we solve the QQ-system in the large-$u$ limit. The first few iterations already provide all relevant constraints, even without imposing the gluing conditions. We find \footnote{One more constraint restricts coefficients $\mathbf{A}_i$ -- it coincides with the leftmost relation in \eqref{AAAA} in the $\delta\to0$ limit.}
\begin{equation}\label{AABBnear}
\mathbb{B}_2 \mathbb{B}_3=\dfrac{1+2\vartheta}{1-2\vartheta}\mathbb{B}_1 \mathbb{B}_4= \dfrac{8 i \cos ^4\left(\frac{\theta
   }{2}\right)}{\vartheta  (1+2 \vartheta )}\,,
\end{equation}
\begin{align}
    a_2=&\frac{a_1 \csc (\theta )}{g}+\frac{a_1^2}{2}-\frac{(2 \vartheta -1) (2 \vartheta +1)}{8 g^2 (\cos (\theta )+1)}-1\,,\nn\\
    b_1=&a_1+\frac{(\cos (\theta )-2) \csc (\theta )}{g}\,,\\
b_3=&a_1 \left(b_2-\frac{\csc ^2(\theta ) \left(4 \vartheta ^2 \cos (\theta )+7
   \cos (\theta )-4 \vartheta ^2+9\right)}{8 g^2}+1\right)\nn\\
   &-\frac{a_1^2 \cot (\theta )}{2 g}-\frac{a_1^3}{2}+a_3+\frac{(2 \vartheta -1)
   (2 \vartheta +1) \sin ^2\left(\frac{\theta }{2}\right) \cot (\theta ) \csc ^2(\theta )}{4 g^3}\,,\nn
\end{align}
   
\begin{align}
   b_5=&a_1^2
   \left(-\frac{a_3}{2}-\frac{3 b_2 \csc (\theta )}{2 g}+\frac{\csc (\theta ) \left(16 \cot (\theta ) \csc (\theta )-4 \vartheta
   ^2+1\right)}{8 g^3}-\frac{(\cos (\theta )+2) \csc (\theta )}{2 g}\right)\nn\\
   &+a_1^3 \left(-\frac{b_2}{2}+\frac{\csc ^2(\theta ) \left(4
   \vartheta ^2 \cos (\theta )+11 \cos (\theta )-4 \vartheta ^2+1\right)}{8 g^2}-1\right)\nn\\
   &+a_3 \left(b_2-\frac{\csc ^2(\theta ) \left(4
   \vartheta ^2 \cos (\theta )+7 \cos (\theta )-4 \vartheta ^2+9\right)}{8 g^2}+1\right)\\
   &+a_1 \left(\frac{a_3 \csc (\theta )}{g}-a_4+b_2
   \left(1-\frac{\csc ^2(\theta ) \left(4 \vartheta ^2 \cos (\theta )-\cos (\theta )-4 \vartheta ^2+9\right)}{8
   g^2}\right)+b_4-\right.\nn\\
   &-\left.\frac{(2 \vartheta -1) (2 \vartheta +1) \csc ^2\left(\frac{\theta }{2}\right) \sec ^4\left(\frac{\theta }{2}\right)
   \left(4 \vartheta ^2 \cos (\theta )+23 \cos (\theta )-4 \vartheta ^2+1\right)}{512 g^4}-\right.\nn\\
   &-\left.\frac{\csc ^2(\theta ) \left(4 \vartheta ^2
   \cos (\theta )+11 \cos (\theta )-4 \vartheta ^2+9\right)}{4 g^2}+1\right)-\frac{a_4 \cot \left(\frac{\theta
   }{2}\right)}{g}+\frac{a_1^4 (\cos (\theta )+2) \csc (\theta )}{4 g}\nn\\
   &+\frac{a_1^5}{4}+a_5+b_2 \left(\frac{(2 \vartheta -1) (2
   \vartheta +1) \sin ^2\left(\frac{\theta }{2}\right) \csc ^3(\theta )}{4 g^3}+\frac{\csc (\theta )}{g}\right)+\frac{b_4 \csc (\theta
   )}{g}\nn\\
   &+\frac{(2 \vartheta -1)^2 (2 \vartheta +1)^2 \sin ^4\left(\frac{\theta }{2}\right) \csc ^5(\theta )}{16 g^5}+\frac{(2 \vartheta
   -1) (2 \vartheta +1) (\cos (\theta )+2) \csc (\theta )}{8 g^3 (\cos (\theta )+1)}+\frac{\cot (\theta )}{g}\,.\nn
\end{align}
Lastly, the gluing condition is modified
\begin{equation}
\tilde{\mathcal{Q}}_i(u)=\mathcal{Q}_i(-u)\,,\qquad i=1,\dots,4\,,
\end{equation}
\beqa
    &&\mathcal{Q}_1=\mathbb{Q}_1 \mathbb{S}+\mathbb{Q}_2\,,\qquad \mathcal{Q}_2=\mathbb{Q}_1 \mathbb{S}\,e^{-2\pi i \vartheta}+\mathbb{Q}_2\,,
\\
\nn&&\mathcal{Q}_3=\mathbb{Q}_3 \mathbb{S}+\mathbb{Q}_4\,,\qquad\mathcal{Q}_4=\mathbb{Q}_3 \mathbb{S}\,e^{-2\pi i \vartheta}+\mathbb{Q}_4\,.
\eeqa
Here we state it without proof; the full derivation is presented in the next appendix. Imposing the gluing conditions on $\mathcal{Q}_1$ and $\mathcal{Q}_2$ alone is sufficient to close the system, much like in the case of a generic cusp angle.

\section{Far-regime QSC discussion}\label{app:BaxFactor}

This appendix is devoted to the technical details behind reducing the far-region limit of the Baxter equation \eqref{4Bax_far} to a second-order differential equation and then matching it with the near-regime. For convenience, we begin by repeating the full fourth-order equation here.
\begin{equation}\label{4Bax_farApp}
    Q(U) \left(\frac{1-4 \Gamma_\text{cusp} ^2-\text{N}_1}{U^2}+\frac{\text{N}_2}{U^4}+1\right)-\frac{2
   \text{N}_1 Q'(U)}{U^3}+\left(\frac{\text{N}_1}{U^2}-2\right) Q''(U)+Q^{(4)}(U)=0\,,
\end{equation}
with parameters $N_1,N_2$ determined by $\bP_a$ expansion coefficients $a_k,b_k$ and angle $\theta$ as follows 
\begin{flalign*}
    &N_1= -2 a_1^2 g^2 (\cos (\theta )+1)+4 a_2 g^2 (\cos (\theta )+1)-4 a_1 g \cot \left(\frac{\theta }{2}\right)+4 g^2 (\cos (\theta
   )+1)+\frac{3}{2}\,,&
\end{flalign*}

{\small \begin{flalign*}
    &N_2=a_3 \left(\frac{1}{4} g^3 \sin ^5(\theta ) \csc ^8\left(\frac{\theta }{2}\right)-\frac{1}{2} b_2 g^5 \sin ^5(\theta ) \csc
   ^6\left(\frac{\theta }{2}\right)\right)+&\end{flalign*}
   \begin{flalign*}&+a_1 \left(a_2 \left(\frac{1}{2} b_2 g^5 \sin ^5(\theta ) \csc ^6\left(\frac{\theta
   }{2}\right)-4 g^3 \cos (\theta ) \cot ^3\left(\frac{\theta }{2}\right)\right)-\frac{1}{2} a_2^2 g^5 \sin ^5(\theta ) \csc^6\left(\frac{\theta }{2}\right)\right.+&\end{flalign*}
   \begin{equation*}+\left.\frac{1}{2} a_4 g^5 \sin ^5(\theta ) \csc ^6\left(\frac{\theta }{2}\right)-a_3 g^4 \sin ^4(\theta )
   \csc ^6\left(\frac{\theta }{2}\right)-\frac{1}{2} b_4 g^5 \sin ^5(\theta ) \csc ^6\left(\frac{\theta }{2}\right)+4 g^3 \cos (\theta )
   \cot ^3\left(\frac{\theta }{2}\right)\right.-\end{equation*}
   \begin{equation*}\left.-9 g \cot \left(\frac{\theta }{2}\right)\right)+a_2 \left(\frac{1}{2} a_3 g^5 \sin ^5(\theta )
   \csc ^6\left(\frac{\theta }{2}\right)+\frac{1}{2} b_2 g^4 \sin ^4(\theta ) \csc ^6\left(\frac{\theta }{2}\right)+\frac{g^4 (-9 \cos
   (\theta )+\cos (3 \theta )-8)}{\cos (\theta )-1}\right.+\end{equation*}\begin{equation*}+\left.9 g^2 (\cos (\theta )+1)\right)-\frac{1}{2} a_5 g^5 \sin ^5(\theta ) \csc
   ^6\left(\frac{\theta }{2}\right)-8 a_2^2 g^4 \cos ^2\left(\frac{\theta }{2}\right) \cos (\theta ) \cot ^2\left(\frac{\theta
   }{2}\right)+\frac{1}{4} a_4 g^4 \sin ^6(\theta ) \csc ^8\left(\frac{\theta }{2}\right)+\end{equation*}\begin{equation*}+a_1^2 \left(\frac{1}{2} a_2 g^4 \sin ^4(\theta
   ) \csc ^6\left(\frac{\theta }{2}\right)-\frac{1}{2} g^4 \sin ^4(\theta ) \csc ^6\left(\frac{\theta }{2}\right)-\frac{9}{2} g^2 (\cos
   (\theta )+1)\right)+\frac{1}{2} b_5 g^5 \sin ^5(\theta ) \csc ^6\left(\frac{\theta }{2}\right)-\end{equation*}\begin{equation}-\frac{1}{2} b_4 g^4 \sin ^4(\theta )
   \csc ^6\left(\frac{\theta }{2}\right)+\frac{g^4 (-9 \cos (\theta )+\cos (3 \theta )-8)}{\cos (\theta )-1}+9 g^2 (\cos (\theta
   )+1)+\frac{45}{16}
\end{equation}}
We then recall that under the condition $\vartheta_1=\vartheta_2$, or alternatively $4N_2=N_1(N_1+6)$,
the fourth order equation factorises
\begin{equation}
    \mathcal{L}_2\mathcal{L}_1Q(U)=0\,,
\end{equation}
where $\mathcal{L}_1$ and $\mathcal{L}_2$ are two second-order differential operators 
\begin{equation}\label{bbax}
    \mathcal{L}_1F(U)=-\frac{F'(U)}{U}+F''(U)+F(U) \left(-\frac{2 \Gamma_\text{cusp} }{U}-1+\frac{\text{N}_1}{2 U^2}\right)\,,
\end{equation}
\begin{equation}
     \mathcal{L}_2F(U)=\frac{F'(U)}{U}+F''(U)+F(U) \left(\frac{2 \Gamma_\text{cusp}}{U}-1+\frac{ \left(\text{N}_1-2\right)}{2 U^2}\right)\,.
\end{equation}
Function $F(U)$ here is arbitrary. This decomposition is not unique: since the original equation (\ref{4Bax_farApp}) is symmetric under $\Gamma_\text{cusp}\to -\Gamma_\text{cusp}$, the same symmetry applies to the decomposition above as well. From now on, we therefore fix a definite sign of $\Gamma_\text{cusp}$ and keep in mind that it is sufficient to find only half of the solutions, namely those satisfying $\mathcal{L}_1 Q = 0$. Fortunately, they can be found analytically 
\begin{equation}\label{Uq1}
    Q_1(U)=B_1 \;\cdot\;\delta^{-\frac{1}{2}-\Gamma_\text{cusp}}2^{-\Gamma_\text{cusp} +\vartheta +\frac{1}{2}}\;\cdot\; e^{-U} U^{\vartheta +1} \times\mathbb{U}\left[-\Gamma_\text{cusp} +\vartheta +\frac{1}{2},2 \vartheta +1,2 U\right]
\end{equation}
\begin{equation*}
    Q_2(U)=B_2\;\cdot\;\delta^{-\frac{1}{2}-\Gamma_\text{cusp}}\dfrac{2^{-\Gamma_\text{cusp} +\vartheta +\frac{1}{2}} \pi\sec (\pi  (\Gamma_\text{cusp} +\vartheta ))}{\Gamma(-\Gamma_\text{cusp}-\vartheta+\frac{1}{2})}\;\cdot\; e^{-U}U^{\vartheta +1} \times\end{equation*}
    \begin{equation}\label{Uq2}\times\left(\frac{\, _1 F_1\left[\Gamma_\text{cusp} +\vartheta +\frac{1}{2};2 \vartheta +1;2 U\right]}{\Gamma \left(1+2 \vartheta\right)}-\frac{i e^{i \pi  (\Gamma_\text{cusp} +\vartheta )} \mathbb{U}\left[\Gamma_\text{cusp} +\vartheta +\frac{1}{2},2 \vartheta +1,2 U\right]}{ \Gamma \left(-\Gamma_\text{cusp} +\vartheta +\frac{1}{2}\right)}\right)\,.
\end{equation}
Here $_1F_1$ denotes the hypergeometric function, while $\mathbb{U}$ is Tricomi’s confluent hypergeometric function. The solutions are chosen to satisfy pure asymptotics consistent with \eqref{qasymp}. We also fix the normalisation coefficients $B_i$ here to be the same as in the general cusp QSC case \eqref{Qbs}   
\begin{equation}
    B_1=B_2=\frac{i (\cos (\theta )+1)}{ \delta  \sqrt{2\Gamma_\text{cusp} }}+O(\delta)\,.
\end{equation}
Obtaining the functions $Q_3$ and $Q_4$ amounts to replacing $\Gamma_\text{cusp} \to -\Gamma_\text{cusp}$ and making sure the constant prefactors are consistent with \eqref{Qbs}   
\begin{equation}\label{Uq34}
    Q_3(U)=i Q_1(U)\Bigl|_{\Gamma_\text{cusp}\to-\Gamma_\text{cusp}}\,,\qquad Q_4(U)=-i Q_2(U)\Bigl|_{\Gamma_\text{cusp}\to-\Gamma_\text{cusp}}\,.
\end{equation}
These four solutions can be expanded at $U\to0$, in which case expansion terms will group into the large $u$ series decomposition of $\mathbb{Q}_i$. To demonstrate, we list below such an expansion for $Q_1$
\begin{align}\label{Q1farexp}
Q_1
&\approx\frac{i 2^{\vartheta -\Gamma_{\text{cusp}} } (\cos (\theta )+1) \Gamma (-2 \vartheta ) \delta ^{-\Gamma_{\text{cusp}} +\vartheta -\frac{1}{2}}}{\sqrt{\Gamma_{\text{cusp}} } \Gamma \left(-\Gamma_{\text{cusp}} -\vartheta +\frac{1}{2}\right)}\times u^{1+\vartheta} \notag\\
&\quad+\frac{i 2^{-\Gamma_{\text{cusp}} -\vartheta } (\cos (\theta )+1) \Gamma (2 \vartheta ) \delta ^{-\Gamma_{\text{cusp}} -\vartheta -\frac{1}{2}}}{\sqrt{\Gamma_{\text{cusp}} } \Gamma \left(-\Gamma_{\text{cusp}} +\vartheta +\frac{1}{2}\right)}\times u^{1-\vartheta}\\
&\quad+\frac{\sqrt{\Gamma_{\text{cusp}} } e^{i \theta } 2^{\vartheta -\Gamma_{\text{cusp}} } \Gamma (1-2 \vartheta ) \delta ^{-\Gamma_{\text{cusp}} +\vartheta +\frac{1}{2}}}{\left(1+e^{i \theta }\right)^2 \Gamma \left(-\Gamma_{\text{cusp}} -\vartheta +\frac{1}{2}\right)}\times \left(\frac{8 i \cos ^4\left(\frac{\theta }{2}\right)}{\vartheta  (1+2 \vartheta)} \;u^{2+\vartheta}\right) \notag\\
&\quad-\frac{\sqrt{\Gamma_{\text{cusp}} } e^{i \theta } 2^{-\Gamma_{\text{cusp}} -\vartheta } \Gamma (2 \vartheta +1) \delta ^{-\Gamma_{\text{cusp}} -\vartheta +\frac{1}{2}}}{\left(1+e^{i \theta }\right)^2 \Gamma \left(-\Gamma_{\text{cusp}} +\vartheta +\frac{1}{2}\right)}\times \left( \frac{8 i \cos ^4\left(\frac{\theta }{2}\right)}{\vartheta  (1-2 \vartheta)}\;u^{2-\vartheta}\right)\,.\nn
\end{align}
Note that the powers of $u$ appearing on the right-hand side above are precisely those of the $\mathbb{Q}_i$ asymptotics (see their reminder in Eq.~\eqref{Qsmall2}). We then extract the relation between $Q_1$ and $\mathbb{Q}_i$ by replacing the asymptotics of $\mathbb{Q}_i$ in the right-hand side of \eqref{Q1farexp} with the full functions $\mathbb{Q}_i$. Repeating this procedure for each component $Q_i$, we obtain the transformation matrix $M$ relating the far and near regions.
\begin{equation}\label{farnearapp}
Q_i(u\delta)=M_{i}^{~j}\mathbb{Q}_j(u)\,,\qquad 1\ll|u|\ll\frac1\delta\,,
\end{equation}
where matrix $M$ is equal to\newline
\begin{equation}
    M=\begin{pmatrix}
        m_{11}&m_{12}\\m_{21}&m_{22}
    \end{pmatrix}\,,
\end{equation}
{\small \begin{align}
    m_{11}&=\left(
\begin{array}{cc}
 \frac{i 2^{\vartheta -\Gamma_{\text{cusp}} } (\cos (\theta )+1) \Gamma (-2 \vartheta ) \delta ^{-\Gamma_{\text{cusp}} +\vartheta -\frac{1}{2}}}{\sqrt{\Gamma_{\text{cusp}} } \Gamma \left(-\Gamma_{\text{cusp}} -\vartheta +\frac{1}{2}\right)} & \frac{i 2^{-\Gamma_{\text{cusp}} -\vartheta } (\cos (\theta )+1) \Gamma (2 \vartheta ) \delta ^{-\Gamma_{\text{cusp}} -\vartheta -\frac{1}{2}}}{\sqrt{\Gamma_{\text{cusp}} } \Gamma \left(-\Gamma_{\text{cusp}} +\vartheta +\frac{1}{2}\right)} \\
 -\frac{i \pi  2^{-\Gamma_{\text{cusp}} +\vartheta +1} e^{i \pi  (\Gamma_{\text{cusp}} +\vartheta )} (\cos (\theta )+1) \delta ^{-\Gamma_{\text{cusp}} +\vartheta -\frac{1}{2}}}{\sqrt{\Gamma_{\text{cusp}} } \left(-1+e^{4 i \pi  \vartheta }\right) \Gamma (2 \vartheta +1) \Gamma \left(-\Gamma_{\text{cusp}} -\vartheta +\frac{1}{2}\right)} & \frac{\left(1+e^{i \theta }\right)^2 2^{-\Gamma_{\text{cusp}} -\vartheta -1} \Gamma (2 \vartheta ) \delta ^{-\Gamma_{\text{cusp}} -\vartheta
   -\frac{1}{2}} e^{i (\pi  (\Gamma_{\text{cusp}} +\vartheta )-\theta )}}{\sqrt{\Gamma_{\text{cusp}} } \Gamma \left(-\Gamma_{\text{cusp}} +\vartheta +\frac{1}{2}\right)} \\
\end{array}
\right)\notag\\
m_{12}&=\left(
\begin{array}{cc}
 \frac{\sqrt{\Gamma_{\text{cusp}} } e^{i \theta } 2^{\vartheta -\Gamma_{\text{cusp}} } \Gamma (1-2 \vartheta ) \delta ^{-\Gamma_{\text{cusp}} +\vartheta +\frac{1}{2}}}{\left(1+e^{i \theta }\right)^2 \Gamma \left(-\Gamma_{\text{cusp}} -\vartheta +\frac{1}{2}\right)} & -\frac{\sqrt{\Gamma_{\text{cusp}} } e^{i \theta } 2^{-\Gamma_{\text{cusp}} -\vartheta } \Gamma (2 \vartheta +1) \delta ^{-\Gamma_{\text{cusp}} -\vartheta +\frac{1}{2}}}{\left(1+e^{i \theta }\right)^2 \Gamma
   \left(-\Gamma_{\text{cusp}} +\vartheta +\frac{1}{2}\right)} \\
 -\frac{\pi  \sqrt{\Gamma_{\text{cusp}} } 2^{-\Gamma_{\text{cusp}} +\vartheta +1} \delta ^{-\Gamma_{\text{cusp}} +\vartheta +\frac{1}{2}} e^{i (\pi  (\Gamma_{\text{cusp}} +\vartheta )+\theta )}}{\left(1+e^{i \theta }\right)^2 \left(-1+e^{4 i \pi  \vartheta }\right) \Gamma (2 \vartheta ) \Gamma \left(-\Gamma_{\text{cusp}} -\vartheta +\frac{1}{2}\right)} & -\frac{i \sqrt{\Gamma_{\text{cusp}} } 2^{-\Gamma_{\text{cusp}} -\vartheta } \Gamma (2 \vartheta +1) \delta ^{-\Gamma_{\text{cusp}}
   -\vartheta +\frac{1}{2}} e^{i (\pi  (\Gamma_{\text{cusp}} +\vartheta )+\theta )}}{\left(1+e^{i \theta }\right)^2 \Gamma \left(-\Gamma_{\text{cusp}} +\vartheta +\frac{1}{2}\right)} \\
\end{array}
\right)\notag\end{align}}{\small \begin{align}
m_{21}&=\left(
\begin{array}{cc}
 -\frac{2^{\Gamma_{\text{cusp}} +\vartheta } (\cos (\theta )+1) \Gamma (-2 \vartheta ) \delta ^{\Gamma_{\text{cusp}} +\vartheta -\frac{1}{2}}}{\sqrt{\Gamma_{\text{cusp}} } \Gamma \left(\Gamma_{\text{cusp}} -\vartheta +\frac{1}{2}\right)} & -\frac{2^{\Gamma_{\text{cusp}} -\vartheta } (\cos (\theta )+1) \Gamma (2 \vartheta ) \delta ^{\Gamma_{\text{cusp}} -\vartheta -\frac{1}{2}}}{\sqrt{\Gamma_{\text{cusp}} } \Gamma \left(\Gamma_{\text{cusp}} +\vartheta +\frac{1}{2}\right)}\,, \\
 -\frac{\pi  2^{\Gamma_{\text{cusp}} +\vartheta +1} e^{-i \pi  (\Gamma_{\text{cusp}} -\vartheta )} (\cos (\theta )+1) \delta ^{\Gamma_{\text{cusp}} +\vartheta -\frac{1}{2}}}{\sqrt{\Gamma_{\text{cusp}} } \left(-1+e^{4 i \pi  \vartheta }\right) \Gamma (2 \vartheta +1) \Gamma \left(\Gamma_{\text{cusp}} -\vartheta +\frac{1}{2}\right)} & -\frac{i \left(1+e^{i \theta }\right)^2 2^{\Gamma_{\text{cusp}} -\vartheta -1} \Gamma (2 \vartheta ) \delta ^{\Gamma_{\text{cusp}} -\vartheta
   -\frac{1}{2}} e^{-i (\pi  (\Gamma_{\text{cusp}} -\vartheta )+\theta )}}{\sqrt{\Gamma_{\text{cusp}} } \Gamma \left(\Gamma_{\text{cusp}} +\vartheta +\frac{1}{2}\right)} \notag\\
\end{array}
\right)\\
m_{22}&=\left(
\begin{array}{cc}
 -\frac{i \sqrt{\Gamma_{\text{cusp}} } e^{i \theta } 2^{\Gamma_{\text{cusp}} +\vartheta } \Gamma (1-2 \vartheta ) \delta ^{\Gamma_{\text{cusp}} +\vartheta +\frac{1}{2}}}{\left(1+e^{i \theta }\right)^2 \Gamma \left(\Gamma_{\text{cusp}} -\vartheta +\frac{1}{2}\right)} & \frac{i \sqrt{\Gamma_{\text{cusp}} } e^{i \theta } 2^{\Gamma_{\text{cusp}} -\vartheta } \Gamma (2 \vartheta +1) \delta ^{\Gamma_{\text{cusp}} -\vartheta +\frac{1}{2}}}{\left(1+e^{i \theta }\right)^2 \Gamma
   \left(\Gamma_{\text{cusp}} +\vartheta +\frac{1}{2}\right)} \\
 -\frac{i \pi  \sqrt{\Gamma_{\text{cusp}} } 2^{\Gamma_{\text{cusp}} +\vartheta +1} \delta ^{\Gamma_{\text{cusp}} +\vartheta +\frac{1}{2}} e^{i (\pi  (\vartheta -\Gamma_{\text{cusp}} )+\theta )}}{\left(1+e^{i \theta }\right)^2 \left(-1+e^{4 i \pi  \vartheta }\right) \Gamma (2 \vartheta ) \Gamma \left(\Gamma_{\text{cusp}} -\vartheta +\frac{1}{2}\right)} & \frac{\sqrt{\Gamma_{\text{cusp}} } 2^{\Gamma_{\text{cusp}} -\vartheta } \Gamma (2 \vartheta +1) \delta ^{\Gamma_{\text{cusp}} -\vartheta
   +\frac{1}{2}} e^{i (\pi  (\vartheta -\Gamma_{\text{cusp}} )+\theta )}}{\left(1+e^{i \theta }\right)^2 \Gamma \left(\Gamma_{\text{cusp}} +\vartheta +\frac{1}{2}\right)} \notag\\
\end{array}
\right)
\end{align}}
The explicit form of $M$ depends on the choice of normalisation for both $Q_i$ and $\mathbb{Q}_i$. We emphasise that in our conventions the large-$u$ behaviour of $\mathbb{Q}_i$ is fixed to be
\begin{equation}\label{Qsmall2}
    \mathbb{Q}_i\sim\{\mathbb{B}_1u^{1+\vartheta},\mathbb{B}_2u^{1-\vartheta},\mathbb{B}_3u^{2+\vartheta},\mathbb{B}_4u^{2-\vartheta}\}\,,
\end{equation}
\begin{equation}\label{BBBs}
    \mathbb{B}_i=\left\{1,1,\frac{8 i \cos ^4\left(\frac{\theta }{2}\right)}{\vartheta  (2 \vartheta +1)},\frac{8 i \cos ^4\left(\frac{\theta }{2}\right)}{(1-2 \vartheta ) \vartheta }\right\}\,.
\end{equation}
Note that this particular choice of $\mathbb{B}_i$ is consistent with \eqref{AABBnear}. At first sight, the relation between $Q_i$ and $\mathbb{Q}_i$ in \eqref{farnearapp} appears rather complicated. However, it simplifies considerably when rewritten in the following form
\begin{equation}\label{farnearappsimple}
    Q_i=M_{i,2}\left(\sum_{j=1}^4\dfrac{M_{i,j}}{M_{i,2}}\;\mathbb{Q}_j\right)\,.
\end{equation}
The vast majority of coefficients in the big brackets above turn out to be related to the function $\mathbb{S}(g)$ appearing in the FME equation, recall \eqref{SandA}. Much like in the ladder model, the FME equation defines $\delta\to0$ scaling of $\Gamma_\text{cusp}$ via \eqref{tower}, letting us expand \eqref{farnearappsimple} near small $\delta$ values:
\begin{align}\label{QfarQnear}
    Q_1=&\mathcal{A}_n\left( \mathbb{Q}_1 \;\mathbb{S}+\mathbb{Q}_2+\delta\;\mathcal{D}_n\left(\mathbb{Q}_3 \;\mathbb{S}+\mathbb{Q}_4\right)\right)\,,\\
    Q_2=&(-1)^n\mathcal{A}_n\left( \mathbb{Q}_1 \;\mathbb{S} e^{-2\pi i\vartheta}+\mathbb{Q}_2-\delta\;\mathcal{D}_n\left(\mathbb{Q}_3\;\mathbb{S} e^{-2\pi i\vartheta}+\mathbb{Q}_4\right)\right)\,,\nn\\
Q_3=&\mathcal{B}_n\left(\mathbb{Q}_2-\delta\;\mathcal{D}_n\mathbb{Q}_4\right)\,,\nn\\
    Q_4=&(-1)^n\mathcal{B}_n\left(\mathbb{Q}_2+\delta\;\mathcal{D}_n\mathbb{Q}_4\right)\,.\nn
\end{align}
 Coefficients $\mathcal{A}_n,\mathcal{B}_n$ and $\mathcal{D}_n$ are defined as follows
\begin{equation}
    \mathcal{A}_n=\frac{i 2^{-n} (\cos (\theta )+1) \delta ^{-n-1} \Gamma (-2 \vartheta )}{S \sqrt{2 n+2 \vartheta +1} \Gamma (-n-2 \vartheta )}\,,\qquad \mathcal{B}_n=-\frac{2^{n+1} (\cos (\theta )+1) \delta ^n \Gamma (2 \vartheta )}{\sqrt{2 n+2 \vartheta +1} \Gamma (n+2 \vartheta +1)}\,,
\end{equation}
\begin{equation}
    \mathcal{D}_n=\frac{1}{8} i  \vartheta  \sec ^4\left(\frac{\theta }{2}\right) (2 n+2 \vartheta +1)\,.
\end{equation}
Finally, to derive the gluing conditions for the functions $\mathbb{Q}_i$, we substitute the relations \eqref{QfarQnear} into \eqref{gluefar1}-\eqref{gluefar4}, using Table~\ref{tab:Qs} to relate $\mathbf{Q}_i$ to $Q_i$. Note that imposing the gluing conditions on $Q_1$ and $Q_2$ alone is sufficient to constrain all four $\mathbb{Q}_i$ components, since these two functions already contain contributions from the entire set of $\mathbb{Q}_i$. In this way, we recover \eqref{THEGLUE}, namely
\begin{equation}\label{QQtapp}
    \mathcal{Q}^{\gamma}_i(u)=\mathcal{Q}_i(-u)\,,\qquad\text{where}\,
\end{equation}
\begin{align}
    \mathcal{Q}_1=&\mathbb{Q}_1 \mathbb{S}+\mathbb{Q}_2\,,\qquad\mathcal{Q}_2=\mathbb{Q}_1 \mathbb{S}e^{-2\pi i \vartheta}+\mathbb{Q}_2\,,\\
    \mathcal{Q}_3=&\mathbb{Q}_3 \mathbb{S}+\mathbb{Q}_4\,,\qquad\mathcal{Q}_4=\mathbb{Q}_3 \mathbb{S}e^{-2\pi i \vartheta}+\mathbb{Q}_4\,.\nn
\end{align}
The same four equations \eqref{QQtapp} written in terms of $\mathbb{Q}_i$ only are
\begin{align}\label{gluenear1}
   \tilde{\mathbb{Q}}_1(u)&=i\mathbb{Q}_1(-u) \csc (\pi  \vartheta ) \sinh (\pi  (2 u-i \vartheta ))+\frac{i\mathbb{Q}_2(-u) \sinh (2 \pi  u) (\cot (\pi  \vartheta )+i)}{\mathbb{S}}\,,\\\label{gluenear2}\tilde{\mathbb{Q}}_2(u)&=\mathbb{S}\;\mathbb{Q}_1(-u) \sinh (2 \pi  u) (-1-i \cot (\pi  \vartheta ))-i\mathbb{Q}_2(-u) \csc (\pi  \vartheta ) \sinh (\pi  (2 u+i \vartheta ))\,,\\\tilde{\mathbb{Q}}_3(u)&=i\mathbb{Q}_3(-u) \csc (\pi  \vartheta ) \sinh (\pi  (2 u-i
   \vartheta ))+\frac{i\mathbb{Q}_4(-u) \sinh (2 \pi  u) (\cot (\pi  \vartheta )+i)}{\mathbb{S}}\,,\\\tilde{\mathbb{Q}}_4(u)&=\mathbb{S}\mathbb{Q}_3(-u) \sinh (2 \pi  u) (-1-i \cot (\pi  \vartheta ))-i\mathbb{Q}_4(-u) \csc (\pi  \vartheta ) \sinh (\pi  (2 u+i \vartheta ))\,.
\end{align}
Imposing the gluing conditions on $Q_3$ and $Q_4$ additionally fixes the previously undetermined coefficients $\mathbf{a}_1$ and $\mathbf{a}_2$ appearing in \eqref{gluefar3} and \eqref{gluefar4}.
\begin{equation}\label{bolda1a2}
    \mathbf{a}_1=\mathbf{a}_2=-\frac{i (-1)^{-n} 4^{n+1} \mathbb{S} e^{2 i \pi  \vartheta } \delta ^{2 n+1} \Gamma (2 \vartheta +1) \Gamma (-n-2 \vartheta )}{\left(-1+e^{2 i \pi  \vartheta }\right) \Gamma (1-2 \vartheta ) \Gamma (n+2 \vartheta +1)}\,.
\end{equation}
Lastly, we would like to emphasise that only half of the gluing conditions on $\mathbb{Q}_i$ need to be imposed in order to solve the system. We had already exploited an analogous property for the functions $\mathbf{Q}_i$, but for $\mathbb{Q}_i$ this appears to be a new observation. We find this to hold both numerically and within the weak-coupling perturbative analysis.

\section{Weak coupling details}\label{app: QSCWeak}
Here we present a step-by-step procedure for solving the fusion-limit QSC perturbatively at weak coupling. Constructing the weak-coupling solution analytically is a nontrivial task, as it requires prior knowledge of how all unknown parameters — namely $a_k$, $b_k$, $\mathcal{S}$, and $\vartheta$ — scale with $g$ in the $g\to0$ limit. This information has to be supplied as an initial input. The different scaling behaviour of these coefficients is, in fact, the key distinction between the even and odd states. 

With a working numerical algorithm, however, this difficulty can be bypassed by fitting all unknown parameters with a power series in $g$. In this way, one can determine both the scaling behaviour of the unknown coefficients and infer the leading terms in their weak-coupling expansions.

\paragraph{Initial data} We expand the Baxter equation \eqref{baxter} in the weak-coupling limit $g\to0$. Its solutions are expected to decompose into a sum over powers of the spectral parameter $u$ and possibly $\eta$-functions defined in \eqref{etas} -- this basis of functions we will refer to as ``untwisted'' in this appendix. At leading order, the solutions can be extracted straightforwardly by making a sufficiently general ansatz. For the odd state, the corresponding set of functions $\mathbbm{q}_i=\mathbb{Q}_i/\sqrt{u}$ is found to be
\begin{align}\label{QLeadodd}
    \bq_1(u)=&u\,,\qquad\bq_2(u)=u^2\,,\nn\\
    \bq_3(u)=&\eta _{2
} u^2-\cos ^2\left(\frac{\theta }{2}\right) \eta _{
1} u+\frac{1}{4} (\cos (\theta )-1)-\frac{i}{4 u}\,,\\
\bq_4(u)=&\frac{1}{2} i \sin ^4\left(\frac{\theta }{2}\right)+\frac{\cos (\theta )-1}{4 u}-u \cos ^2\left(\frac{\theta }{2}\right) \eta _{
2}+u^2 \eta _{
3}
-\frac{i}{4 u^2}\,.\nn
\end{align}
Analogously, Q-functions for the even state are 
\begin{align}\label{QLeadeven}
    \bq_2(u)=&1\,,\qquad\bq_3(u)=u^2\,,\\
    \bq_1(u)=&u+\left(u \left(2 i (\cos (\theta )+1) \eta _{
1}-i \pi  (\cos (\theta )+1)\right)-i (\cos (\theta )+1)\right) g^2\,,\nn\\
    \bq_4(u)=&\frac{i u (\cos (\theta )+1)}{g^2}-4 \cos ^4\left(\frac{\theta }{2}\right)-\frac{2 i (\cos (\theta )+1)}{u}\nn\\
    &+u \left(8 \cos ^4\left(\frac{\theta }{2}\right) \eta _{
1}-\frac{2}{3} \cos ^2\left(\frac{\theta }{2}\right) (3 (-2 i+\pi ) \cos (\theta )+(3+4 i \pi ) \pi -6 i)\right)\,.\nn
\end{align}
For $\bq_1$ and $\bq_4$, the next order in the coupling has been included, since otherwise they coincide up to an overall constant factor. It is important to note that solutions of the Baxter equation are not uniquely defined. The specific sets of Q-functions \eqref{QLeadodd} and \eqref{QLeadeven} are picked to have pure asymptotics.

We additionally reconstruct the functions $Q_{a|i}$ from the QQ-relations \eqref{fm}. We will need them to build higher orders in the weak-coupling expansion. For convenience, we repeat \eqref{fm} here.
\begin{equation}\label{rmRep}
    Q_{a|i}^+-Q_{a|i}^-=-\mathbf{P}_a\mathbf{P}^bQ_{b|i}^{+}\,,\qquad e^{\pm \pi u}\mathbb{Q}_{i}=-\mathbf{P}^{a}Q_{a|i}^{\pm}\,,
\end{equation}
Fortunately, we find that all of the $Q_{a|i}$ components decompose over the untwisted basis. This is, in fact, expected: consider the rightmost equation in \eqref{rmRep}. The functions $\bP_a$ appearing there are given by expansions in inverse powers of $u$, while the functions $\mathbb{Q}_i$ involve only regular $\eta$-functions. Hence, the functions $Q_{a|i}$ cannot introduce any nontrivial twisted structures into the equation.

\paragraph{Iterations} 
We now describe how to proceed to higher orders. Following \cite{Gromov:2015vua}, let $Q_{a|i}^{(0)}$ denote the leading-order term in the small-$g$ expansion of $Q_{a|i}$. We then expand $\bP_a$ to the next order in $g$ and define
\begin{equation}
    dS_{a|i}=Q_{a|i}^{(0)+}-Q_{a|i}^{(0)-}+\bP_a\bP^bQ_{b|i}^{(0)+}\,.
\end{equation}
The quantity $dS_{a|i}$ measures the mismatch in the QQ-relation \eqref{rmRep} induced by the higher-order correction in $g$. The next approximation to $Q_{a|i}$ is then sought in the form
\begin{equation}
    Q_{a|i}=Q_{a|i}^{(0)}+b_i^{j+}Q_{a|j}^{(0)}\,.
\end{equation}
Substituting this ansatz back into \eqref{rmRep} determines the correction $b_i^{\;k}$ through the first-order finite-difference equation
\begin{equation}\label{weakitereq}
b_i^{k\;[+2]}-b_i^{k}=-dS_{a|i}Q^{(0)a|k+}\,.
\end{equation}

The right-hand side of this equation has a special structure at large $u$: its components either scale as a power of $u$ or contain factors of the form $e^{\pm2\pi u}$. For components with purely power-like behaviour, it is well known that the solutions to \eqref{weakitereq} belong to the untwisted basis. The exponential factors do not alter this conclusion, since they are invariant under the finite-difference shift, $e^{\pm2\pi (u+i)}=e^{\pm2\pi u}$, and therefore factor out of the equation without modifying its structure. As a result, the basis of polynomials and $\eta$-functions \eqref{etas} remains sufficient for solving the system. The resulting $Q_{a|i}$ can then be used as the new starting approximation, and the procedure repeated to reach any desired order in $g$. Finally, the $\mathbb{Q}_i$ is reconstructed using the rightmost equation in \eqref{rmRep}.

\paragraph{Purification} Since $Q_{a|i}$ is obtained from a finite-difference equation \eqref{rmRep}, it is fixed only up to a linear transformation $C^{~j}_{i}Q_{a|j}$. The same ambiguity is then inherited by $\mathbb{Q}_i$. As a result, each new order in $Q_{a|i}$ typically spoils the pure-asymptotics condition for the $\mathbb{Q}_i$ functions. This can be corrected by solving the QQ-system independently in the large-$u$ limit to the first few orders, where the pure asymptotics are much easier to maintain. The weak-coupling solution can then be adjusted to match the large-$u$ behaviour, as illustrated in the diagram below.

\begin{center}
\begin{tikzpicture}[
    >=stealth,
    every node/.style={font=\small},
    lab/.style={font=\scriptsize}
]

\matrix (m) [matrix of nodes,
             row sep=1.8cm,
             column sep=3.2cm,
             nodes={align=center}] {

QQ-system & $\mathbb{Q}_i$ \\
$\mathbb{Q}^{\text{pure}}_i$ & $\mathbb{Q}^{\text{pure}}_i = C_i^{~j}\mathbb{Q}_j$ \\
};

\draw[->] (m-1-1) -- node[lab, above]{Solve at $g\to0$} (m-1-2);
\draw[->] (m-1-1) -- node[lab, left]{Solve at $u\to\infty$} (m-2-1);
\draw[->] (m-1-2) -- node[lab, right]{Expand $u\to\infty$} (m-2-2);
\draw[->] (m-2-1) -- node[lab, below]{Expand $g\to0$} (m-2-2);

\end{tikzpicture}
\end{center}
Matrix $C_i^{~j}$ here is a function of coupling $g$ only. It adjusts the weak coupling expansion of $\mathbb{Q}_i$ to fit pure asymptotics order by order in $g$.

\paragraph{Gluing} Finally, the system of equations is closed by imposing the gluing conditions. Knowing $\mathbbm{q}_i=\mathbb{Q}_i/\sqrt{u}$, we first determine their analytic continuation using \eqref{THEGLUE}. Then, since in the vicinity of $u\to0$ the functions $\mathbbm{q}_i$ possess only a single square-root branch cut along $(-2g,2g)$, certain combinations of them are expected to be regular. In particular, the following combinations must be free of singularities:
\begin{equation} 
    \mathbbm{q}_i+\mathbbm{q}^\gamma_i=\text{regular}\,,\qquad\dfrac{\mathbbm{q}_i-\mathbbm{q}^\gamma_i}{\sqrt{u^2-4g^2}}=\text{regular}\,.
\end{equation}
We expand the left-hand side of these relations in the $g\to0$ limit and set the coefficients in front of any negative powers of $u$ to zero. This provides sufficient constraints to solve the system. In particular, imposing the gluing conditions only for $\mathbb{Q}_1$ and $\mathbb{Q}_2$ is already sufficient, although using all four gluing conditions fixes the unknown parameters faster. 

\section{Higher order numerics and weak coupling expansion.}\label{app: weak}
In this appendix, we collect our numerical and perturbative results with the highest precision we were able to achieve. We begin by listing the critical coupling for the various values of angle $\theta$, see Table \ref{tab:gcritFull}.

\begin{table}[H]
    \centering
    \small
    \setlength{\tabcolsep}{9pt}
    \renewcommand{\arraystretch}{1.15}

    \textbf{Critical coupling
    $\boldsymbol{g_{\mathrm{crit}}}$ at
    $\boldsymbol{\vartheta=0}$}

    \vspace{0.4em}

    \begin{tabular}{lcc}
        \toprule
        & \multicolumn{2}{c}{State} \\
        \cmidrule(lr){2-3}
        \textbf{$~\theta$}
        & Even
        & Odd \\
        \midrule

        $0$
        & 0.207242228
        & -- \\

        $\frac{\pi}{10}$
        & 0.2091799545014922491
        & 0.2828533059200995339107482750 \\

        $\frac{3\pi}{10}$
        & 0.22573598514908932
        & 0.318667583776879 \\

        $\frac{5\pi}{10}$
        & 0.2657882332063967
        & 0.417350557962263 \\

        $\frac{7\pi}{10}$
        & 0.354421337459177
        & 0.69863878068128556 \\

        $\frac{9\pi}{10}$
        & 0.650836216716
        & -- \\
        \bottomrule
    \end{tabular}

        \caption{
        Critical coupling value $g_{\mathrm{crit}}$.
        The columns correspond to the even and odd states, while the rows
        correspond to different values of the angle $\theta$.
        Dashes indicate cases for which no critical value was obtained.
    }
    \label{tab:gcritFull}

\end{table}

Find below the highest order we managed to get for the weak coupling expansion of $\vartheta$ and $\mathbb{S}$. Denoting $\cos(\theta)\equiv2\mathcal{C}$ we find the following\\
1. Even state:\\
\begin{figure}[H]
    \centering
    \includegraphics[width=0.72\linewidth]{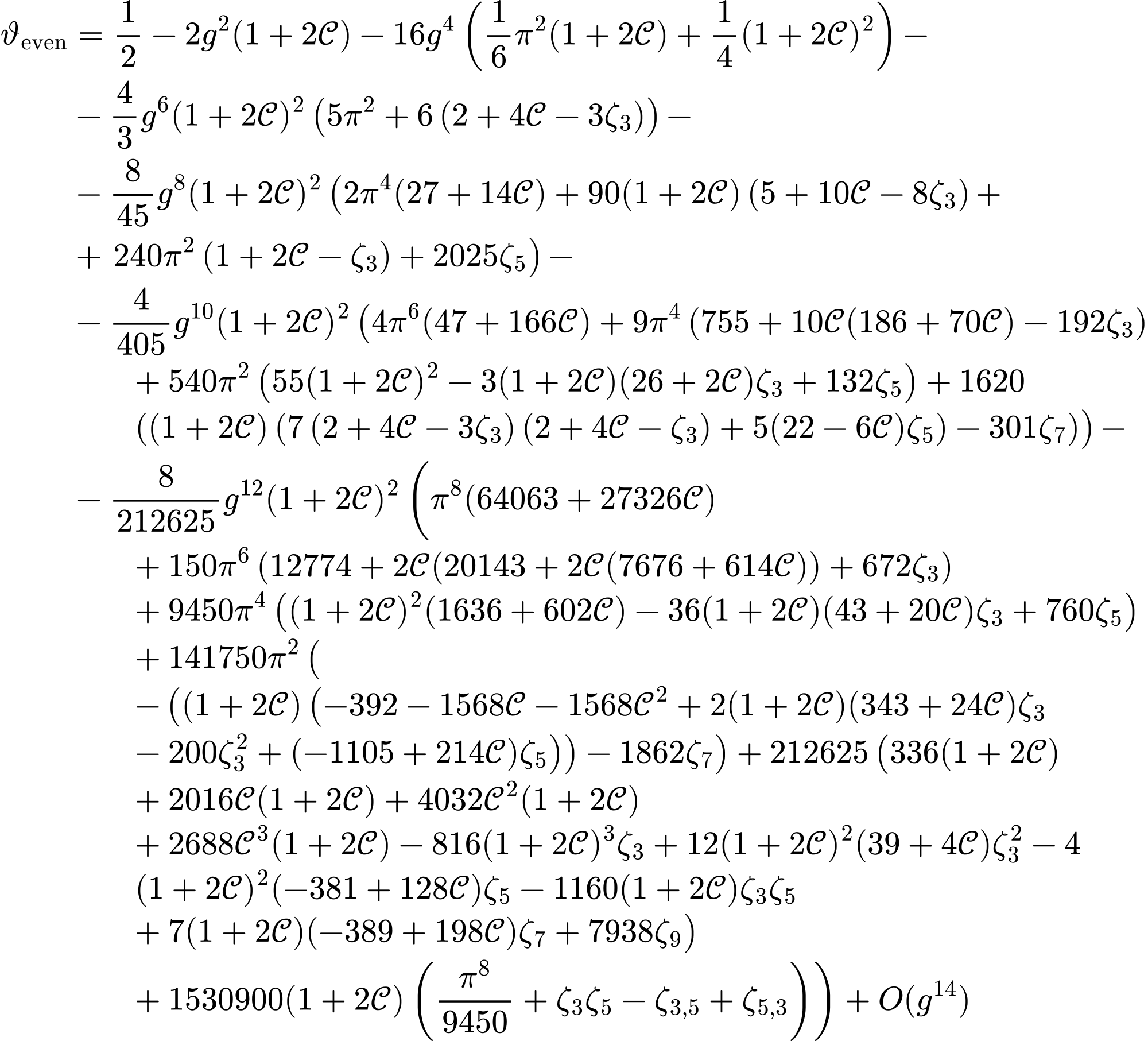}
\end{figure}
\begin{figure}[H]
    \centering
    \includegraphics[width=0.9\linewidth]{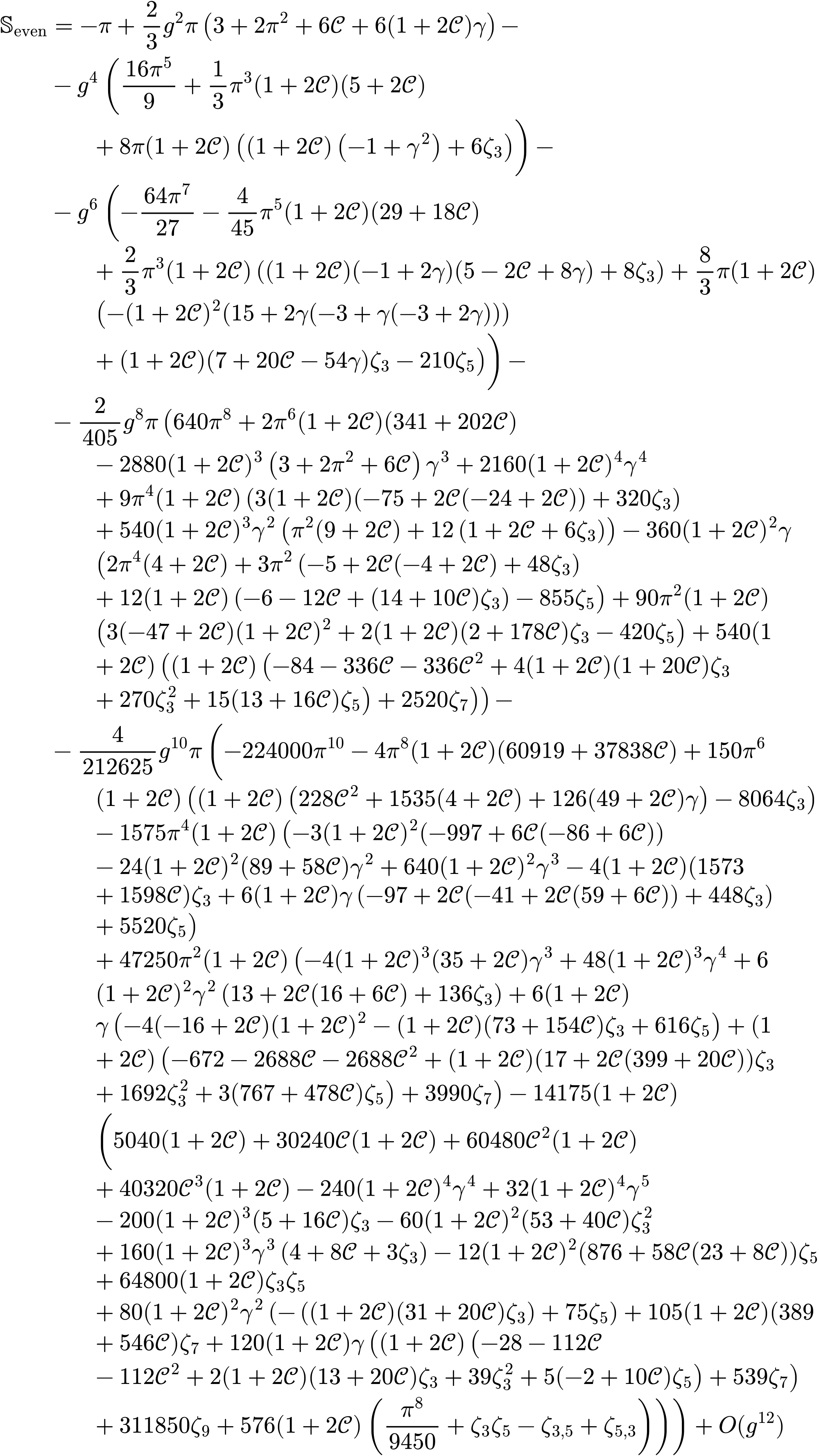}
\end{figure}
\newpage
2. Odd state:\\
\begin{figure}[H]
    \centering
    \includegraphics[width=0.9\linewidth]{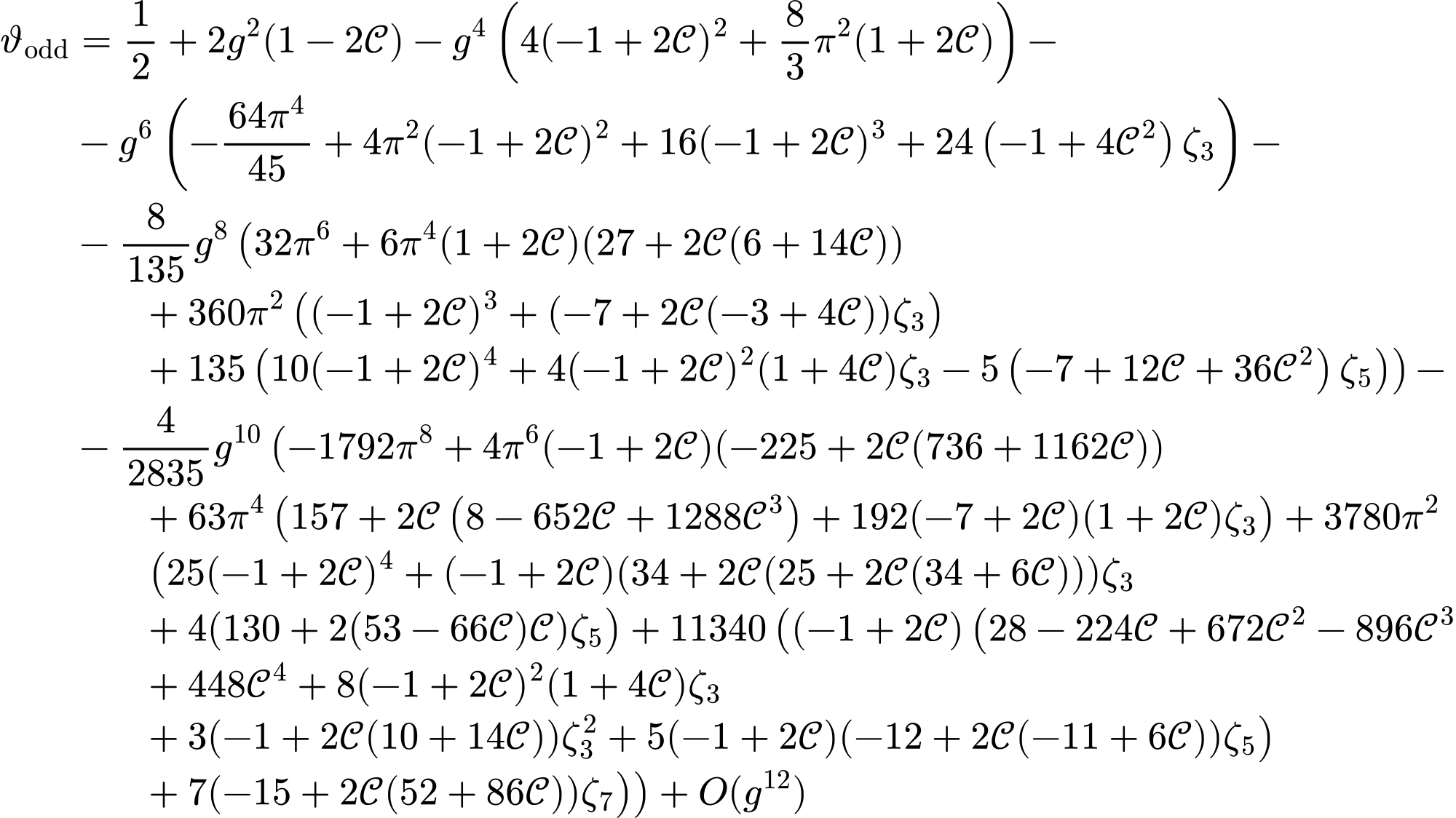}
\end{figure}
\begin{figure}[H]
    \centering
    \includegraphics[width=0.9\linewidth]{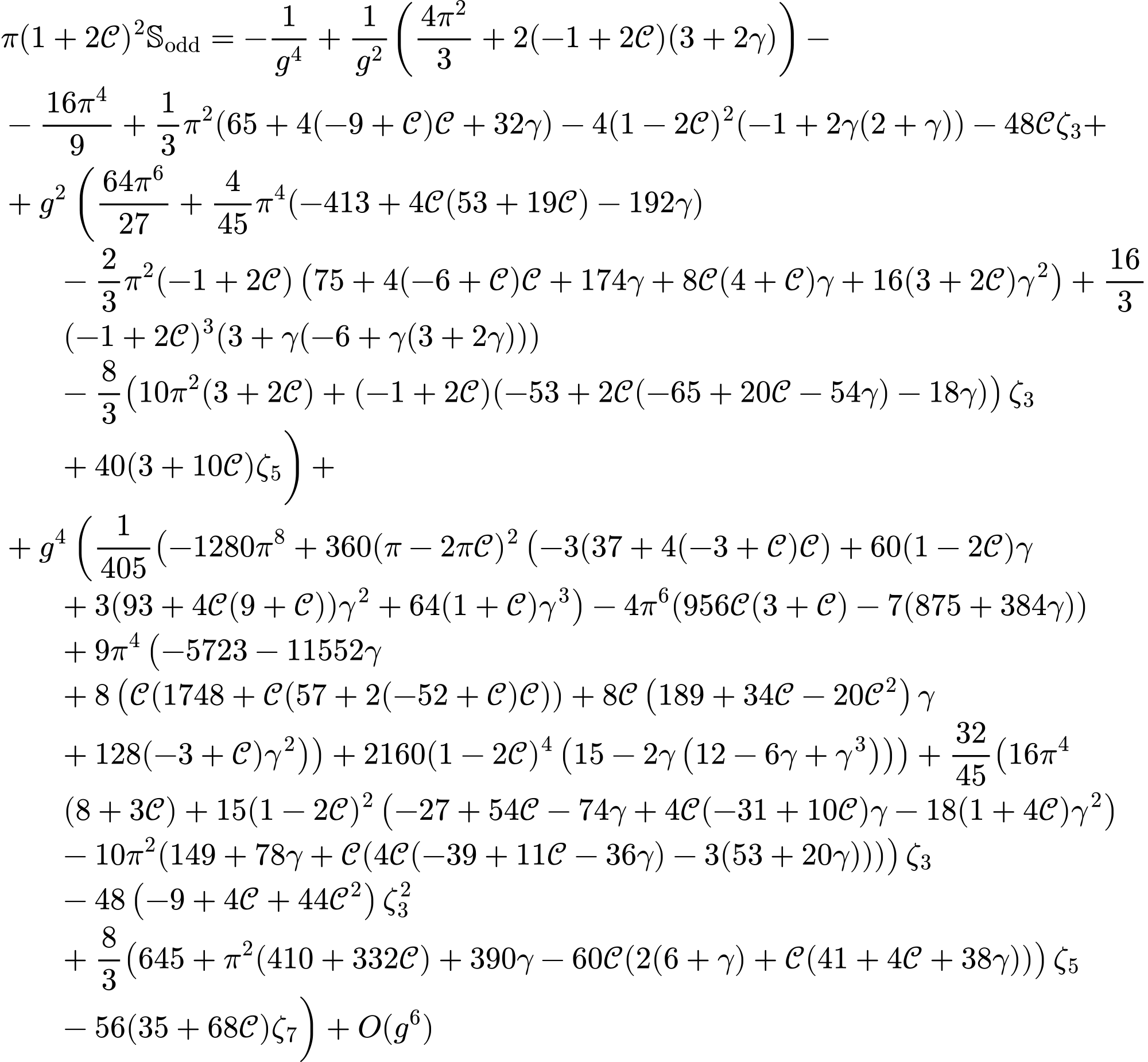}
\end{figure}

 We have verified the above series expansions against numerical data. Running the numerical solver down to values as small as $g\sim0.004$, we fitted $\vartheta(g)$ and $\mathbb{S}(g)$ for the case $\theta=7\pi/10$ and found that the results listed above agree with the numerical predictions to the following accuracy:\newline
1. Even state:
\begin{align}
    \vartheta_\text{even}^{\text{(Fit)}}-\vartheta_\text{even}&=-1\cdot10^{-29}+6\cdot10^{-25}\;g^2-4\cdot10^{-21}\;g^4+9\cdot10^{-17}\;g^6-\notag\\
    &-2\cdot10^{-14}\;g^8+1\cdot10^{-11}g^{10}-1\cdot10^{-8}\;g^{12}+O(g^{14})\,,
    \end{align}
\begin{align}
    \mathbb{S}_\text{even}^{\text{(Fit)}}-\mathbb{S}_\text{even}&=-6\cdot10^{-24}+3\cdot10^{-19}\;g^2-2\cdot10^{-15}\;g^4+5\cdot10^{-12}\;g^6-\notag\\
    &-7\cdot10^{-9}\;g^8+6\cdot10^{-6}g^{10}+O(g^{12})\,.
\end{align}
2. Odd state:
\begin{align}
    \vartheta_\text{odd}^{\text{(Fit)}}-\vartheta_\text{odd}&=4\cdot10^{-38}-2\cdot10^{-30}\;g^2+5\cdot10^{-23}\;g^4-6\cdot10^{-16}\;g^6+\notag\\
    &+4\cdot10^{-9}\;g^8-2\cdot10^{-2}g^{10}+O(g^{12})\,,\\
\mathbb{S}_\text{odd}^{\text{(Fit)}}-\mathbb{S}_\text{odd}&=-8\cdot10^{-37}\;g^{-4}+5\cdot10^{-29}\;g^{-2}-1\cdot10^{-21}+\notag\\
    &+1\cdot10^{-14}\;g^2-1\cdot10^{-7}\;g^4+O(g^{6})\,.
\end{align}
The fit for the odd state is visibly less accurate than for the even state, due to the fact that $\mathbb{S}_{\text{odd}}$ diverges as $g^{-4}$ in the $g\to0$ limit. As a result, the fit had to be performed for numerically large values, reaching absolute magnitudes of order $10^{14}$.

\bibliographystyle{JHEP}


\providecommand{\href}[2]{#2}\begingroup\raggedright\endgroup

\end{document}